\documentclass[twocolumn,english,aps,prd,nofootinbib,hidelinks]{revtex4-2}
\usepackage{lmodern}
\usepackage[LGR,T1]{fontenc}
\usepackage[latin9]{inputenc}
\usepackage{xcolor}
\usepackage{babel}
\usepackage{varioref}
\usepackage{amsmath}
\usepackage{amssymb}
\usepackage{graphicx}
\usepackage[unicode=true]
 {hyperref}

\makeatletter

\DeclareRobustCommand{\greektext}{%
  \fontencoding{LGR}\selectfont\def\encodingdefault{LGR}}
\DeclareRobustCommand{\textgreek}[1]{\leavevmode{\greektext #1}}
\ProvideTextCommand{\~}{LGR}[1]{\char126#1}

\usepackage{enumerate}
\usepackage{slashed}

\makeatother

\begin{document}
\title{A mechanism to generate varying speed of light via Higgs-dilaton coupling:\vskip1pt
Theory and cosmological applications}
\author{Hoang Ky Nguyen$\,$}
\email[\ ]{hoang.nguyen@ubbcluj.ro}

\affiliation{{\vskip2pt}Department of Physics, Babe\c{s}-Bolyai University, Cluj-Napoca
400084, Romania\vskip1ptInstitute for Interdisciplinary Research
in Science and Education,~\linebreak{}
 ICISE, Quy Nhon 55121, Vietnam}
\date{April 9, 2025}
\begin{abstract}
We probe into a class of scale-invariant actions, which allow the
Higgs field $\Phi$ to interact with a dilaton field $\chi$ of the
background spacetime through the term $\chi^{2}\,\Phi^{\dagger}\Phi$.
Upon spontaneous gauge symmetry breaking, the vacuum expectation value
(VEV) of the Higgs field becomes proportional to $\chi$. Although
this linkage is traditionally employed to make the Planck mass and
particle masses dependent on $\chi$, we present an \emph{alternative}
mechanism: the Higgs VEV will be used to \emph{construct} Planck's
quantum of action $\hbar$ and speed of light $c$. Specifically,
each open set vicinity of a given point $x^{*}$ on the spacetime
manifold is equipped with a replica of the Glashow--Weinberg--Salam
action operating with \emph{its own effective values of $\hbar_{*}$
and $c_{*}$} per $\hbar_{*}\propto\chi^{-1/2}(x^{*})$ and $c_{*}\propto\chi^{1/2}(x^{*})$,
causing these ``fundamental constants'' to vary alongside the dynamical
field $\chi$. Moreover, in each open set around $x^{*}$, the prevailing
value $\chi(x^{*})$ determines the length and time scales for physical
processes occurring in this region as $l\propto\chi^{-1}(x^{*})$
and $\tau\propto\chi^{-3/2}(x^{*})$. This leads to an \emph{anisotropic}
relation $\tau^{-1}\propto l^{-3/2}$ between the rate of clocks and
the length of rods, resulting in a distinct set of novel physical
phenomena. For late-time cosmology, the variation of $c$ along the
trajectory of light waves from distant supernovae towards the Earth-based
observer necessitates modifications to the Lema\^itre redshift formula,
the Hubble law, and the luminosity distance--redshift relation. These
modifications are capable of: (1) Accounting for the Pantheon Catalog
of Type Ia supernovae \emph{through a declining speed of light in
an expanding Einstein--de Sitter universe}, thus avoiding the need
for dark energy; (2) Revitalizing Blanchard--Douspis--Rowan-Robinson--Sarkar's
CMB power spectrum analysis that bypassed dark energy {[}\textcolor{blue}{\href{https://arxiv.org/abs/astro-ph/0304237}{A\&A 412, 35 (2003)}}{]};
and (3) Resolving the $H_{0}$ tension without requiring a dynamical
dark energy component.
\end{abstract}
\maketitle

\section{Motivation}

In light of several outstanding issues and tensions in cosmology,
growing interest has emerged in promoting the parameters of physical
models to be variable scalar fields in spacetime \citep{Perivolaropoulos-2022,Bull,diValentino-2021,Secrest-2022}.
For example, the cosmological constant $\Lambda$ is equipped with
a kinetic term that allows it to evolve and `relax' to its (small)
value in the current epoch, thus potentially offering a relief to
the fine-tuning and coincidence problems in late-time cosmology \citep{Peccei-1987,Zhao-2017}.
\vskip4pt

Against this backdrop, the accustomed Planck $cGh$ unit system (1899-1900)
solidifies the role of the speed light $c$, the (Newton) gravitational
constant $G$, and the quantum of action $\hbar$ as ``fundamental
constants'' which encompass the realms of special relativity, gravitation,
and quantum mechanics, respectively \citep{Okun-1991,Okun-2004}.
The trio $\{c,\,G,\,\hbar\}$ are viewed as \emph{fixed} cornerstones,
convertible to the three ``fundamental units''---the Planck mass,
the Planck length, and the Planck time, defined as
\begin{equation}
M_{P}:=\sqrt{\frac{\hbar c}{G}};\ \ l_{P}:=\sqrt{\frac{\hbar G}{c^{3}}};\ \ \tau_{P}:=\sqrt{\frac{\hbar G}{c^{5}}}\label{eq:Planck-scales-def}
\end{equation}
These units span the basis for describing all physical phenomena known
to date.\vskip4pt

This orthodox view has faced challenges, however. The most well-known
example is the concept of variable $G$, inspired by Dirac in 1937
\citep{Dirac-LNH}, and culminating in the generally covariant Brans-Dicke
(BD) theory of gravity in 1961 \citep{BransDicke-1961}. In this theory,
a dynamical scalar degree of freedom $\chi$ is introduced alongside
the metric tensor $g_{\mu\nu}$ as
\begin{equation}
\int d^{4}x\,\frac{\sqrt{-g}}{16\pi}\left[\chi^{2}\,\mathcal{R}-4\omega\,\nabla_{\mu}\chi\nabla^{\mu}\chi\right]\label{eq:BD-action}
\end{equation}
where $\mathcal{R}$ is the Ricci scalar, and $\omega$ a dimensionless
(BD) parameter. The BD theory has grown into a family of scalar--tensor
theories, a very popular theme of research nowadays. The field $\chi$,
often referred to as a `dilaton', also naturally arises in string
theory and other contexts \citep{Fujii-book}. \vskip4pt

Comparing Eq. \eqref{eq:BD-action} with the Einstein--Hilbert (EH)
action
\begin{equation}
\mathcal{S}_{\text{EH}}=\int d^{4}x\,\frac{\sqrt{-g}}{16\pi}\frac{c^{3}}{\hbar\,G}\,\mathcal{R}
\end{equation}
where the speed of light $c$ and quantum of action $\hbar$ are explicitly
restored, the (Newton) gravitational `constant' in BD theory becomes
a scalar field via the following formal identification
\begin{equation}
\chi^{2}:=\frac{c^{3}}{\hbar\,G}\label{eq:a-1}
\end{equation}

Under the \emph{canonical assumption} that $c$ and $\hbar$ do not
depend on $\chi$, Eqs. \eqref{eq:Planck-scales-def} and \eqref{eq:a-1}
lead to the relationships
\begin{equation}
G=\frac{c^{3}}{\hbar}\chi^{-2};\ \ \ M_{P}=\frac{\hbar}{c}\chi;\ \ \ l_{P}=\chi^{-1};\ \ \ \tau_{P}=\frac{1}{c}\chi^{-1}\label{eq:Planck-scaling}
\end{equation}
The dilaton field $\chi$ hence directly determines the Planck length.\vskip4pt

It is imperative to stress that, in BD theory, the field $\chi$ itself
serves as a (dynamical) length scale that is used to construct $G$,
a \emph{dimensionful} quantity. However, the identity expressed in
Eq. \eqref{eq:a-1} permits an \emph{alternative} mechanism: The field
$\chi$, a quantity equal to $\left(c^{3}/(\hbar G)\right)^{1/2}$,
can also be used to \emph{construct $c$ and $\hbar$, rather than
$G$}. The key objective of our paper is to establish the relations
of $\hbar$ and $c$ with $\chi$ \emph{via the matter sector}.\vskip4pt

In the original BD theory, the matter sector coupled minimally to
the gravitational sector to preserve Einstein's Equivalence Principle
\citep{BransDicke-1961}. However, around the same time, Utiyama and
DeWitt showed that even if one starts with a minimal coupling action
between matter and gravity, radiative corrections would generate non-minimal
coupling between them \citep{Utiyama-1962}. Since the 1980s, efforts
to introduce non-minimal coupling directly at tree level have been
actively pursued in the literature. For example, in \citep{Wetterich-1988a,Wetterich-1988b,Wetterich-2013a,Wetterich-2013b,Wetterich-2014}
Wetterich embedded the BD action in a broader framework that projects
a dilatation symmetry, with the non-minimal coupling in the form \footnote{Wetterich referred the field $\chi$ a `cosmon' \citep{Wetterich-1988a}.
In place of Eq. \eqref{eq:Wetterich-potential}, he considered a more
general form $V(\chi,\Phi)=\Phi^{4}\,w\left(\chi^{2}/\Phi^{2}\right)$
where $w$ is a function of the ratio $x:=\chi^{2}/\Phi^{2}$. If
$w(x)$ adopts an affine form $\lambda-\mu^{2}\,x$, then Eq. \eqref{eq:Wetterich-potential}
would ensue.}
\begin{equation}
V(\chi,\Phi)=\lambda\,\Phi^{4}-\mu^{2}\,\chi^{2}\,\Phi^{2}\label{eq:Wetterich-potential}
\end{equation}
where $\lambda\in\mathbb{R^{+}}$ and $\mu\in\mathbb{R}$ are dimensionless
parameters. In \citep{Bezrukov:2011ghu}, a similar form was also
considered in the context of Higgs--dilaton cosmology. Preceding
to this, in \citep{Fujii-1982}, in place of $\chi^{2}$, Fujii let
the Higgs doublet couple with the Ricci scalar in the form $\mathcal{R}\,\Phi^{\dagger}\Phi$.\vskip4pt

In the presence of Higgs--dilaton coupling, when the electroweak
gauge symmetry of the action undergoes spontaneous breaking, the field
$\chi$ influences the vacuum expectation value (VEV) of the Higgs
doublet $\Phi$, which in turn affects the mass of the fermions, the
gauge vector bosons, and the Higgs bosons in the Glashow--Weinberg--Salam
(GWS) model. Specifically, Fujii and Wetterich allowed the particle
mass $m$ to scale proportional to $\chi$, thereby maintaining a
constant ratio $m/M_{P}$ at classical level. We shall refer to their
practice collectively as \emph{the Fujii--Wetterich (FW) scheme}.
Moreover, the length scale $l$ and the time scale $\tau$ of \emph{any
given physical process} also become dependent on $\chi$ as $l\propto\chi^{-1}$
and $\tau\propto\chi^{-1}$, in exact proportion to $l_{P}$ and $\tau_{P}$
respectively. Importantly, the field $\chi$ affects not only the
Planck time per Eq. \eqref{eq:Planck-scaling} but also the rate of
physical clocks (viz. the revolution rate of a mechanical clock, the
oscillation rate of an atomic clock, or the decay rate of an unstable
quantum system) which are made of matter governed by the GWS model.
Quantitatively, \emph{the FW scheme thus predicts that the evolution
rate $\tau^{-1}$ of physical processes is proportional to $\chi$.
}\footnote{\label{fn:quotes}We find it illuminating to quote Fujii \citep{Fujii-1982}:
\emph{``... the time and length in the microscopic unit frame are
measured in units of $m^{-1}(t)$, in agreement with the physical
situation that the time scale of atomic clocks, for example, is provided
by the atomic levels which are determined by the Rydberg constant
$(me^{4})^{-1}$''} and Wetterich \citep{Wetterich-2013a}: \emph{``The
clock provided by the Hubble expansion in the standard description
is now replaced by a clock associated to the increasing value of $\chi$''}.}\vskip4pt

As stated earlier, our approach constitutes a major departure from
the FW scheme: Instead of imposing that $\hbar$ and $c$ be constant,
we require the charge and inertial mass of particles to be independent
of $\chi$. This unambiguously leads to the relations $c\propto\chi^{1/2}$
and $\hbar\propto\chi^{-1/2}$, while $G$ remains constant. Furthermore,
via the time evolution of quantum states $i\hbar\frac{\partial}{\partial t}\left|\psi\right\rangle =\hat{H}\left|\psi\right\rangle $,
the dependency $\hbar\propto\chi^{-1/2}$, along with $\hat{H}\propto\chi$,
causes \emph{the evolution rate of physical processes to scale as
$\tau^{-1}\propto\chi^{3/2}$}, in decisive distinction from the FW
scheme (which posits $\tau^{-1}\propto\chi$).\vskip4pt

\emph{The anisotropic time scaling enabled in our approach, characterized
by its anomalous $3/2$--exponent, leads to new physics with a distinct
set of phenomenology and\linebreak predictions. }We will discuss
these significant implications in this paper, with further details
developed in Ref. \citep{VSL2024-Pantheon}.\vskip12pt

\subsection*{In brief: $\:$History of variable $\fontsize{11pt}{0pt}\selectfont{\bf\textit{c}}$
and $\fontsize{11pt}{0pt}\selectfont{\bf\textit{$\hbar$}}$ }

Prior to the development of Brans-Dicke theory, Dicke also briefly
considered variability in the speed of light $c$ in 1957 \citep{Dicke-1957}
although he did not pursue it further. A relatively little-known fact
is that the concept of variable speed of light (VSL) was initially
explored by Einstein in 1911, published in three obscure papers (originally
in German) \citep{Einstein-1911,Einstein-1912-a,Einstein-1912-b}
\footnote{It seems that Dicke in 1957 \citep{Dicke-1957} was not aware of Einstein's
1911 VSL proposal \citep{Einstein-1911}.} during his quest for a generally covariant theory of gravity. In
\citep{Einstein-1912-a,Einstein-1912-b} Einstein emphasized that
the Michelson--Morley experimental results and Lorentz symmetry are
meant to hold only \emph{locally}; while $c$ is invariant with respect
to local boosts, its value needs \emph{not} be universal. Einstein
envisioned the possibility that the gravitational potential affected
the clock rate, hence generating a varying $c$ in spacetime. Such
a variation in $c$ could then produce a refraction effect on light
rays around the Sun's disc as a result of Huygens's principle \citep{Einstein-1911}.
The later successes of General Relativity (GR) quickly overshadowed
Einstein's 1911 paper on VSL, leading him to neglect the idea, however.
In the 1990s, the VSL concept was revived independently by Moffat
and by Albrecht and Magueijo to tackle issues in early-time cosmology,
such as the horizon paradox \citep{Moffat-1992,Albrecht-1999}. Since
then, several researchers actively explore various aspects of VSL
\citep{Zhang-2014,Qi-2014,Ravanpak-2017,Barrow-2000,Bassett-2000,Drummond-1980,Drummond-1999,Liberati-2000,Magueijo-2002,Magueijo-2003,Magueijo-2008,Novello-1989,Volovik-2023,Ellis-2005,Ellis-2007,Zou-2018,Zhang-2024,Wang-2019,Uzan-2011,Uzan-2003,Santos-2024,Salzano-2017a,Salzano-2016b,Salzano-2016a,Salzano-2015,Rodrigues-2022,Pan-2020,Mukherjee-2024,Moffat-2016,Mendonca-2021,Martins-2017,Magueijo-2000,Liu-2023,Liu-2021,Lee-2023b,Lee-2023a,Lee-2021b,Lee-2021a,Lang-2018,HWAC-2018,HAWC-2020,Gupta-2021,Gupta-2020,Franzmann-2017,Eaves-2022,Cuzinatto-2023,Cuzinatto-2022,Cruz-2018,Cruz-2012,Costa-2019,Colaco-2022,Clayton-2002,Clayton-2000,Clayton-1999,Cao-2018,Cao-2017,Cai-2016,Barrow-1999b,Barrow-1999a,Barrow-1998b,Barrow-1998a,Balcerzak-2014b,Balcerzak-2014a,Avelino-1999,Agrawal-2021,Abdo-2009,Salzano-2016c,Liu-2018,Xu-2016a,Xu-2016b,Zhu-2021,Avelino-2000,Salzano-2016d,Buchalter-2004}.\vskip4pt

The possibility of a varying quantum of action $\hbar$ has been much
less explored, with a rare exception of \citep{Mangano-2015}. One
important empirical guidance is that, although the fine-structure
constant $\alpha:=e^{2}/(\hbar c)$ is known to `run' in the renormalization
group (RG) flow of quantum field theory, there is no clear experimental
evidence supporting $\alpha$ varying in spacetime \footnote{In \citep{Webb-1999,Webb-2001} Webb et al. reported observational
evidence of a varying $\alpha$ with respect to the redshift, but
this result has been in contention; e.g., see a recent paper \citep{Jiang-2024}.
In addition, it violates the tight theoretical bound given in \citep{Banks-2002}.}. If $\alpha$ is to remain independent of the field $\chi$ while
$c$ is allowed to vary, it would require a co-variation of $e$,
$\hbar$, and $c$. One such scenario, as briefly mentioned earlier,
posits $\hbar\propto\chi^{-1/2}$ and $c\propto\chi^{1/2}$, which
would insulate $\alpha$ from depending on $\chi$.

\section{\label{sec:Higgs=002013dilaton-coupling}Higgs--dilaton coupling}

Consider the full action in a 4-dimentional spacetime:
\begin{align}
\mathcal{S} & =\int d^{4}x\,\sqrt{-g}\,\mathcal{L}_{\text{mat}}+\int d^{4}x\,\frac{\sqrt{-g}}{16\pi}\,\mathcal{L}_{\text{grav}}\label{eq:full-action}
\end{align}
Let us first focus on the matter sector. As a prototype, we consider
the following Lagrangian, representing a massless spinor field $\psi$
coupled with a massless $U(1)$ gauge field $A_{\mu}$, and a charge-neutral
Higgs singlet $\Phi$ coupled with a dilaton $\chi$:
\begin{align}
\mathcal{L}_{\text{mat}} & =i\,\bar{\psi}\gamma^{\mu}(\nabla_{\mu}-i\sqrt{\alpha}A_{\mu})\psi+f\,\bar{\psi}\psi\,\Phi\nonumber \\
 & +\frac{\mu^{2}}{2}\,\chi^{2}\,\Phi^{2}-\frac{\lambda}{4}\,\Phi^{4}+g^{\mu\nu}\partial_{\mu}\Phi\,\partial_{\nu}\Phi-\frac{1}{4}\,F_{\mu\nu}F^{\mu\nu}\label{eq:L-mat}
\end{align}
\vskip4pt

All parameters $\alpha$, $f$, $\mu$, and $\lambda$ are dimensionless.
It is important to remark that the \emph{adimensional} action of the
matter sector, $\int d^{4}x\sqrt{-g}\,\mathcal{L}_{\text{mat}}$,
does \emph{not} involve the (dimensionful) quantum of action $\hbar$
and speed of light $c$ at its outset. The length dimensions of $\{\psi,\,A_{\mu},\,\Phi$\}
are $\{-3/2,-1,-1\}$ respectively. The gamma matrices satisfy $\gamma^{\mu}\gamma^{\nu}+\gamma^{\nu}\gamma^{\mu}=2g^{\mu\nu}$,
and the spacetime covariant derivative $\nabla_{\mu}$ acts on the
spinor via vierbein and spin connection.\vskip4pt

The Lagrangian $\mathcal{L}_{\text{mat}}$ in Eq. \eqref{eq:L-mat}
is invariant under the local $U(1)$ gauge transformation:
\begin{align}
\psi(x) & \rightarrow e^{i\sqrt{\alpha}\sigma(x)}\,\psi(x)\label{eq:gauge-1}\\
A_{\mu}(x) & \rightarrow A_{\mu}(x)-\partial_{\mu}\sigma(x)\label{eq:gauge-2}
\end{align}
where $\sigma(x)$ is an arbitrary scalar function. In addition, $\mathcal{L}_{\text{mat}}$
is also invariant under the discrete $\mathbb{Z}_{2}$ symmetry of
the Higgs singlet
\begin{equation}
\Phi\leftrightarrow-\Phi\label{eq:Z2}
\end{equation}
\vskip4pt

Regarding the gravitation Lagrangian $\mathcal{L}_{\text{grav}}$,
most extended theories of gravitation naturally host a scalar degree
of freedom \citep{Clifton-review,Capo-review,Odintsov-review,Fujii-book,Bohmer-review,CANTANA-2021}.
Several possibilities exist, one being a massive Brans-Dicke action
\begin{equation}
\mathcal{L}_{\text{grav}}=\chi^{2}\,\mathcal{R}-4\omega\,\nabla^{\mu}\chi\nabla_{\mu}\chi-V(\chi)\label{eq:L-grav}
\end{equation}
When $\omega=0$, $\mathcal{L}_{\text{grav}}$ in Eq. \eqref{eq:L-grav}
belongs to the $f(\mathcal{R})$ family of gravity \citep{Buchdahl-1970,deFelice-review,Sotiriou-review}.
It can also be embedded within a broader family of scale invariant
gravity, such as those considered in \citep{Blas-2011,Ghilencea-2023,Ferreira-2018}
or ``agravity'' which includes the Weyl term, $\mathcal{C}^{\mu\nu\lambda\rho}\,\mathcal{C}_{\mu\nu\lambda\rho}$
\citep{Einhorn-2015,Salvio-2014,Edery-2014}. Alternatively, it can
represent Horndeski gravity, the most general theory in four dimensions
constructed out of the metric tensor and a scalar field that leads
to second-order equations of motion \citep{Horndeski-1974}. Another
viable option would be Lyra geometry \citep{Cuzinatto-2021}, where
the inherent scale function of this geometry can play the role of
the dilaton field $\chi$. However, the exact details of the gravitational
sector are \emph{not} the focus of our paper. Our work concerns \emph{the
matter sector}, and we require only the existence of a `dilaton' singlet
$\chi$ that couples with the Higgs field $\Phi$ as described in
Eq. \eqref{eq:L-mat}.\vskip4pt

It has been established in \citep{Ferreira-2016,Blas-2011} that a
scale-invariant action, such as the one described in Eqs. \eqref{eq:full-action},
\eqref{eq:L-mat}, and \eqref{eq:L-grav}, can evade observational
constraints on the fifth force. Among the several possible choices
for $\mathcal{L}_{\text{grav}}$, the pure $\mathcal{R}^{2}$ theory
$\mathcal{L}_{\text{R2}}=\chi^{2}\,\mathcal{R}-f_{0}\,\chi^{4}$ is
a promising candidate \citep{Alvarez-2018,Donoghue-2021,Salvio-2018,AlvarezGaume-2016}.
At the classical level, the dilaton and the Ricci scalar are one-to-one
related as $\mathcal{R}=2f_{0}\,\chi^{2}$. The Newtonian limit has
been established \citep{Nguyen-Newtonian,AlvarezGaume-2016}. The
theory is known to be Ostrogradsky stable and free of ghosts \citep{AlvarezGaume-2016},
and properties of its graviton propagators have also been investigated
\citep{Karananas-2024,Alvarez-2018,AlvarezGaume-2016}.

\section{\label{sec:A-new-mechanism}A new mechanism to generate variable
$\fontsize{12pt}{0pt}\selectfont{\textit{$\hbar$}}$ and $\fontsize{12pt}{0pt}\selectfont{\textit{c}}$ }

Assuming that $\chi$ is a slowly varying background field, at a given
point $x^{*}$ on the manifold, its value is $\chi_{*}:=\chi(x^{*})$.
In the open set vicinity of the point $x^{*}$, the terms $\frac{\mu^{2}}{2}\,\chi_{*}^{2}\,\Phi^{2}-\frac{\lambda}{4}\,\Phi^{4}$
in Eq. \eqref{eq:L-mat}, evaluated for the prevailing $\chi_{*}$,
induce a spontaneous breaking of the \emph{discrete $\mathbb{Z}_{2}$
symmetry} of the Higgs singlet, given by Eq. \eqref{eq:Z2} \footnote{Note: The $U(1)$ gauge symmetry described by Eqs. \eqref{eq:gauge-1}--\eqref{eq:gauge-2}
remains unbroken in this process.}.\linebreak This process, traditionally known as the Higgs mechanism
\citep{SSB-Anderson,SSB-Englert,SSB-Guralnik,SSB-Higgs}, results
in a non-zero vacuum expectation value (VEV) of the Higgs field
\begin{equation}
\left\langle \Phi(x^{*})\right\rangle =\frac{\mu}{\sqrt{\lambda}}\chi_{*}\label{eq:Higgs-VEV}
\end{equation}
which is proportional to $\chi_{*}$. In a local reference frame tangent
to the manifold at the point $x^{*}$ (i.e, an inertial frame at $x^{*}$),
the action for the matter sector (excluding the Higgs excitation above
the vacuum) becomes \footnote{If we adopt the analysis presented in Ref. \citep{Gorbar-2003}, the
Higgs VEV $v(x):=\left\langle \Phi(x)\right\rangle $ would obey the
equation $-2\square v+\mu^{2}\chi^{2}v-\lambda v^{3}=0$. However,
we treat $v$ as a slowly varying background field; under this condition,
the projection of the matter sector onto the local inertial frame,
viz. Eq. \eqref{eq:S-mat-eff}, is justified.}
\begin{equation}
\int d^{4}x\,\biggl[i\bar{\psi}\gamma^{\mu}\partial_{\mu}\psi+\sqrt{\alpha}\,\bar{\psi}\gamma^{\mu}A_{\mu}\psi+\frac{f\mu}{\sqrt{\lambda}}\,\chi_{*}\,\bar{\psi}\psi-\frac{1}{4}F_{\mu\nu}F^{\mu\nu}\biggr]\label{eq:S-mat-eff}
\end{equation}
Varying the spinor field and the gauge vector field, we obtain two
equations of motion
\begin{align}
\biggl(i\gamma^{\mu}\partial_{\mu}+\sqrt{\alpha}\,\gamma^{\mu}A_{\mu}+\frac{f\mu}{\sqrt{\lambda}}\,\chi_{*}\biggr)\,\psi & =0\label{eq:my-Dirac}
\end{align}
and
\begin{equation}
\partial_{\nu}F^{\nu\mu}=j^{\mu}:=\sqrt{\alpha}\,\bar{\psi}\gamma^{\mu}\psi\label{eq:my-Maxwell}
\end{equation}
The first equation resembles the Dirac equation for the `electron'
$\psi$, coupled with a $U(1)$ gauge vector $A_{\mu}$
\begin{equation}
\biggl(i\gamma^{\mu}\partial_{\mu}+\frac{e}{\sqrt{\hbar c}}\,\gamma^{\mu}A_{\mu}+m\,\frac{c}{\hbar}\biggr)\,\psi=0\label{eq:Dirac-original}
\end{equation}
The second equation resembles the Maxwell equation for $A_{\mu}$
sourced by a $U(1)$--charged current $j^{\mu}$
\begin{equation}
\partial_{\nu}F^{\nu\mu}=j^{\mu}:=\frac{e}{\sqrt{\hbar c}}\,\bar{\psi}\gamma^{\mu}\psi\label{eq:Maxwell-original}
\end{equation}

In Eq. \eqref{eq:Dirac-original}, the parameters $e$ and $m$ represent
the $U(1)$ gauge charge and inertial mass of the electron, respectively.
As they are \emph{intrinsic} properties of the electron, we then require
them to be \emph{parameters rather than fields}, namely, they are
independent of $\chi_{*}$. (Note: We must emphasize that $e$ and
$m$ are not constants, but they can `run' in the renormalization
group flow when radiative corrections involving $\psi$ and $A_{\mu}$
are included.) By comparing Eqs. \eqref{eq:my-Dirac}--\eqref{eq:my-Maxwell}
against Eqs. \eqref{eq:Dirac-original}--\eqref{eq:Maxwell-original},
we can identify
\begin{equation}
m:=\frac{f\mu}{\sqrt{\lambda}};\ \ \ e:=\sqrt{\alpha}\label{eq:m-e-def}
\end{equation}
These identities then lead to
\begin{equation}
\hbar\,c:=1;\ \ \ \frac{c}{\hbar}:=\chi_{*}
\end{equation}
which unambiguously yield the following relations
\begin{equation}
\hbar_{*}:=\chi_{*}^{-1/2};\ \ \ c_{*}:=\chi_{*}^{1/2}\label{eq:scalings}
\end{equation}
Here, the subscript $*$ in $\hbar_{*}$ and $c_{*}$ signifies the
dependence of $\hbar$ and $c$ as functions of $\chi$. It should
be noted that Newton's constant in independent of $\chi_{*}$ because
$G:=\frac{c_{*}^{3}}{\hbar_{*}\chi_{*}^{2}}=1$.\vskip4pt

Conversely, in the tangent---i.e., inertial---frame at the point
$x^{*}$, the action in Eq. \eqref{eq:S-mat-eff} can be recast as
an action of Quantum Electrodynamics (QED) with a Lagrangian expressed
in terms of $\hbar_{*}$, $c_{*}$, $m$, and $e$, viz.
\begin{equation}
\mathcal{L}_{\text{QED}}=i\bar{\psi}\gamma^{\mu}\partial_{\mu}\psi+\frac{e}{\sqrt{\hbar_{*}c_{*}}}\bar{\psi}\gamma^{\mu}A_{\mu}\psi+m\frac{c_{*}}{\hbar_{*}}\bar{\psi}\psi-\frac{1}{4}F_{\mu\nu}F^{\mu\nu}\label{eq:QED-Lagrangian}
\end{equation}
Therefore, \emph{each open set enclosing a given point $x^{*}$ on
the manifold is equipped with a replica of the QED action} \eqref{eq:QED-Lagrangian},
operating with an \emph{effective} Planck constant $h_{*}$ and an
\emph{effective} speed of light $c_{*}$. Both of these effective
parameters are determined by the prevailing value $\chi_{*}$ of the
\emph{background} dilaton field, per Eq. \eqref{eq:scalings}. As
a component of the gravitational sector, $\chi_{*}$ can vary across
the manifold, leading to corresponding variations in $\hbar_{*}$
and $c_{*}$ throughout the manifold. \vskip4pt

Within each open set, the effective speed of light $c_{*}$ governs
the propagation of the (massless \footnote{Note: Only the discrete $\mathbb{Z}_{2}$ symmetry \eqref{eq:Z2}
of the Higgs singlet is broken. The local $U(1)$ gauge symmetry in
Eqs. \eqref{eq:gauge-1} and \eqref{eq:gauge-2} remains unbroken,
leaving the gauge vector boson $A_{\mu}$ massless.}) gauge vector field $A_{\mu}$, whereas the effective Planck constant
$\hbar_{*}$ regulates the quantization of the fields $\psi$, $A_{\mu}$,
and $\Phi$. We should note that our mechanism for generating variable
$\hbar$ and $c$ is readily applicable to the generalization of $\mathcal{L}_{\text{mat}}$
in Eq. \eqref{eq:L-mat} to the full GWS model of particle physics,
where the gauge group is enlarged to $SU(3)\times SU(2)\times U(1)$.
This extension is a relatively straightforward exercise. A simplified
version of our mechanism is also provided in Ref. \citep{VSL2024-dilaton}
where we allowed the dilaton to interact directly with the fermion
field in the form $\chi\,\bar{\psi}\psi$. We obtained results that
are essentially identical to those presented above.\vskip4pt

\emph{The role of $\hbar_{*}$ is instrumental:} from Eq. \eqref{eq:my-Dirac}
(and using $\hbar_{*}c_{*}=1$ per Eq. \eqref{eq:scalings}), it can
be shown that the time evolution of the electron wavefunction is given
by
\begin{align}
i\hbar_{*}\frac{\partial}{\partial t}\psi(t) & =\hat{H}(t)\,\psi(t)\label{eq:time-op-1}\\
\hat{H}(t) & :=-i\alpha^{k}\frac{\partial}{\partial x^{k}}+\sqrt{\alpha}\left(\alpha^{k}A_{k}-\beta A_{0}\right)+\frac{f\mu}{\sqrt{\lambda}}\,\chi_{*}\,\beta\label{eq:time-op-2}
\end{align}
Here, the matrices are $\alpha^{k}:=\gamma^{0}\gamma^{k}$ and $\beta:=\gamma^{0}$.
The time evolution of $\psi(t)$ within the open set enclosing $x^{*}$
thus depends on the behavior of $\hbar_{*}$ with regard to $\chi_{*}$.
This crucial connection between $\hbar_{*}$ and the evolution rate
of quantum states leads to a concrete prediction, which we shall elaborate
in the subsequent sections.

\section{A prediction: $\ $Anisotropic scaling in the clock rate}

The effective QED action given in Eq. \eqref{eq:S-mat-eff} possesses
a dilatation symmetry; namely, it remains invariant under the transformation
\begin{equation}
dx^{\mu}\rightarrow\chi_{*}\,dx^{\mu};\ \psi\rightarrow\chi_{*}^{-3/2}\psi;\ A_{\mu}\rightarrow\chi_{*}^{-1}A_{\mu}
\end{equation}
The prevailing value of the dilaton field $\chi$ at the point $x^{*}$
thus sets the length scale for a given physical process within the
open set surrounding $x^{*}$, i.e.
\begin{equation}
l\propto\chi_{*}^{-1}\label{eq:length-scaling}
\end{equation}
justifying the term ``dilaton'' for $\chi$. However, due to the dependence
of $c$ on $\chi$ as given in Eq. \eqref{eq:scalings}, the timescale
for the physical process in the open set exhibits an anisotropic behavior
\begin{equation}
\tau:=\frac{\chi_{*}^{-1}}{c_{*}}\propto\chi_{*}^{-3/2}\label{eq:tau-scaling}
\end{equation}
This behavior can also be understood through the time evolution operator
\eqref{eq:time-op-1}--\eqref{eq:time-op-2}. Since $dx^{k}\propto\chi_{*}^{-1}$
and $A_{\mu}\propto\chi_{*}$, the Hamiltonian in \eqref{eq:time-op-2}
scales as $\chi_{*}$. The rate of evolution is thus
\begin{equation}
\tau\simeq\frac{\hbar_{*}}{\hat{H}}\propto\chi_{*}^{-3/2}
\end{equation}
which is compatible with the anomalous time scaling found in Eq. \eqref{eq:tau-scaling}.\vskip4pt

To illustrate the behavior of the length and time scales with respect
to $\chi_{*}$, we will consider the Hydrogen atom as an example.
The Bohr radius is
\begin{align}
a_{B} & =\frac{\hbar_{*}}{\alpha\,mc_{*}}=\frac{\sqrt{\lambda}}{\alpha f\mu}\,\chi_{*}^{-1}\propto\chi_{*}^{-1}
\end{align}
which is consistent with Eq. \eqref{eq:length-scaling}. The energy
level of (relativistic) electron in a quantum state $\left|n,\,j\right\rangle $
of a hydrogen atom is a well-established result \citep{Greiner-RQM-book}
\begin{align}
E_{n}^{j} & =\mathcal{N}_{n}^{j}\,mc_{*}^{2}=\mathcal{N}_{n}^{j}\,\frac{f\mu}{\sqrt{\lambda}}\,\chi_{*}
\end{align}
in which $\mathcal{N}_{n}^{j}\!:=\!{\scriptstyle {\scriptstyle \left(1+\alpha^{2}\left(n-j-\frac{1}{2}+\sqrt{\left(j+\frac{1}{2}\right)^{2}-\alpha^{2}}\right)^{-2}\right)^{-1/2}}}$
with $n=1,2,3,...$ and $j=\frac{1}{2},\frac{3}{2},\frac{5}{2},...$.
Note that energy is proportional to $\chi_{*}$. The groundstate $\left|n=1,\,j=1/2\right\rangle $
and the excited state $\left|n=2,\,j=3/2\right\rangle $ have energy
levels
\begin{align}
E_{n=1}^{j=1/2} & =\sqrt{1-\alpha^{2}}\,\frac{f\mu}{\sqrt{\lambda}}\,\chi_{*}\\
E_{n=2}^{j=3/2} & =\sqrt{1-\text{\scriptsize\ensuremath{\frac{1}{4}}}\alpha^{2}}\,\frac{f\mu}{\sqrt{\lambda}}\,\chi_{*}
\end{align}
A transition from the (initial) excited state $\left|i\right\rangle =\left|n=2,\,j=3/2\right\rangle $
to the (final) groundstate $\left|f\right\rangle =\left|n=1,\,j=1/2\right\rangle $,
induced by electric dipole, is allowed as it satisfies the selection
rule $\Delta j=\pm1$. The energy of the photon emitted is $E_{n=2}^{j=3/2}-E_{n=1}^{j=1/2}$,
and the frequency of the emitted photon is
\begin{align}
\nu & =\frac{E_{n=2}^{j=3/2}-E_{n=1}^{j=1/2}}{2\pi\,\hbar_{*}}\label{eq:nu-1}\\
 & =\frac{f\mu}{2\pi\sqrt{\lambda}}\Bigl(\sqrt{1-\text{\scriptsize\ensuremath{\frac{1}{4}}}\alpha^{2}}-\sqrt{1-\alpha^{2}}\Bigr)\,\chi_{*}^{3/2}\label{eq:nu-2}
\end{align}
Therefore, the propagation of the photon has the time scale that behaves
as
\begin{equation}
\tau:=\frac{1}{\nu}\propto\chi_{*}^{-3/2}\label{eq:tau-photon}
\end{equation}
which is in perfect agreement with Eq. \eqref{eq:tau-scaling}.\vskip4pt

The time scaling \eqref{eq:tau-scaling} implies that the evolution
rate of a clock---regardless of whether it is a mechanical clock,
an electronic clock, or an atomic clock---varies in spacetime, as
a function of $\chi$, in an \emph{anisotropic} fashion. In principle,
this effect can be measured experimentally: Prepare two identical
clocks at a location $A$. Keep one clock at location $A$ and send
the second clock to a location $B$. Suppose that the background dilaton
field has different values $\chi_{A}$ and $\chi_{B}$ at the two
locations $A$ and $B$, respectively. At their respective locations,
the clocks would run at different rates per
\begin{equation}
\tau_{A}\propto\chi_{A}^{-3/2};\ \ \ \tau_{B}\propto\chi_{B}^{-3/2}\label{eq:two-clocks}
\end{equation}
When the clock from location $B$ is brought back to location $A$,
it will show \emph{a different elapsed time} compared to the clock
that resided at location $A$ during the whole experiment.\vskip4pt

We emphasize that this \emph{predicted} effect is physical, meaning
it is, in principle, \emph{measurable by comparing the time lapses
of two clocks situated at two separate locations with different values
of the dilaton field}. This time dilation effect differs from the
time dilation effect in GR which is associated with the $g_{00}$
component of the spacetime metric, arises from the dependence of the
clock rate on the dilaton field $\chi$, viz. Eq. \eqref{eq:two-clocks}.
This new phenomenon is distinct and shall be referred to as a \emph{``time
dilation effect of the Third kind''}, to be investigated in future
work. \footnote{In addition to the time dilation effect in GR, there is a well-known
effect in Special Relativity where two twice-intersecting time-like
paths can have different total amounts of proper time in between.
Therefore, we refer our predicted time dilation effect as ``the Third
kind''.}

\subsection*{Decay rate of unstable quantum systems}

The prediction that we have just made can be validated via a different
setup. Reconsider the Hydrogen atom. Induced by electric dipole perturbation
in the vacuum, the electron in the excited state $\left|n=2,\,j=3/2\right\rangle $
can spontaneously transition to the groundstate $\left|n=1,\,j=1/2\right\rangle $
(allowed by the selection rule $\Delta j=\pm1$). According to quantum
mechanics \citep{QM-book}, the decay rate (i.e. Einstein's coefficient)
is known to be 
\begin{equation}
A=\frac{4}{3}\frac{\omega_{if}^{3}e^{2}\left\langle \vec{r}_{if}\right\rangle ^{2}}{\hbar_{*}c_{*}^{3}}
\end{equation}
where the effective $\hbar_{*}$ and $c_{*}$ are made explicit. The
angular frequency (with $h_{*}:=2\pi\hbar_{*}$) is
\begin{equation}
\omega_{if}=\frac{1}{h_{*}}\left(E_{n=2}^{j=3/2}-E_{n=1}^{j=1/2}\right)
\end{equation}
and the matrix element of the electric dipole between the initial
state $\left|i\right\rangle :=\left|n=2,\,j=3/2\right\rangle $ and
the final state $\left|f\right\rangle :=\left|n=1,\,j=1/2\right\rangle $
is $\left\langle \vec{r}_{if}\right\rangle :=\left\langle f|\vec{r}|i\right\rangle $.
Using $e^{2}=\alpha\,\hbar_{*}c_{*}$, we get
\begin{equation}
A=\frac{4\alpha}{3}\frac{m^{3}c_{*}^{4}}{h_{*}^{3}}\left\langle \vec{r}_{if}\right\rangle ^{2}\label{eq:decay-1}
\end{equation}
Given that $\left\langle \vec{r}_{if}\right\rangle \propto\chi_{*}^{-1}$,
$c_{*}\propto\chi_{*}^{1/2}$, $h_{*}\propto\chi_{*}^{-1/2}$, we
find
\begin{equation}
A\propto\frac{\chi_{*}^{2}}{\chi_{*}^{-3/2}}\,\chi_{*}^{-2}\propto\chi_{*}^{3/2}\label{eq:decay-2}
\end{equation}
Thus, the lifetime of the decay process for this unstable quantum
system behaves as
\begin{equation}
\tau:=\frac{1}{A}\propto\chi_{*}^{-3/2}
\end{equation}
which perfectly conforms with Eq. \eqref{eq:tau-scaling}. Consequently,
the clock involved in our experiment proposed earlier in this section
can be \emph{any} generic timekeeping device. It can be a mechanical
clock, an electronic clock, an atomic clock, or even a radioactive
quantum system. \footnote{The Hafele--Keating experiment \citep{HafeleKeating-1972-a,HafeleKeating-1972-b}
was carried out using atomic clocks.}

\section{Comparison of our {\fontsize{11pt}{0pt}\selectfont{VSL}} scheme
with the Fujii--Wetterich scheme}

We must stress that the dependence of the clock rate on $\chi_{*}$
has been \emph{documented} in the works of Fujii and Wetterich \citep{Fujii-1982,Wetterich-1988a,Wetterich-1988b,Wetterich-2013a,Wetterich-2013b,Wetterich-2014},
although it was not a focal point in their analysis; see Footnote
\vref{fn:quotes}. However, their findings diverge from ours, particularly
in how the clock rate scales with $\chi_{*}$. Specifically, they
predicted a relation
\begin{equation}
\tau_{\text{FW}}^{-1}\propto\chi_{*}
\end{equation}
where ``FW'' refers to Fujii--Wetterich. This contrasts with our
prediction $\tau^{-1}\propto\chi_{*}^{3/2}$, given in Eq. \eqref{eq:tau-scaling}.\vskip4pt

The distinction arises due to a different set of conditions used in
Fujii and Wetterich's treatments \citep{Fujii-1982,Wetterich-1988a,Wetterich-1988b,Wetterich-2013a,Wetterich-2013b,Wetterich-2014}.
These authors \emph{judiciously} kept $\hbar$ and $c$ independent
of $\chi_{*}$, setting $\hbar_{\text{FW}}=1$ and $c_{\text{FW}}=1$.
Instead, they allowed the electron mass $m$ to be a \emph{field},
while the electron charge $e$ remained a constant. With regard to
Eqs. \eqref{eq:my-Dirac}--\eqref{eq:my-Maxwell} and Eqs. \eqref{eq:Dirac-original}--\eqref{eq:Maxwell-original},
their conditions then produce the following identities
\begin{equation}
m_{\text{FW}}:=\frac{f\mu}{\sqrt{\lambda}}\,\chi_{*};\ \ \ e:=\sqrt{\alpha}\label{eq:FW-def}
\end{equation}
Consequently, Newton's ``constant'' varies as $G_{\text{FW}}:=\frac{c_{\text{FW}}^{3}}{\hbar_{\text{FW}}\chi_{*}^{2}}=\chi_{*}^{-2}$.
Instead of Eq. \eqref{eq:time-op-1}, the time evolution of the electron
wavefunction in the FW scheme is governed by
\begin{align}
i\,\hbar_{FW}\,\frac{\partial}{\partial t}\psi(t) & =\hat{H}(t)\,\psi(t)
\end{align}
with the Hamiltonian $\hat{H}(t)$ given in Eq. \eqref{eq:time-op-2}.
Here\emph{, $\hbar_{FW}$ is constant} (equal to $1$). Therefore,
the time scale for evolution in the FW scheme scales as
\begin{equation}
\tau_{\text{FW}}\simeq\frac{\hbar_{\text{FW}}}{\hat{H}}\propto\chi_{*}^{-1}
\end{equation}
This result can be verified with the concrete examples we used previously.
In the FW scheme, the frequency of emitted photon from the excited
state $\left|n=2,\,j=3/2\right\rangle $ to the groundstate $\left|n=1,\,j=1/2\right\rangle $
is
\begin{align}
\nu_{\text{FW}} & =\frac{E_{n=2}^{j=3/2}-E_{n=1}^{j=1/2}}{2\pi\,\hbar_{\text{FW}}}\\
 & =\frac{f\mu}{2\pi\sqrt{\lambda}}\Bigl(\sqrt{1-\text{\scriptsize\ensuremath{\frac{1}{4}}}\alpha^{2}}-\sqrt{1-\alpha^{2}}\Bigr)\,\chi_{*}
\end{align}
instead of the result \eqref{eq:nu-2} of our scheme. Likewise, in
their scheme, the decay rate of spontaneous emission from the excited
state $\left|n=2,\,j=3/2\right\rangle $ to the groundstate $\left|n=1,\,j=1/2\right\rangle $
reads
\begin{equation}
A=\frac{4\alpha}{3}\frac{m_{\text{FW}}^{3}\,c_{\text{FW}}^{4}}{h_{\text{FW}}^{3}}\left\langle \vec{r}_{if}\right\rangle ^{2}\propto\frac{\chi_{*}^{3}.1^{4}}{1^{3}}\,\chi_{*}^{-2}\propto\chi_{*}
\end{equation}
instead of Eqs. \eqref{eq:decay-1}--\eqref{eq:decay-2} of our scheme.\vskip4pt

Therefore, despite having an \emph{identical} matter action, Eq. \eqref{eq:S-mat-eff},
our scheme and the FW scheme are \emph{not physically equivalent}.
They result in decisively \emph{different} predictions for the behavior
of the clock rate. Future technologies may be able to distinguish
the two predictions, and hence the validity of each scheme.

\subsection*{Properties of our mechanism}

Our approach offers two distinct benefits:\vskip8pt

1. The FW scheme treats inertial mass and gauge charge at \emph{disparity}:
while mass is promoted to fields dependent on $\chi_{*}$, charge
remains as a fixed parameter (see Eq. \eqref{eq:FW-def}). Given that
inertial mass and gauge charge are \emph{intrinsic} properties of
particles, our scheme treats them on \emph{equal footing}: particle
mass and charge are parameters but not fields. (Note: in our scheme,
$m$ and $e$ can `run' in the RG flow, but they are independent of
the background dilaton $\chi$.)\vskip8pt

2. The FW scheme only applies to massive particles, leaving massless
particles unaffected. In contrast, the varying $c$ and $\hbar$ in
our scheme impact \emph{all} particles---massive and massless alike.
Specifically, the variation in $c$ influences the propagation of
(massless) photon in an expanding intergalactic space in cosmology,
a feature absent in the FW scheme.\vskip12pt

In conclusion of our derivation, the $3/2$--exponent in our time
scaling \eqref{eq:tau-scaling} is a novel discovery, leading to \emph{new
physics} with a unique set of previously unexplored phenomena and
predictions. Specifically, it induces the variability of the speed
of light in spacetime, causing lightwaves traveling through the expanding
universe to undergo an additional refraction effect. This effect thence
necessitates a reanalysis of the Pantheon Catalog of Type Ia supernovae,
a task we carry out in the next section.

\section{\label{sec:Phenomenology}Phenomenology of {\fontsize{11pt}{0pt}\selectfont{VSL}}}

This section focuses on our application of variable speed of light
(VSL) for late-time cosmology, bypassing the need for dark energy
(DE). The detailed work containing full technicality is presented
in Refs. \citep{VSL2024-Pantheon}. \vskip12pt

\subsection{\label{subsec:VSL-cosmography-and}{\fontsize{10pt}{0pt}\selectfont{VSL}}
cosmology and late-time cosmography}

Our starting point is to generalize the Friedmann--Lema\^itre--Robertson--Walker
(FLRW) metric for the cosmic scale factor (with $\kappa=\{1,0,-1\}$)
\begin{align}
ds^{2} & =c^{2}\,dt^{2}-a^{2}(t)\,\biggl(\frac{dr^{2}}{1-\kappa\,r^{2}}+r^{2}\,d\Omega^{2}\biggr)\label{eq:gen-RW-metric}\\
d\Omega^{2} & =d\theta^{2}+\sin^{2}\theta\,d\phi^{2}
\end{align}
by allowing the speed of light $c$ to vary alongside the dilaton
field $\chi$ (per $c\propto\chi^{1/2}$ as specified in Eq. \eqref{eq:scalings}).
Since the dilaton directly determines the lengthscale for physical
processes, $\chi\propto l^{-1}$, as in Eq. \eqref{eq:length-scaling},
and the cosmic scale factor plays the role of a lengthscale in the
FLRW metric, it is reasonable to make the following ansatz:\vskip8pt

\textbf{\emph{Ansatz \#1:}}\emph{\vspace{-.25cm}
\begin{equation}
\chi\propto a^{-1}
\end{equation}
}Furthermore, since the dilaton also determines the timescale for
physical processes, $\chi\propto\tau^{-2/3}$, as in Eq. \eqref{eq:tau-scaling},
and the cosmic time $t$ plays the role of a timescale in the FLRW
metric, it is reasonable to make the following ansatz on the evolution
of the cosmic scale factor:\vskip8pt

\textbf{\emph{Ansatz \#2:}}\emph{\vspace{-.25cm}}
\begin{equation}
a=a_{0}\left(t/t_{0}\right)^{\,2/3}\label{eq:evolution}
\end{equation}

This ansatz is identical to the evolution of the scale factor in the
Einstein--de Sitter (EdS) universe. Obviously, the vanilla EdS model---by
itself---cannot account for the late-time cosmic acceleration observed
in the Hubble diagram of Type Ia supernova (SNeIa). This failure forms
the basis to replace the EdS model by the ``concordance'' cosmological
model in which a $\Lambda$ component of density $\Omega_{\Lambda}\approx0.7$.\vskip4pt

However, in the rest of this paper, we consider the EdS universe in
conjunction with a varying speed of light. From $c\propto\chi^{1/2}$
and Ansatz \#1, we deduce that, \emph{in the intergalactic space},
the following relation holds 
\begin{equation}
c\propto\chi^{1/2}\propto a^{-1/2}
\end{equation}
The \emph{modified} FLRW metric suitable for our VSL cosmology is
thus expressed as
\begin{equation}
ds^{2}=\frac{c_{0}^{2}}{a(t)}\,dt^{2}-a^{2}(t)\,\biggl(\frac{dr^{2}}{1-\kappa\,r^{2}}+r^{2}\,d\Omega^{2}\biggr)\label{eq:mod-RW-metric}
\end{equation}
where the cosmic scale factor at our current time $t_{0}$ is set
equal to 1 (i.e., $a_{0}=1$), and $c_{0}$ is the speed of light
measured \emph{in the intergalactic space} at $t_{0}$. It is important
to note that several authors have previously applied a VSL cosmology
to SNeIa standard candles, while circumventing dark energy \citep{Barrow-2000,Rodrigues-2022,Ravanpak-2017,Cai-2016,Salzano-2016d,Salzano-2016a,Liu-2023,Zhang-2014,Qi-2014}.
The consensus among these studies is that VSL alone could not adequately
account for the SNeIa data. However, these conclusions warrant reconsideration,
as all previous analyses implicitly assumed the speed of light depends
\emph{solely} on cosmic time $t$. This assumption fails to hold for
our VSL cosmology whereby $c$ varies in both space and time, making
it location-dependent. This is because the dilaton field $\chi$ can
vary in spacetime, with its value in the intergalactic space (which
is subject to cosmic expansion) differing from that in gravitationally
bound galaxies (which host SNeIa and are resistant to cosmic expansion).
Hence, as $c\propto\chi^{1/2}$, $c$ is not only time-dependent but
also location-dependent. A thorough treatment of this intricate issue
is provided in Ref. \citep{VSL2024-Pantheon}.\vskip4pt

In the presence of VSL, the classic Lema\^itre redshift formula,
$1+z=a^{-1}$, in standard cosmology is \emph{no longer applicable}.
Previous analyses of SNeIa using VSL overlooked this crucial distinction
and continued to use the classic formula, \emph{resulting in incorrect
conclusions}. The dependence of $c$ on the dilaton field $\chi$
introduces two fundamental modifications to the right hand side of
the Lema\^itre redshift relation, as follows:
\begin{enumerate}
\item The standard $a^{-1}$ term is replaced with $a^{-3/2}$, reflecting
the $3/2$--exponent associated with the anisotropic time scaling,
Eq. \eqref{eq:tau-scaling}.
\item An additional correction accounts for the variation of the dilaton
field among galaxies as the function of redshift $z$. This is because
galaxies hosting SNeIa may possess different values for the dilaton
field due to their evolution in the expanding cosmic background. We
model this physically motivated effect by introducing a monotonic
function $F(z)$ that smoothly interpolates between two extremes:
$F(z\rightarrow0)=1$ and $F(z\rightarrow\infty)=F_{\infty}$ (with
$F_{\infty}$ being an adjustable parameter). Specifically, we find
the following formulation to be suitable 
\begin{equation}
F(z)=1+(1-F_{\infty})\bigl(1-(1+z)^{-2}\bigr)^{2}\label{eq:F-functional}
\end{equation}
By virtue of the length scaling \eqref{eq:length-scaling}, $F(z)$
thus signifies the evolution in the typical size of galaxies in response
to the expansion of the intergalactic space. This \emph{astronomical}
effect will have implications for the $H_{0}$ tension, which will
be shown later in Section \ref{subsec:Resolving-H0-tension}.
\end{enumerate}
Derivations of these modifications are detailed in Ref. \citep{VSL2024-Pantheon}.
Here, we only quote the results. In our VSL cosmology, the \emph{modified}
Lema\^itre redshift formula reads
\begin{equation}
1+z=a^{-3/2}\,F(z)\label{eq:mod-Lemaitre}
\end{equation}

At very low $z$, with $a(t)=1-H_{0}\,d/c_{0}+\dots$ and $F(z)\approx1+4(1-F_{\infty})z^{2}+\dots$,
the relation \eqref{eq:mod-Lemaitre} yields a \emph{modified} Hubble
law
\begin{equation}
z\approx\frac{3}{2}H_{0}\,\frac{d_{L}}{c_{0}}\label{eq:mod-Hubble-law}
\end{equation}
This result contrasts with the Hubble law $z\approx H_{0}\,d/c$\linebreak
in standard cosmology. The multiplicative $3/2$--factor in Eq. \eqref{eq:mod-Hubble-law}
will have crucial implications in the estimation of $H_{0}$, as we
shall see in Sections \ref{subsec:A-reduced-value} and \ref{subsec:Resolving-the-age}.\vskip4pt

Combining Eq. \eqref{eq:mod-Lemaitre} and Ansatz \#2 \eqref{eq:evolution},
we obtain the \emph{modified} luminosity distance--redshift relation
applicable for our VSL cosmology \citep{VSL2024-Pantheon}
\begin{equation}
\frac{d_{L}}{c_{0}}=\frac{1+z}{H_{0}\,F(z)}\,\ln\frac{(1+z)^{2/3}}{F(z)}\label{eq:dL-z-VSL}
\end{equation}
For comparison, this formula fundamentally differs from the canonical
ones in the spatially flat $\Lambda$CDM model ($\Omega_{M}+\Omega_{\Lambda}\!=\!1$)
\begin{equation}
\frac{d_{L}}{c}=\frac{1+z}{H_{0}}\int_{0}^{z}\frac{dz'}{\sqrt{\Omega_{M}(1+z')^{3}+\Omega_{\Lambda}}}\label{eq:dL-z-LCDM}
\end{equation}
and in the EdS universe (i.e. $\Omega_{M}=1,\,\Omega_{\Lambda}=0$)
\begin{equation}
\frac{d_{L}}{c}=2\,\frac{1+z}{H_{0}}\,\biggl(1-\frac{1}{\sqrt{1+z}}\biggr)\label{eq:dL-z-EdS}
\end{equation}

\subsection{\label{subsec:Analyzing-Pantheon-Catalog}Analyzing Pantheon Catalog
of SNeIa using {\fontsize{10pt}{0pt}\selectfont{VSL}}}

\begin{figure}[!t]
\begin{centering}
\includegraphics[scale=0.65]{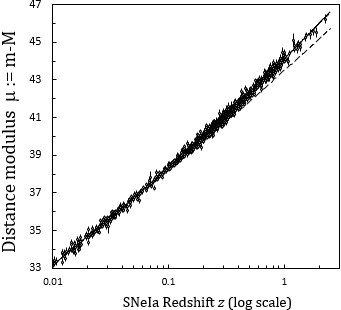}\vskip8pt
\par\end{centering}
\begin{centering}
\includegraphics[scale=0.65]{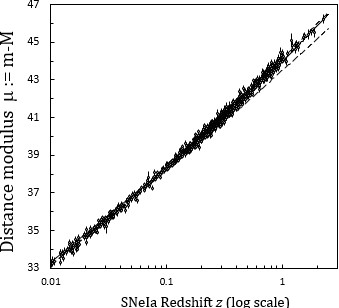}
\par\end{centering}
\caption{Hubble diagram of SNeIa in the Pantheon Catalog. Open circles: 1,048
data points with error bars, listed in Ref. \citep{Scolnic-2018}.
In both panels, long-dashed line is the $\Lambda$CDM Formula \eqref{eq:dL-z-LCDM}
with $H_{0}=70.2$, $\Omega_{M}=0.285$, $\Omega_{\Lambda}=0.715$;
dotted line is the EdS Formula \eqref{eq:dL-z-EdS} with $H_{0}=70.2$.
Upper panel: Solid line is our VSL Formula \eqref{eq:dL-z-VSL} with
$H_{0}=47.2$ and $F_{\infty}=0.93$. Lower panel: Solid line is our
VSL Formula \eqref{eq:dL-z-VSL} with $F(z)\equiv1\ \forall z$ with
$H_{0}=44.4$.}

\label{fig:fit}
\end{figure}

The \emph{modified} $d_{L}$-vs-$z$ relation \eqref{eq:dL-z-VSL},
supplemented with \eqref{eq:F-functional}, now stands ready for application
to the Pantheon Catalog of SNeIa. Our detailed analysis is provided
in Ref. \citep{VSL2024-Pantheon}. The Pantheon dataset, as provided
in Ref. \citep{Scolnic-2018}, is shown as open circles in the Hubble
diagram in Fig. \ref{fig:fit}. It comprises $N=1,048$ data points
given in terms of redshift and luminosity distance $\{z,\,\mu\}$,
together with an error bar $\sigma$ of $\mu$, for $z$ spanning
the range $(0,\,2.26)$. For a given theoretical model, one would
minimize the error
\begin{equation}
\mu:=m-M=5\log_{10}(d_{L}/\text{Mpc})+25\label{eq:modulus}
\end{equation}
\begin{equation}
\chi^{2}:=\frac{1}{N}\sum_{i=1}^{N}\biggl(\frac{\mu_{i}^{\text{model}}-\mu_{i}^{\text{Pantheon}}}{\sigma_{i}^{\text{Pantheon}}}\biggr)^{2}
\end{equation}

Let us first review the canonical interpretation of SNeIa data in
standard cosmology. For very low $z$, both formulae \eqref{eq:dL-z-LCDM}
and \eqref{eq:dL-z-EdS} yield $\frac{d_{L}}{c}\approx\frac{z}{H_{0}}$.
However, for positive $\Omega_{\Lambda}$, given the same $H_{0}$,
the flat $\Lambda$CDM formula \eqref{eq:dL-z-LCDM} produces a higher
value for $d_{L}$ compared to  the EdS formula \eqref{eq:dL-z-EdS}.
The crossover occurs at a redshift $z_{*}$ which approximately satisfies
$\Omega_{M}(1+z_{*})^{3}\approx\Omega_{\Lambda}$. We fit Formula
\eqref{eq:dL-z-LCDM} to the Pantheon Catalog and obtain $H_{0}=70.2$,
$\Omega_{M}=0.285$, $\Omega_{\Lambda}=0.715$ with an error $\chi_{\text{min}}^{2}=0.98824$.
Hence, starting at $z\gtrsim z_{*}\approx0.36$, the spatially flat
$\Lambda$CDM model begins to show an \emph{excess} in $d_{L}$, meaning
high-$z$ SNeIa appear dimmer than the EdS model would predict. This
behavior, manifest in the high-$z$ portion of the Hubble diagram
(Fig. \ref{fig:fit}), has been interpreted as evidence for late-time
cosmic accelerating expansion \citep{Riess-1998,Perlmutter-1999}
and for the existence of dark energy.\vskip4pt

\begin{figure}[t]
\begin{centering}
\hskip-6pt\includegraphics[scale=0.8]{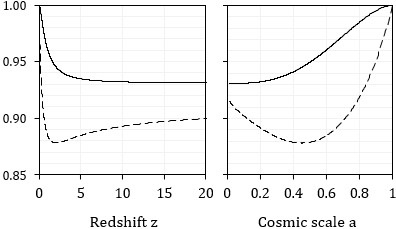}
\par\end{centering}
\caption{Solid curves: $F(z)$ and $F(a)$. Dashed curves: $H_{0}(z)/H_{0}(z\text{=}0)$
and $H_{0}(a)/H_{0}(a=1)$.}

\noindent \label{fig:eff-F-and-H0-functional}\vskip12pt
\noindent \begin{centering}
\hskip-6pt\includegraphics[scale=0.48]{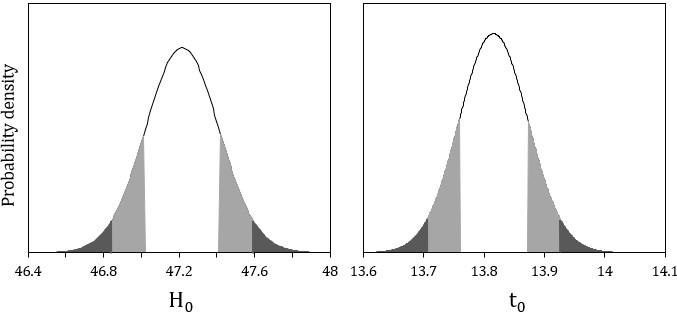}
\par\end{centering}
\caption{Posterior distributions of $H_{0}$ (left panel) and $t_{0}$ (right
panel), with 68\% CL and 95\% CL bands shown.}

\label{fig:H0-and-t0}
\end{figure}

Our VSL formula \eqref{eq:dL-z-VSL} involves two adjustable parameters
$H_{0}$ and $F_{\infty}$. Its best fit to the Pantheon Catalog yields
$H_{0}=47.2$, $F_{\infty}=0.93$ with an error $\chi_{\text{min}}^{2}=0.98556$,
which is highly competitive in quality to the $\Lambda$CDM fit. Note
that in the upper panel of Fig. \ref{fig:fit}, the two curves---VSL
and $\Lambda$CDM---are indistinguishable in the range $z\lesssim2.26$
available for the Pantheon data \footnote{Light sources with even higher $z$, such as quasars \citep{Lusso-2020},
might help distinguish Formula \eqref{eq:dL-z-VSL} versus Formula
\eqref{eq:dL-z-LCDM}.}.\linebreak Both curves manifest an excess in $d_{L}$ in the high-$z$
segment of the Hubble diagram. For completeness, we also show $F(z)$
and $F(a)$ as solid curves in Fig. \ref{fig:eff-F-and-H0-functional}.\vskip4pt

Of the two modifications to our \emph{modified} Lema\^itre redshift
formula $1+z=a^{-3/2}\,F(z)$ given in \eqref{eq:mod-Lemaitre}, the
$a^{-3/2}$ term is expected to be of primary significance, while
the $F(z)$ plays a secondary role. To test this intuition, we disable
$F(z)$ by setting $F_{\infty}=1$ (making $F(z)\equiv1\ \forall z\in\mathbb{R}$).
A refit of \eqref{eq:dL-z-VSL} to the Pantheon data with just \emph{one}
parameter (i.e. $H_{0}$) yields $H_{0}=44.4$ with $\chi_{\text{min}}^{2}=1.25366$.
Despite a substantial decrease in quality of fit, the $a^{-3/2}$
term alone still produces a reasonable fit, as shown in the lower
panel of Fig. \ref{fig:fit}. Most importantly, this simplified fit
continues to exhibit the excess in $d_{L}$ prominent in the high-$z$
segment of the Hubble diagram.\vskip4pt

This indicates that our analysis practically---and \emph{parsimoniously}---requires
only one parameter $H_{0}$. The function $F(z)$ plays a role in
resolving the $H_{0}$ tension, which will be explained in Section
\ref{subsec:Resolving-H0-tension} below.

\subsection{\label{subsec:A-reduced-value}A reduced value of $\fontsize{10pt}{0pt}\selectfont{\textit{H$_0$}}$
by a factor of $3/2$}

A striking result deduced from our SNeIa analysis thus far is a \emph{low}
value $H_{0}=47.2$, which starkly contrasts with the local measurement
$H_{0}=73$ widely accepted in the $\Lambda$CDM framework. Note that
this low value is inherent to the \emph{modified} $d_{L}$-vs-$z$
formula \eqref{eq:dL-z-VSL} \emph{regardless} of the function $F(z)$.
When $F(z)$ is disabled, the formula still produces $H_{0}=44.4$,
a value comparable to that obtained when $F(z)$ is enabled. The posterior
distribution of $H_{0}$ is shown in the left panel of Fig. \ref{fig:H0-and-t0}
and corresponds to $H_{0}=47.2\pm0.4$ (at 95\% confidence level).\vskip4pt

We should note that a method of ``redshift remapping'' was introduced
in \citep{Bassett-2013,Wojtak-2016,Wojtak-2017}. Our modified Lema\^itre
redshift formula could be viewed as effectively a ``redshift remapping''.
Interestingly, in \citep{Wojtak-2017}, based on this method, a joint
analysis of all primary cosmological probes including the local measurement
of the Hubble constant, SNeIa, Baryon Acoustic Oscillations (BAO),
Planck observations of the CMB power spectrum and cosmic chronometers
yields $H_{0}=48\pm2$, a range in good agreement with our result
$H_{0}=47.2\pm0.4$.\vskip4pt

The drastic deviation in the current-time Hubble value between our
$H_{0}=47.2$ from the local measurement $H_{0}=73$ can be understood
by using the modified Hubble law \eqref{eq:mod-Hubble-law} in VSL
cosmology. For very low $z$, the Hubble law for our VSL cosmology
is \emph{modified} to $z\approx\frac{3}{2}H_{0}\,d/c_{0}$, as opposed
to the standard Hubble law $z\approx H_{0}\,d/c$. Thus, for our VSL
cosmology, the numerical value (empirically known to be $\sim71$)
for the slope of the $z$-vs-$d$ relation is \emph{not $H_{0}$ but
}$\frac{3}{2}H_{0}$, which promptly leads to $H_{0}\sim47$. In conclusion,
it is the $3/2-$exponent in the $a^{-3/2}$ term in our \emph{modified}
Lema\^itre redshift formula \eqref{eq:mod-Lemaitre} that is responsible
for \emph{the reduction in the $H_{0}$ value by a factor of $3/2$
compared to the canonical range}.\vskip4pt

Surprisingly, this reduced value of $H_{0}$ finds corroboration in
a remarkable proposal made by Blanchard et al in 2023 \citep{Blanchard-2003},
when these authors analyzed the CMB power spectrum. We shall review
their work below.

\subsection{\label{subsec:Revitalizing-Blanchard}Revitalizing Blanchard et al
(BDRS)'s analysis of CMB power spectrum, bypassing DE}

It is well established that the $\Lambda$CDM, with the primordial
fluctuations spectrum $P(k)\propto k^{n}$, successfully accounts
for Planck's thermal power spectrum of the Cosmic Microwave Background
(CMB) with parameters $n\approx0.96$, $\Omega_{\Lambda}\approx0.7$,
$H_{0}\approx67$. However, in 2003, Blanchard, Douspis, Rowan-Robinson,
and Sarkar (BDRS) made a surprising discovery \citep{Blanchard-2003}.
Instead of adhering to the standard primordial fluctuation spectrum
$P(k)\propto k^{n}$, they considered two separate power-law forms
for the low $k$ and high $k$ regimes, viz.
\begin{equation}
P(k)=\begin{cases}
A_{1}\,k^{n_{1}} & \text{for }k\leqslant k_{*}\\
A_{2}\,k^{n_{2}} & \text{for }k\geqslant k_{*}
\end{cases}\label{eq:BDRS}
\end{equation}
where $k_{*}$ is a break point, and continuity across $k_{*}$ is
enforced by imposing $A_{1}\,k_{*}^{n_{1}}=A_{2}\,k_{*}^{n_{2}}$.
Most importantly, they deliberately set $\Omega_{\Lambda}=0$, thereby
avoiding the dark energy hypothesis. Remarkably, their model achieved
an excellent fit to WMAP's CMB data available at the time, with the
optimal parameter values including $H_{0}=46$, $k_{*}=0.0096$ Mpc$^{-1}$,
$n_{1}=1.015$, $n_{2}=0.806$ \citep{Blanchard-2003}. Of these parameters,
the most striking aspect was the \emph{low} Hubble value $H_{0}=46$.
A subsequent in-depth study by Hunt and Sarkar \citep{Hunt-2007},
utilizing a supergravity-based multiple inflation scenario, also successfully
accounted for the CMB power spectrum \emph{without} invoking dark
energy, while requiring a similar value of $H_{0}\approx43.5$. (Interestingly,
Shanks in 2004 \citep{Shanks-2004} also ventured that if $H_{0}\lesssim50$
then a simpler, inflationary model with $\Omega_{baryon}=1$ might
be allowed with no need for dark energy or cold dark matter.)\vskip4pt

Remarkably, the low values of $H_{0}\sim43\text{--}46$ that BDRS
and Hunt and Sarkar derived from the CMB power spectrum closely align
with the value $H_{0}=47.2$ that we obtained in our VSL reanalysis
of the SNeIa standard candles, as presented in Section \ref{subsec:Analyzing-Pantheon-Catalog}
and elaborated on in \ref{subsec:A-reduced-value}. Both sets of results---ours
and those of BDRS/Hunt/Sarkar---markedly diverge from the canonical
range $H_{0}\sim67\text{--}73$. Importantly, both sets of analyses
\emph{exclude} the necessity for dark energy. It is noteworthy that
the datasets involved are of ``orthogonal'' natures: the (early-time)
CMB data provides a 2-dimensional snapshot \emph{across the sky} at
the recombination event, while the (late-time) SNeIa data represents
a tracking of evolution \emph{along the cosmic time direction}.\vskip4pt

Hence, despite the fundamentally distinct natures of the data in use,
BDRS's 2003 analysis of the CMB and our current VSL-based analysis
of the Hubble diagram of SNeIa converge on two crucial points: (i)
\emph{the avoidance of the dark energy hypothesis}, and (ii) \emph{the
reduced value of $H_{0}$$\,\sim\,$$47$}. In the light of our VSL
framework, the canonical estimates $\sim\,$$67\text{--}73$ suffer
from an upward systematic bias by a factor of $3/2$ due to the anomalous
$3/2$-exponent in the scaling behavior of the clock rate, as described
in Eq. \eqref{eq:tau-scaling}. Remarkably, the new (reduced) value
$H_{0}=47.2$ also provides a natural resolution to the age problem,
which we shall explain in Section \ref{subsec:Resolving-the-age}
below.\vskip4pt

It is also important to note that the perceived low $H_{0}$ value
$\sim46$ eventually led BDRS to abandon their original findings in
a follow-up study related to SDSS's two-point correlation data of
luminous red galaxies \citep{Blanchard-2006}. However, their abandonment
may have been premature, since in the VSL framework the anomalous
$3/2$--exponent in the anisotropic time scaling influences the observation
of distant light sources along the time direction. Therefore, a reassessment
of BDRS's work of SDSS data, incorporating this effect, would be warranted.
Additionally, future investigations should explore the impacts of
VSL on phenomena such as lensing, BAO, and the Etherington distance
duality relation.\vskip4pt

As we stated earlier in Section \ref{subsec:A-reduced-value}, using
a ``redshift remapping'' method for all primary cosmological probes,
the authors in \citep{Wojtak-2017} obtained $H_{0}=48\pm2$, which
is compatible with our value $H_{0}=47.2$ and the value $H_{0}=46$
obtained by BDRS.

\subsection{\label{subsec:Resolving-the-age}Resolving the age problem, bypassing
DE}

The spatially flat $\Lambda$CDM model gives the age formula\vspace{-.1cm}
\begin{equation}
t_{0}^{\Lambda\text{CDM}}=\frac{2}{3\sqrt{\Omega_{\Lambda}}H_{0}}\,\text{arcsinh}\sqrt{\frac{\Omega_{\Lambda}}{\Omega_{M}}}\label{eq:age-LCDM}
\end{equation}
With $H_{0}=70.2$, $\Omega_{M}=0.285$, $\Omega_{\Lambda}=0.715$,
it yields an age of $13.6$ billion years, an accepted figure in standard
cosmology. If dark energy is absent ($\Omega_{\Lambda}=0,\,\Omega_{M}=1$),
Eq. \eqref{eq:age-LCDM} simplifies to $t_{0}^{\text{EdS}}=2/(3H_{0})$
which then, with $H_{0}\sim71$, would result in an age of $9.2$
Gy which would be too short to accommodate the existence of the oldest
stars---a paradox commonly referred to as the age problem.\vskip4pt

However, our VSL cosmology naturally resolves the age problem without
requiring dark energy. Using the definition $H_{0}:=\left.\frac{1}{a}\frac{da}{dt}\right|_{t=t_{0}}$
and the evolution law \eqref{eq:evolution}, the age of the our VSL
universe is
\begin{equation}
t_{0}^{\text{VSL}}=\frac{2}{3H_{0}}\label{eq:age-VSL}
\end{equation}
A crucial point is that $H_{0}$ is reduced by a factor of $3/2$,
as detailed in Section \ref{subsec:A-reduced-value}. The \emph{reduced}
value $H_{0}=47.2\pm0.4$ (95\% CL) promptly yields $t_{0}=13.8\pm0.1$
billion years (95\% CL), consistent with the accepted age value (with
the posterior distribution of $t_{0}$ shown in the right panel of
Fig. \ref{fig:H0-and-t0}), thereby successfully resolving the age
paradox. Notably, our resolution does \emph{not} require dark energy,
but rather a reduction in the value of $H_{0}$ by a factor of $3/2$
compared to the canonical range $\sim67\text{--}73$.
\begin{figure*}[!t]
\begin{centering}
\includegraphics[scale=0.65]{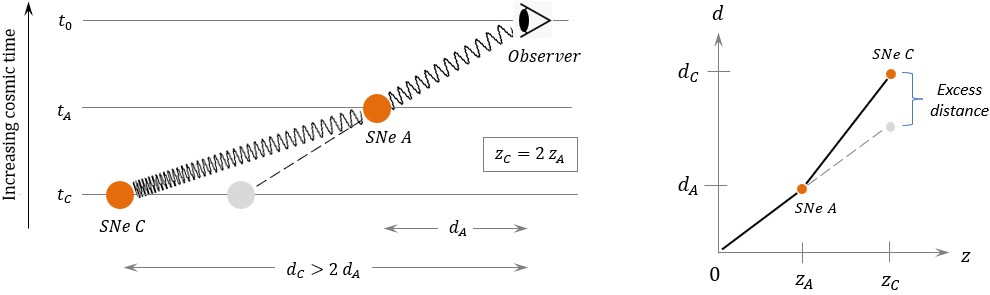}
\par\end{centering}
\caption{The physical intuition of late-time acceleration based on VSL as explained
in Section \ref{subsec:Physical-intuition}. In the left panel, photons
from SNe A and SNe C were emitted at times $t_{A}$ and $t_{C}$ with
$t_{C}-t_{0}=2\,(t_{A}-t_{0})$; thus, their redshifts satisfy $z_{C}=2\,z_{A}$.
However, for the photons emitted from SNe C, the earlier segment of
their trajectory had a higher speed of light than the later segment,
allowing them to cover a longer distance. This results in $d_{C}>2\,d_{A}$,
leading in an excess distance modulus in the Hubble diagram for high-$z$
SNe, as depicted in the right panel.}

\label{fig:intuition}
\end{figure*}

\subsection{\label{subsec:Obtaining-late-time-acceleration}Obtaining late-time
acceleration via {\fontsize{10pt}{0pt}\selectfont{VSL}}, bypassing
DE}

It is an empirical fact that high-$z$ SNeIa's appear fainter than
what would be predicted by the EdS model \citep{Riess-1998,Perlmutter-1999}.
This phenomenon is reflected in the distance modulus excess observed
in the high-$z$ segment of the Hubble diagram in Fig.$\ $\ref{fig:fit}.
Standard cosmology attributes this behavior to late-time cosmic acceleration,
necessitating the introduction of a cosmological constant $\Lambda$
into the Friedmann equations, resulting in the $\Lambda$CDM model
where $\Omega_{\Lambda}$ quantifies the amount of dark energy.\vskip4pt

However, the analysis presented in Sections \ref{subsec:VSL-cosmography-and}
and \ref{subsec:Analyzing-Pantheon-Catalog} demonstrates the efficacy
of our VSL cosmology in accounting for the excess in $d_{L}$ without
invoking dark energy. Its success can be quantitatively understood
as follows. At high $z$, the luminosity distance formula of the EdS
model \eqref{eq:dL-z-EdS} behaves as
\begin{equation}
d_{L}\propto z\label{eq:asymp-EdS}
\end{equation}
In contrast, the \emph{modified} formula for our VSL cosmology \eqref{eq:dL-z-VSL}
behaves as\vspace{-.1cm}
\begin{equation}
d_{L}\simeq z\,\ln z\label{eq:asym-VSL}
\end{equation}
which gains an \emph{additional} $\ln z$ term compared to \eqref{eq:asymp-EdS}.
Consequently, according to our VSL Formula, $d_{L}$ grows faster
and exhibits a \emph{steeper} upward slope in the high-$z$ section
in the Hubble diagram. Furthermore, this explanation can be strengthened
through a qualitative account, as we will illustrate below.

\subsection{\label{subsec:Physical-intuition}Physical intuition of late-time
acceleration\vskip1ptvia {\fontsize{10pt}{0pt}\selectfont{VSL}},
bypassing DE}

Consider two supernovae, $A$ and $B$, located at distances $d_{A}$
and $d_{B}$ from the Earth, with $d_{B}=2\,d_{A}$. In standard cosmology,
their redshift values $z_{A}$ and $z_{B}$ are related as $z_{A}\approx2\,z_{A}$
(to first-order approximation). However, this relationship breaks
down in our VSL cosmology. Since $c\propto a^{-1/2}$, light traveled
faster in the distant past (when the cosmic factor $a\ll1$) than
in the more recent epoch (when $a\lesssim1$). Therefore, photons
emitted from supernova $B$ could cover twice the distance in less
than double the time required by photons emitted from supernova $A$.
Having spent less time in transit than predicted by standard cosmology,
the $B$--photons experience less cosmic expansion than expected,
resulting in a lower redshift such that:
\begin{equation}
z_{B}<2\,z_{A}\ \ \text{for}\ \ d_{B}=2\,d_{A}
\end{equation}
Conversely, consider a supernova $C$ with $z_{C}=2\,z_{A}$. For
photons emitted from supernova $C$ to experience twice the redshift
of the $A$--photons, they must travel a distance greater than twice
that of the $A$--photons:
\begin{equation}
d_{C}>2\,d_{A}\ \ \text{for}\ \ z_{C}=2\,z_{A}
\end{equation}
This occurs because the $C$--photons traveled faster at the beginning
of their journey toward Earth and must originate from a farther distance
(thus appearing fainter than expected) to accumulate sufficient redshift.
In either case, the VSL-based Hubble diagram exhibits a steeper upward
slope in the high-$z$ section. This intuition is illustrated in Fig.
\ref{fig:intuition}.\vskip4pt

An \emph{alternative} explanation for late-time acceleration is \emph{a
decrease in the speed of light} in the intergalactic space as the
universe expands. In summary, VSL cosmology can account---qualitatively
and quantitatively---for the distance modulus excess observed in
high-$z$ SNeIa without the need for dark energy. This approach overcomes
the coincidence problem associated with the concordance $\Lambda$CDM
model. Furthermore, it bypasses the enigmatic nature of dark energy
and aligns with BDRS's 2003 groundbreaking analysis of the CMB power
spectrum which also negates the dark energy hypothesis \citep{Blanchard-2003}.

\subsection{\label{subsec:Resolving-H0-tension}{\fontsize{10pt}{0pt}\selectfont{\textit{H$_0$}}}
tension: $\ $A potential resolution with an astronomical origin}

In recent years, interest in the $H_{0}$ tension has intensified
\citep{Perivolaropoulos-2022,diValentino-2021,Vagnozzi-2023}. One
popular approach to address this issue involves treating $H_{0}$
as a ``running'' value, meaning that its value depends on the redshift
of the data used to deduce it; e.g., see Refs. \citep{Dainotti-2020,Krishnan-2020,Krishnan-2021}.\vskip4pt

Our VSL cosmology naturally produces a ``running'' $H_{0}$ through
the function $F(z)$. As mentioned in Section \ref{subsec:VSL-cosmography-and},
$F(z)$ represents the change in the typical size of galaxies as they
evolve in an expanding intergalactic space (see also Ref. \citep{VSL2024-Pantheon}
for clarifications). From the \emph{modified} formula \eqref{eq:dL-z-VSL},
we can define the ``running'' value for $H_{0}$ as
\begin{equation}
\frac{d_{L}}{c_{0}}=\frac{1+z}{H_{0}(z)}\,\ln(1+z)^{2/3}\label{eq:dL-vs-z-sim}
\end{equation}
By comparing this with Eq. \eqref{eq:dL-z-VSL} itself, we find 
\begin{equation}
H_{0}(z)=H_{0}\,F(z)\,\biggl(1-\frac{3}{2}\frac{\ln F(z)}{\ln(1+z)}\biggr)^{-1}
\end{equation}
If $F(z)\equiv1\ \forall z>0$, then $H_{0}(z)\equiv H_{0}$, a constant.
Otherwise, as $z\rightarrow\infty$, $H_{0}(z)\rightarrow H_{0}F_{\infty}$;
namely, at very high $z$, its ``running'' value is reduced by a factor
$F_{\infty}$.\vskip4pt

Using \eqref{eq:F-functional} and the value $F_{\infty}=0.93$ obtained
from the fit in Section \ref{subsec:Analyzing-Pantheon-Catalog},
the computed ``running'' $H_{0}(z)$ is depicted as dashed curves
in Fig. \ref{fig:eff-F-and-H0-functional}. At the time of recombination,
the ``running'' value for $H_{0}$ would be $H_{0}\,F_{\infty}=47.2\times0.93=43.9$,
representing a \emph{$7\%$ reduction} from its value estimated using
low-redshift SNeIa. The function $F(z)$ hence shed important light
on the $H_{0}$ tension. Furthermore, in \citep{Hunt-2007} Hunt and
Sarkar obtained $H_{0}\approx44$ from the CMB power spectrum using
a supergravity-based multiple inflation scenario. Our result of $H_{0}(z\gg1)\approx43.9$
is thus in agreement with that of the Hunt--Sarkar team.\vskip4pt

Importantly, the function $F(z)$ is a well-defined and physically
motivated concept: it models the evolution of galaxies in an expanding
intergalactic space. Consequently, the ``running'' $H_{0}$ in our
approach is rooted in an \emph{astronomical} origin, rather than a
cosmic origin.

\subsection{\label{subsec:cosmo-horizon}An infinite cosmological horizon using
{\fontsize{10pt}{0pt}\selectfont{VSL}}}

In the \emph{modified }FLRW metric \eqref{eq:gen-RW-metric}, the
cosmological horizon at the recombination time $t_{rec}$ is given
by
\begin{align}
l_{H}\left(t_{rec}\right) & =a(t_{rec})\int_{0}^{t_{rec}}\frac{c}{a(t')}dt'
\end{align}
Since $c=c_{0}\,a^{-1/2}$ and $a=\left(t/t_{0}\right)^{2/3}$, we
find
\begin{align}
l_{H}\left(t_{rec}\right) & =c_{0}\,t_{0}\,\left(\frac{t_{rec}}{t_{0}}\right)^{2/3}\int_{0}^{t_{rec}}\frac{dt'}{t'}=\infty\label{eq:cosmo-horizon}
\end{align}
The cosmological horizon is thus (logarithmically) divergent, suggesting
that the entire universe was \emph{causally connected} at the recombination
event. This result dovetails with the VSL proposals previously made
by Moffat \citep{Moffat-1992} and Albrecht and Magueijo \citep{Albrecht-1999}
which sought to explain the near isotropy of the CMB temperatures---an
empirical fact commonly known as the horizon problem.

\section{\label{sec:Discussion}Discussion and Summary}

\textbf{\emph{On scale symmetry}}---Symmetries plays a fundamental
role in constructing modern physical theories, with notable examples
including general covariance in GR and gauge invariance in the Glashow--Weinberg--Salam
(GWS) model of particle physics. In his seminal 1995 report \citep{Bardeen-1995},
Bardeen proposed employing scale symmetry as a viable alternative
to supersymmetry to cure the naturalness problem of the Higgs mass.
It is well established that the Higgs particles, as quanta of scalar
fields, suffer a quadratic divergence in their loop amplitudes, which
destabilizes their ultraviolet-limit behavior and requires fine tuning.
Bardeen suggested that \emph{classical} scale invariance could protect
the Higgs mass from this divergence. Theories that impose scale invariance
in the matter and gravity sectors have been actively pursued in both
particle physics and gravitational physics \citep{Adak:2023vhd,Adler:1980bx,Adler:2021fgr,Adler:2023thu,Adler:2024dtn,Ahriche:2015loa,Ahriche:2016cio,Ahriche:2016ixu,Ahriche:2022ngh,AlexanderNunneley:2010nw,Allison:2014hna,Allison:2014zya,Ametani:2015jla,Antipin:2013exa,Aoki:2024kty,Banados:2024bnh,Banerjee:2018ngt,Banik:2020lrt,Barman:2022htu,Benic:2014aga,Bertini:2024dfr,Bezrukov-2015,Borissova:2023dth,Boudet:2023hrd,Braathen:2021ful,Braathen:2022ngy,Burrage:2018krt,Callan:1970ze,Carone:2013wla,Carone:2015jra,Casas:2019mku,Cerioni:2010ke,Chang:2007ki,Chivukula:2013xka,Chun:2013soa,Chway:2013pta,Coleman:1970je,Cooper:1982du,Das:2015nwk,Das:2016zue,Davoudiasl:2014pya,DelCima:2022knr,Devecioglu:2018jut,Dias:2022ent,Dirgantara:2023jhu,Einhorn-2017,Einhorn:2015lzy,Einhorn:2016mws,Einhorn:2019lin,Endo:2015ifa,Endo:2015nba,Englert:2013gz,Farnsworth:2022guk,Farnsworth:2024dft,Farzinnia:2013pga,Farzinnia:2014xia,Farzinnia:2015fka,Farzinnia:2015uma,Fernandes:2021adr,Ferreira-2017,Ferreira:2016jhu,Ferreira:2018rth,Ferreira:2020ghj,Finelli:2007wb,Foot:2007as,Foot:2007ay,Foot:2007iy,Foot:2010av,Foot:2010et,Foot:2011et,Fortin:2011sz,Frasca:2024pon,Gabrielli:2013hma,GarciaBellido-2011,Ghilencea-2019,Ghilencea:2015mza,Ghilencea:2016ckm,Ghilencea:2017afg,Ghilencea:2019iuj,Ghilencea:2019nhg,Ghilencea:2023hyt,Ghilencea:2024kuy,Ghorbani:2015xvz,Ghoshal:2023jtv,Gildener:1976ih,Girmohanta:2024ijn,Gueorguiev:2025njh,Guo:2014bha,Guo:2015lxa,Haba:2015lka,Haba:2015nwl,Haba:2015qbz,Haba:2015rha,Hambye:2013sna,Hashimoto:2013hta,Hatanaka:2016rek,Heikinheimo:2013fta,Heikinheimo:2014xza,Helmboldt:2016mpi,Hempfling:1996ht,HerreroValea:2020dht,Hill:2014mqa,Holthausen:2009uc,Holthausen:2013ota,Humbert:2015epa,Hur:2011sv,Ishida:2016ogu,Ishiwata:2011aa,Iso:2009nw,Iso:2009ss,Jack:1990eb,Jinno:2016knw,Kaiser:2010ps,Kamenshchik:2012rs,Kang:2014cia,Kang:2015aqa,Kannike-2017,Kannike:2014mia,Kannike:2015apa,Kannike:2015kda,Kannike:2016bny,Kannike:2021nrd,Karam:2015jta,Karam:2016rsz,Karam:2017rty,Karananas-2016,Karananas:2016kyt,Karananas:2017ikj,Karananas:2018oiu,Karananas:2021ner,Karananas:2023dth,Khoze:2013oga,Khoze:2013uia,Khoze:2014xha,Khoze:2016zfi,Khoze:2023sth,Kobakhidze:2017grt,Koivisto:2021rhy,Kubo:2014ida,Kubo:2014ova,Kubo:2015cna,Kugo:20220htd,Kurkov:2016nhj,Lee:2012jn,Lin:2014mua,Lindner:2014oea,Litim:2017nru,Maeder:2016kyn,Maeder:2024fgd,Maeder:2024kid,Maeder:2024ngh,Maitiniyazi:2025thg,Martins:2021olp,Marzola:2016xgb,Meissner:2006zh,Minkowski:1977aj,Mooij:2019rgh,Myung:2018fty,Nishino-2011,Nishino:2007wer,Nishino:2009nju,Oda:2013uca,Oda:2015gna,Oda:2021lun,Oda:2022mhu,Okada:2015gia,Ota:2023kuy,Papadopoulos:2024jhg,Pelaggi:2014wba,Plascencia:2015xwa,Polchinski:1987dy,Quiros:2014ngh,Quiros:2024fgd,Radovcic:2014rea,Rothman:1982lki,Rubio-2014,Rubio-2017,Safari:2022kum,Salvio:2017hju,Salvio:2018lkj,Sannino:2015wka,Shaposhnikov:2008xb,Shaposhnikov:2008xi,Shaposhnikov:2018gul,Shaposhnikov:2020oln,Shimon:2021lkt,Shimon:2022rth,Smolin:1979uz,Tang:2020ktu,Tronconi:2010pq,Tronconi:2025thy,Utiyama:1973mnk,Utiyama:1975egy,vanDeBruck:2021juk,Wang:2015cda,Wang:2015sxe,Wang:2023tfh,Weisswange:2023dft,Wetterich:2002wm,Wetterich:2019lmi,Wetterich:2021kil,Wu:2016jdo,Zee:1978wi,Blas-2011,Ghilencea-2023,Ferreira-2018,Salvio-2014,Einhorn-2015,Edery-2014,Horndeski-1974,Cuzinatto-2021,Ferreira-2016,AlvarezGaume-2016,Alvarez-2018,Donoghue-2021,Salvio-2018,Nguyen-Newtonian,Karananas-2024,Iso:2009prn,Altmannshofer:2015ppp}.
Scale invariance typically necessitates (non-minimal) coupling of
the Higgs field with gravity via a scalar field $\chi$---commonly
referred to as a ``dilaton''---which arises naturally in various
theoretical contexts, such as Kaluza--Klein, string theory, and braneworld
scenarios \citep{Fujii-book}.\vskip4pt

In Section \ref{sec:Higgs=002013dilaton-coupling}, we consider a
\emph{scale-invariant} action of gravity and matter, as described
by Eqs. \eqref{eq:full-action}, \eqref{eq:L-mat}, and \eqref{eq:L-grav}
with $V(\chi)\propto\chi^{4}$. This action is generally covariant,
\emph{locally} Lorentz invariant, and $U(1)$ gauge invariant (for
the matter sector). Notably, the scale invariance of the action allows
it to evade the observational bounds on the fifth force \citep{Ferreira-2016,Blas-2011}.
The Higgs field couples with the dilaton field $\chi$ in the form
$\chi^{2}\,\Phi^{2}$. In addition, the matter Lagrangian $\mathcal{L}_{\text{mat}}$,
given in Eq. \eqref{eq:L-mat} as a \emph{prototype}, is invariant
with respect to a discrete $\mathbb{Z}_{2}$ symmetry of the Higgs
field, viz. $\Phi\leftrightarrow-\Phi$, although it can be readily
generalized to encompass the $SU(3)\times SU(2)\times U(1)$ gauge
group of the GWS model.\vskip8pt

\textbf{\emph{Our mechanism to generate variable $\hbar$ and $c$}}---Because
of scale invariance, all parameters in the action are \emph{dimensionless}.
This crucial fact means that the Planck constant $\hbar$ and the
speed of light $c$---as \emph{dimensionful} quantities---must be
\emph{absent} from the outset of this \emph{adimensional} action.
\vskip4pt

In Section \ref{sec:A-new-mechanism}, we developed---for the first
time---a\emph{ }recipe to \emph{construct} $\hbar$ and $c$ from
the Higgs-dilaton coupling in the \emph{matter} sector $\mathcal{L}_{\text{mat}}$
of Eq. \eqref{eq:L-mat}. The metric $g_{\mu\nu}$ and the dilaton
$\chi$ are treated as slowly varying background fields. Within a
local set surrounding a given point $x^{*}$, the discrete $\mathbb{Z}_{2}$
symmetry of the Higgs field is spontaneously broken due to the quartic
``potential'' $\frac{1}{2}\mu^{2}\,\chi^{2}\,\Phi^{2}-\frac{1}{4}\lambda\,\Phi^{4}$
(for $\lambda>0$). The Higgs field then acquires a non-zero VEV $\left\langle \Phi(x^{*})\right\rangle $
which is proportional to $\chi(x^{*})$; see Eq. \eqref{eq:Higgs-VEV}.
The non-zero Higgs VEV gives rise to mass for both the fermion and
the Higgs scalar boson (while the gauge vector boson remains massless
since the $U(1)$ gauge remains unbroken). Importantly, the Higgs
VEV allows us to construct a Planck constant $\hbar_{*}$ and a speed
of light $c_{*}$ to be effective within the local set around $x^{*}$.\vskip4pt

It is worth noting that in Ref. \citep{VSL2024-dilaton}, we presented
a ``short-cut'' approach. Instead of invoking the Higgs-dilaton coupling
$\chi^{2}\,\Phi^{2}$ and the Yukawa coupling between the fermion
and the Higgs field $\bar{\psi}\psi\,\Phi$ as we do in Eq. \eqref{eq:L-mat}
in this current paper, therein we intentionally suppressed the Higgs
field and let the fermion directly couple with the dilaton via the
term $\chi\,\bar{\psi}\psi$. In doing so, we also obtained a construction
for $\hbar_{*}$ and $c_{*}$ similar to that presented in this paper.
However, the current paper aims to explicitly incorporate the Higgs
field and utilize the mechanism of spontaneous symmetry breaking.
so that a connection to the GWS model for the matter sector can be
established.\vskip4pt

In either approach, the end result is that each open set vicinity
of a given point $x^{*}$ on the spacetime manifold is equipped with
a replica of the Quantum Electrodynamics (QED) action for matter fields
(i.e. fermion, $U(1)$ gauge vector boson, and Higgs scalar boson),
operating with its own effective values of $\hbar_{*}$ and $c_{*}$.
Lorentz symmetry is valid \emph{locally} within each open set. These
QED replicas are isomorphic (i.e. structurally identical), differing
only in the values of $\hbar_{*}$ and $c_{*}$, which are determined
by the prevailing value of the background dilaton field $\chi(x^{*})$
per $\hbar_{*}=\chi^{-1/2}(x^{*})$ and $c_{*}=\chi^{1/2}(x^{*})$.
With the gravitational Lagrangian $\mathcal{L}_{\text{grav}}$ in
Eq. \eqref{eq:L-grav} governing the dynamics of $\chi$, \emph{the
values of $\hbar_{*}$ and $c_{*}$ thus vary from one open set to
another across spacetime}.\vskip4pt

Crucially, within each open set, the length and time scales of physical
processes are governed by the dilaton field as $l\propto\chi^{-1}(x^{*})$
and $\tau\propto\chi^{-3/2}(x^{*})$. This leads to an \emph{anisotropic}
relationship between the rate of clocks and the length of measuring
rods, per $\tau^{-1}\propto l^{-3/2}$. This novel relationship allows
us to predict a new ``time dilation effect of the Third kind'', to
be reported in future work.\vskip8pt

\textbf{\emph{The role of VSL in late-time cosmology}}\emph{---}The
anisotropic relationship $\tau^{-1}\propto l^{-3/2}$ has immediate
cosmological consequences. Section \ref{sec:Phenomenology} presents
key findings regarding the impacts of variable speed of light (VSL)
on the Hubble diagram of SNeIa, with full technical details of our
analysis provided in a companion paper, Ref. \citep{VSL2024-Pantheon}.\vskip4pt

In the presence of VSL, the standard cosmography is no longer applicable.
Section \ref{subsec:VSL-cosmography-and} produces a new cosmography
that accommodates VSL: \vskip4pt

(i) Under the assumption that the dilaton is related to the cosmic
scale factor as $\chi\propto a^{-1}$, the FLRW metric is modified
to
\[
ds^{2}=\frac{c_{0}^{2}}{a(t)}\,dt^{2}-a^{2}(t)\,\left[dr^{2}+r^{2}d\Omega^{2}\right]\tag{c.f. \eqref{eq:mod-RW-metric}}
\]
which leads to a \emph{modified} Lema\^itre redshift formula (where
the function $F(z)$ accounts for variations in the dilaton value
among individual galaxies)
\[
1+z=a^{-3/2}\,F^{3/2}(z)\tag{c.f. \eqref{eq:mod-Lemaitre}}
\]
and a \emph{modified} Hubble law
\[
z=\frac{3}{2}H_{0}\,\frac{d_{L}}{c_{0}}\ \ \ \ \ \tag{c.f. \eqref{eq:mod-Hubble-law}}
\]

(ii) Further assuming an EdS-type evolution for the cosmic scale factor,
i.e. $a\propto t^{\,2/3}$, a \emph{modified} luminosity distance--redshift
formula is obtained
\[
\frac{d_{L}}{c_{0}}=\frac{1+z}{H_{0}\,F(z)}\,\ln\frac{(1+z)^{2/3}}{F(z)}\tag{c.f. \eqref{eq:dL-z-VSL}}
\]
Based on this formula, in Section \ref{subsec:Analyzing-Pantheon-Catalog},
we (re)analyzed the Pantheon Catalog of SNeIa. The key findings are
as follows:\vskip8pt

{*} We achieved a fit to the Pantheon data that surpasses the quality
of the $\Lambda$CDM model. The function $F(z)$ is parametrized as
$F(z)=1-(1-F_{\infty}).\left(1-(1+z)^{-2}\right)^{2}$ with $F_{\infty}=0.931\pm0.11$
(95\% CL).\vskip4pt

{*} We obtained a new Hubble value $H_{0}=47.2\pm0.4$ (95\% CL),
which is approximately $2/3$ of the $H_{0}\approx70$ value derived
from SNeIa when relying on the $\Lambda$CDM model (with $\Omega_{\Lambda}\approx0.7$).
The reduction in the $H_{0}$ value arises from the $3/2$--multiplicative
factor in the modified Hubble law, Eq. \eqref{eq:mod-Hubble-law}.
See Section \ref{subsec:A-reduced-value}.\vskip4pt

{*} The reduced value of $H_{0}=47.2$ yields a value of $t_{0}=13.9\pm0.1$
Gy (95\% CL) for the age of an EdS universe, effectively resolving
the age paradox without requiring dark energy. See Section \ref{subsec:Resolving-the-age}.\vskip4pt

{*} Additionally, the function $F(z)$ offers a potential resolution
of the $H_{0}$ tension. See Section \ref{subsec:Resolving-H0-tension}.\vskip4pt

Importantly, Sections \ref{subsec:Obtaining-late-time-acceleration}
and \ref{subsec:Physical-intuition} provide a \emph{new perspective}:
the Hubble diagram of SNeIa does \emph{not} necessarily indicate an
accelerating universe. \emph{Instead, it supports an alternative interpretation
in favor of an EdS universe characterized by a declining speed of
light.}\vskip8pt

\textbf{\emph{On the Blanchard--Douspis--Rowan-Robinson--Sarkar
(BDRS) analysis of the CMB power spectrum, avoiding dark energy}}---Section
\ref{subsec:Revitalizing-Blanchard} revisits an important---but
long-overlooked---proposal from 2003 \citep{Blanchard-2003}, where
BDRS shed new light on the CMB power spectrum. By adopting a double-power
form for the primordial fluctuation spectrum as given Eq.$\ $\eqref{eq:BDRS},
they achieved an excellent fit to WMAP's CMB power spectrum within
an EdS model, namely, without invoking dark energy. A surprising outcome
of their analysis was a value of $H_{0}\approx46$, which, while at
odds with the canonical range of $H_{0}\sim67-73$, aligns remarkably
well with the $H_{0}=47.2$ value derived from our analysis of SNeIa.
The nearly perfect agreement in the values of $H_{0}$, obtained from
two datasets with different natures and covering separate epochs,
indicates a consistent cosmological framework based on the EdS model
with a variable speed of light, while eliminating the need for dark
energy.

\section{Conclusion}

From the Higgs-dilaton coupling, we have derived variations in Planck's
quantum of action and the speed of light in \emph{curved} spacetime.
Our work hence---for the first time---fulfills Einstein's original
vision, dating back to 1911-1912 \citep{Einstein-1911,Einstein-1912-a,Einstein-1912-b},
by integrating the \emph{local} validity of Lorentz symmetry, the
variability of the speed of light, and the (Riemannian) geometric
nature of gravity into a unified framework.\vskip6pt

Importantly, the variable speed of light provides an alternative explanation
for the Hubble diagram of SNeIa:\emph{ Rather than being attributed
to a late-time cosmic acceleration, the excess in the distance modulus
observed in high-redshift SNeIa is a result of a declining speed of
light in an expanding Einstein--de Sitter universe.}\vskip6pt

Strikingly, the reduced value of $H_{0}=47.2$ derived herein from
the SNeIa data aligns remarkably well with the $H_{0}\approx46$ value
that BDRS deduced from the CMB data, all while bypassing the need
for dark energy \citep{Blanchard-2003}. Taken together, our work
and that of BDRS diminish the necessity for the Dark Energy hypothesis,
along with its associated fine-tuning and coincidence problems.\vskip6pt

In addition, potential implications of scale invariance and our mechanism
in realms beyond late-time cosmology are briefly outlined in Appendix
\ref{sec:Outlook}.\vskip6pt

Lastly, as $\hbar$ and $c$ become variable in scale-invariant gravity,
they can no longer serve as ``fundamental units'' within Planck's
$cGh$ system \citep{Okun-1991,Okun-2004}. Appendix \ref{sec:Unit-systems}
introduces a new unit system suitable for variable $\hbar$ and $c$.\vskip12pt
\begin{acknowledgments}
I thank Clifford Burgess, Tiberiu Harko, Robert Mann, Anne-Christine
Davis, Eoin \'O Colg\'ain, Subir Sarkar, Leandros Perivolaropoulos,
Demosthenes Kazanas, John Moffat, Alberto Salvio, Ilya L. Shapiro,
Gregory Volovik and Bruce Bassett for their constructive and insightful
feedback. I wish to thank the anonymous Referee for his or her comments,
which significantly helped improve the paper.\vspace{.5cm}
\end{acknowledgments}

\noindent \begin{center}
---------------$\infty$---------------\newpage
\par\end{center}

\appendix

\section{\label{sec:Outlook}$\ $\fontsize{8pt}{0pt}\selectfont{\bf{AN OUTLOOK BEYOND LATE-TIME COSMOLOGY}}}

Here, we \emph{speculate} on the potential ramifications of scale
invariance, as well as variable $c$ and $\hbar$, in realms beyond
late-time cosmology.\vskip6pt

{*} \emph{Early-time cosmology:} Section \ref{subsec:cosmo-horizon}
``An infinite cosmological horizon using VSL'' explicitly shows that
an EdS universe with declining speed of light exhibits an \emph{infinite}
cosmological horizon; see Eq. \eqref{eq:cosmo-horizon}. This result
could help resolve the horizon paradox, which was one of the key motivations
for the VSL proposals made by Moffat and the Albrecht--Magueijo team
in the 1990s \citep{Moffat-1992,Albrecht-1999}.\vskip4pt

{*} \emph{On the flattening galactic rotation curves:} For $\mathcal{L}_{\text{grav}}$
in Eq. \eqref{eq:L-grav}, the case of $\omega=0$ and $V(\chi)\propto\chi^{4}$
corresponds to quadratic gravity. It has been shown \citep{Nguyen-Newtonian,AlvarezGaume-2016}
that when expanding around a de Sitter background metric (rather than
the flat background conventionally used), the \emph{massless} rank-2
tensor graviton mode produces a long-range gravitational potential
with the correct Newtonian tail, viz. $\sim1/r$. Furthermore, it
has been shown \citep{AlvarezGaume-2016} that the \emph{massless}
rank-0 scalar graviton mode yields a long-range potential which grows
linearly with distance, viz. $\sim r$. A superposition of these modes
could give rise to the Mannheim--Kazanas potential (previously derived
for conformal gravity \citep{MannheimKazanas-1989,MannheimKazanas-1994}),
$-GM/r+\gamma\,r$, where $GM$ and $\gamma$ are two adjustable parameters.
In \citep{Kazanas-2021,MannheimOBrien-2012}, this potential has been
employed to successfully account for the observed flattening of rotation
curves across a wide range of galaxies, without invoking dark matter.\vskip4pt

{*} \emph{Variable $\hbar$: }A Planck constant that varies in spacetime
could, in principle, alter---both qualitatively and quantitatively---the
radiative behavior of black holes in scale-invariant gravity.\vskip4pt

{*} \emph{On the Hierarchy Problem and the Cosmological Constant Problem:}
In quantum field theories of matter in \emph{flat }spacetime, the
loop corrections to the Higgs mass and vacuum energy manifest as quadratic
and quartic divergences, respectively \citep{Martin-2012,Weinberg-2000,Padilla-2015,Burgess-2013,Csaki-2013}.
However, it is plausible to suspect that these divergences only appear
problematic \emph{when considered in the context of flat spacetime}---namely,
in isolation from the otherwise \emph{curved} background spacetime.
That is to say, these loop corrections currently account for contributions
solely from the matter sector, while neglecting those from the gravitational
sector. One intuitive way to understand this is that in scale-invariant
gravity, as $\hbar$ and $c$ become dependent on the dilaton per
$\hbar\propto\chi^{-1/2}$ and $c\propto\chi^{1/2}$, the dilaton---as
a component of the gravitational sector---could, in turn, modulate
these loop corrections and keep them in check. If so, scale invariance,
as inspired by Bardeen's seminal idea \citep{Bardeen-1995}, might
open a pathway toward protecting the Higgs mass and vacuum energy
from divergence and fine tuning.\vskip6pt

These intriguing possibilities, albeit \emph{speculative} at this
stage, represent open avenues for future investigation.\vskip4pt

\section{\label{sec:Unit-systems}$\ $\fontsize{8pt}{0pt}\selectfont{\bf{ON PLANCK'S AND HARTREE'S UNIT SYSTEMS}}}

In our mechanism, the electron mass $m$ and charge $e$ are \emph{dimensionless}
quantities derived from the parameters $f$, $\mu$, $\lambda$, and
$\alpha$ via Eq. \eqref{eq:m-e-def}. The dilaton field $\chi$ is
the single entity that carries a scale, which then serves as \emph{the
only unit available}. Consequently, all physical quantities can be
expressed \emph{exclusively} in terms of the magnitude of $\chi$.
For instance, $\hbar$ and $c$ have the units of $\chi^{-1/2}$ and
$\chi^{1/2}$, respectively, whereas $G$ is dimensionless. The rest
mass energy $E=mc^{2}$ has the unit of $\chi$, owing to $m$ being
dimensionless and $c\propto\chi^{1/2}$. Similarly, the Coulomb potential
of a point charge $e$, given by $e^{2}/r$, also carries the unit
of $\chi$, by virtue of $e$ being dimensionless and $r\propto\chi^{-1}$.
In general, a theory with dilatation symmetry would require \emph{only
one `unit'} (associated with the dilaton field $\chi$), rendering
the canonical three-unit system (mass $M$, length $L$, time $T$)
\emph{unnecessary}. It is worth noting that this conclusion agrees
with Ref. \citep{Matsas-2023}, which deduced from a spacetime perspective
that the number of fundamental constants equals one in relativistic
spacetimes.\vskip4pt

Nevertheless, to distinguish between mass and charge, which represent
two distinct aspects of the electron (viz., inertia and $U(1)$ gauge
coupling strength), we introduce two `labels' $m_{0}$ and $e_{0}$
into their definitions, Eq. \eqref{eq:m-e-def}
\begin{equation}
m:=\frac{f\mu}{\sqrt{\lambda}}\,m_{0};\ \ \ e:=\sqrt{\alpha}\,e_{0}
\end{equation}
This redefinition leads to
\begin{equation}
\hbar_{*}:=e_{0}\sqrt{m_{0}}\,\chi_{*}^{-1/2};\ \ c_{*}:=\frac{e_{0}}{\sqrt{m_{0}}}\,\chi_{*}^{1/2};\ \ G:=\frac{e_{0}^{2}}{m_{0}^{2}}\label{eq:app-1}
\end{equation}

Our new unit system based on the trio $\{e_{0},\,m_{0},\,\chi_{*}\}$
deviates from the Planck $cGh$ unit system, which relies on the trio
of constants $\{c,\,G,\,\hbar\}$. Note that our new unit system is
akin to the unit system introduced by Hartree in 1928 \citep{Hartree-1928},
based on the electron charge $e$, mass $m$, and the Bohr radius
$a_{B}$ (the typical size of the hydrogen atom in its ground state).
In the Hartree system, the Planck constant and speed of light are
defined via $\{e,\,m,\,a_{B}\}$ as 
\begin{align}
\hbar & =e\sqrt{m\,a_{B}};\ \ \ c=\frac{e}{\alpha\sqrt{ma_{B}}}
\end{align}
in close resemblance to the first two identities in Eq. \eqref{eq:app-1}.


\begin{thebibliography}{10}
\bibitem{Perivolaropoulos-2022}L. Perivolaropoulos and F. Skara,
\emph{Challenges for \textgreek{L}CDM: An update}, New Astron. Rev.
\textbf{95}, 101659 (2022), \textcolor{purple}{\href{https://arxiv.org/abs/2105.05208}{\tt 2105.05208 [astro-ph.CO]}}

\bibitem{diValentino-2021}E. Di Valentino, O. Mena, S. Pan, L. Visinelli,
W. Yang, A. Melchiorri, D. F. Mota, A. G. Riess and J. Silk,\emph{
In the realm of the Hubble tension---a review of solutions}, Class.
Quant. Grav. \textbf{38} (2021) 153001, \textcolor{purple}{\href{https://arxiv.org/abs/2103.01183}{\tt 2103.01183 [astro-ph.CO]}}

\bibitem{Secrest-2022}N. Secrest, S. von Hausegger, M. Rameez, R.
Mohayaee and S. Sarkar, \emph{A Challenge to the Standard Cosmological
Model}, Astrophys. J. Lett. \textbf{937} (2022) L31, \textcolor{purple}{\href{https://arxiv.org/abs/2206.05624}{\tt 2206.05624 [astro-ph.CO]}}

\bibitem{Bull}P. Bull, Y. Arkrami et al, \emph{Beyond \textgreek{L}CDM:
Problems, solutions, and the road ahead}, Physics of the Dark Universe
\textbf{12} (2016) 56, \textcolor{purple}{\href{https://arxiv.org/abs/1512.05356}{\tt 1512.05356 [astro-ph.CO]}}

\bibitem{Peccei-1987}R.D. Peccei, J. Sola and C. Wetterich, \emph{Adjusting
the Cosmological Constant Dynamically: Cosmons and a New Force Weaker
Than Gravity}, Phys. Lett. B \textbf{195} (1987) 183-190

\bibitem{Zhao-2017}G-B. Zhao et al., \emph{Dynamical dark energy
in light of the latest observations}, Nature Astronomy \textbf{1},
627-632 (2017), \textcolor{purple}{\href{https://arxiv.org/abs/1701.08165}{\tt 1701.08165 [astro-ph.CO]}}

\bibitem{Okun-1991}L. B. Okun, \emph{The fundamental constants of
physics}, Usp. Fiz. Nauk \textbf{161}, 177-194 (1991)

\bibitem{Okun-2004}L. B. Okun, \emph{Fundamental units: Physics and
Metrology}, in ``Astrophysics, Clocks and Fundamental Constants'',
S.G. Karshenboim and E. Peik (Eds.), Springer (2004)

\bibitem{Dirac-LNH}P. A. M. Dirac, \emph{The Cosmological Constants},
Nature \textbf{139} (3512): 323

\bibitem{BransDicke-1961}C. H. Brans and R. Dicke, \emph{Mach's Principle
and a relativistic theory of gravitation}, Phys. Rev. \textbf{124},
925 (1961)

\bibitem{Fujii-book}Y. Fujii and K. Maeda, \emph{The Scalar-Tensor
Theory of Gravitation}, Cambridge University Press (2007)

\bibitem{Utiyama-1962}R. Utiyama and B. S. DeWitt, \emph{Renormalization
of a classical gravitational field interacting with quantized matter
fields}, J. Math. Phys. \textbf{3}, 608 (1962)

\bibitem{Wetterich-1988a}C. Wetterich, \emph{Cosmologies with variable
Newton's `constant'}, Nucl. Phys. B \textbf{302}, 645 (1988)

\bibitem{Wetterich-1988b}C. Wetterich, \emph{Cosmology and the fate
of dilatation symmetry}, Nucl. Phys. B \textbf{302}, 668 (1988), \textcolor{purple}{\href{https://arxiv.org/abs/1711.03844}{\tt 1711.03844 [hep-th]}}

\bibitem{Wetterich-2013a}C. Wetterich, \emph{Variable gravity Universe},
Phys. Rev. D \textbf{89}, 024005 (2014), \textcolor{purple}{\href{https://arxiv.org/abs/1308.1019}{\tt 1308.1019 [astro-ph.CO]}}

\bibitem{Wetterich-2013b}C. Wetterich, \emph{Universe without expansion},
Phys. Dark Univ. \textbf{2}, 184 (2013), \textcolor{purple}{\href{https://arxiv.org/abs/1303.6878}{\tt 1303.6878 [astro-ph.CO]}}

\bibitem{Wetterich-2014}C. Wetterich, \emph{Eternal Universe}, Phys.
Rev. D \textbf{90}, 043520 (2014), \textcolor{purple}{\href{https://arxiv.org/abs/1404.0535}{\tt 1404.0535 [gr-qc]}}

\bibitem{Bezrukov:2011ghu}F. Bezrukov, G. K. Karananas, J. Rubio
and M. Shaposhnikov, \emph{Higgs-Dilaton Cosmology: an effective field
theory approach}, Phys. Rev. D \textbf{87}, 096001 (2013), \textcolor{purple}{\href{https://arxiv.org/abs/1212.4148}{\tt 1212.4148 [hep-ph]}}

\bibitem{Fujii-1982}Y. Fujii, \emph{Origin of the gravitational constant
and particle masses in scale invariant scalar--tensor theory}, Phys.
Rev. D \textbf{26}, 2580 (1982)

\bibitem{VSL2024-Pantheon}H. K. Nguyen, \emph{New analysis of SNeIa
Pantheon Catalog: Variable speed of light as an alternative to dark
energy and late-time cosmic acceleration}, JCAP \textbf{04} (2025) 005,
\textcolor{purple}{\href{https://arxiv.org/abs/2412.05262}{\tt 2412.05262 [astro-ph.CO]}}

\bibitem{Dicke-1957}R. H. Dicke, \emph{Gravitation without a Principle
of Equivalence}, Rev. Mod. Phys. \textbf{29}, 363 (1957) 

\bibitem{Einstein-1911}A. Einstein, \emph{On the influence of gravitation
on the propagation of light,} Annalen der Physik \textbf{35}, 898-908
(1911), \textcolor{purple}{\href{http://einsteinpapers.press.princeton.edu/vol3-trans/393}{einsteinpapers.press.princeton.edu/vol3-trans/393}}

\bibitem{Einstein-1912-a}A. Einstein, \emph{The speed of light and
the statics of the gravitational field,} Annalen der Physik \textbf{38},
355-369 (1912), \textcolor{purple}{\href{http://einsteinpapers.press.princeton.edu/vol4-trans/107}{einsteinpapers.press.princeton.edu/vol4-trans/107}}

\bibitem{Einstein-1912-b}A. Einstein, \emph{Relativity and gravitation:
Reply to a comment by M. Abraham,} Annalen der Physik \textbf{38},
1059-1064 (1912), \textcolor{purple}{\href{http://einsteinpapers.press.princeton.edu/vol4-trans/142}{einsteinpapers.press.princeton.edu/vol4-trans/142}}

\bibitem{Moffat-1992}J. W. Moffat, \emph{Superluminary universe:
A possible solution to the initial value problem in cosmology}, Int.
J. Mod. Phys. D \textbf{2}, 351 (1993), \textcolor{purple}{\href{https://arxiv.org/abs/gr-qc/9211020}{\tt gr-qc/9211020}}

\bibitem{Albrecht-1999}A. Albrecht and J. Magueijo, \emph{Time varying
speed of light as a solution to cosmological puzzles,} Phys. Rev.
D \textbf{59}, 043516 (1999), \textcolor{purple}{\href{https://arxiv.org/abs/astro-ph/9811018}{\tt astro-ph/9811018}}

\bibitem{Barrow-2000}J. D. Barrow and J. Magueijo, \emph{Can a changing
$\alpha$ explain the Supernovae results?}, Astrophy. J. \textbf{532},
87 (2000), \textcolor{purple}{\href{https://arxiv.org/abs/astro-ph/9907354}{\tt astro-ph/9907354} }

\bibitem{Zhang-2014}{[}12{]} P. Zhang and X. Meng, \emph{SNe data
analysis in variable speed of light cosmologies without cosmological
constant}, Mod. Phys. Lett. A \textbf{29}, 1450103 (2014), \textcolor{purple}{\href{https://arxiv.org/abs/1404.7693}{\tt 1404.7693 [astro-ph.CO]}}

\bibitem{Qi-2014}J-Z. Qi, M-J. Zhang, and W-B. Liu, \emph{Observational
constraint on the varying speed of light theory}, Phys. Rev. D \textbf{90},
063526 (2014), \textcolor{purple}{\href{https://arxiv.org/abs/1407.1265}{\tt 1407.1265 [gr-qc]}}

\bibitem{Ravanpak-2017}A. Ravanpak, H. Farajollahi, and G. F. Fadakar,
\emph{Normal DGP in varying speed of light cosmology}, Res. Astron.
Astrophys. \textbf{17}, 26 (2017),\textcolor{purple}{{} \href{https://arxiv.org/abs/1703.09811}{\tt 1703.09811}}

\bibitem{Salzano-2016a}V. Salzano, \emph{Recovering a redshift-extended
VSL signal from galaxy surveys}, Phys. Rev. D \textbf{95}, 084035
(2017), \textcolor{purple}{\href{https://arxiv.org/abs/1604.03398}{\tt 1604.03398 [astro-ph.CO]}}

\bibitem{Salzano-2016d}V. Salzano and M. P. D\c{a}browski, \emph{Statistical
hierarchy of varying speed of light cosmologies}, Astrophys. J. \textbf{851},
97 (2017), \textcolor{purple}{\href{https://arxiv.org/abs/1612.06367}{\tt 1612.06367}}

\bibitem{Rodrigues-2022}G. Rodrigues and C. Bengaly, \emph{A model-independent
test of speed of light variability with cosmological observations},
JCAP \textbf{07}, 029 (2022),\textcolor{purple}{{} \href{https://arxiv.org/abs/2112.01963}{\tt 2112.01963 [astro-ph.CO]}}

\bibitem{Cai-2016}R. G. Cai, Z. K. Guo and T. Yang, \emph{Dodging
the cosmic curvature to probe the constancy of the speed of light},
JCAP \textbf{08} (2016), 016, \textcolor{purple}{\href{https://arxiv.org/abs/1601.05497}{\tt 1601.05497 [astro-ph.CO]}}

\bibitem{Liu-2023}Y. Liu, S. Cao, M. Biesiada, Y. Lian, X. Liu and
Y. Zhang, \emph{Measuring the Speed of Light with Updated Hubble Diagram
of High-redshift Standard Candles}, Astrophys. J. \textbf{949} (2023)
no.2, 57, \textcolor{purple}{\href{https://arxiv.org/abs/2303.14674}{\tt 2303.14674 [astro-ph.CO]}}

\bibitem{Barrow-1998a}J. D. Barrow, \emph{Cosmologies with varying
light speed}, Phys. Rev. D \textbf{59}, 043515 (1999), \textcolor{purple}{\href{https://arxiv.org/abs/astro-ph/9811022}{\tt astro-ph/9811022}}

\bibitem{Barrow-1998b}J. D. Barrow and J. Magueijo, \emph{Varying-$\alpha$
theories and solutions to the cosmological problems}, Phys. Lett.
B \textbf{443}, 104 (1998), \textcolor{purple}{\href{https://arxiv.org/abs/astro-ph/9811072}{\tt astro-ph/9811072}}

\bibitem{Magueijo-2002}J. Magueijo and L. Smolin, \emph{Lorentz invariance
with an invariant energy scale}, Phys. Rev. Lett. \textbf{88}, 190403
(2002), \textcolor{purple}{\href{https://arxiv.org/abs/hep-th/0112090}{\tt hep-th/0112090}}

\bibitem{Magueijo-2003}J. Magueijo, \emph{New varying speed of light
theories}, Rept. Prog. Phys. \textbf{66}, 2025 (2003), \textcolor{purple}{\href{https://arxiv.org/abs/astro-ph/0305457}{\tt astro-ph/0305457}}

\bibitem{Barrow-1999a}J. D. Barrow and J. Magueijo, \emph{Solutions
to the Quasi-flatness and Quasi-lambda Problems}, Phys. Lett. B \textbf{447},
246 (1999), \textcolor{purple}{\href{https://arxiv.org/abs/astro-ph/9811073}{\tt astro-ph/9811073}}

\bibitem{Barrow-1999b}J. D. Barrow and J. Magueijo, \emph{Solving
the Flatness and Quasi-flatness Problems in Brans-Dicke Cosmologies
with a Varying Light Speed}, Class. Quant. Grav.\textbf{ 16}, 1435
(1999), \textcolor{purple}{\href{https://arxiv.org/abs/astro-ph/9901049}{\tt astro-ph/9901049}}

\bibitem{Clayton-1999}M. A. Clayton and J. W. Moffat, \emph{Dynamical
mechanism for varying light velocity as a solution to cosmological
problems}, Phys. Lett. B \textbf{460}, 263 (1999), \textcolor{purple}{\href{https://arxiv.org/abs/astro-ph/9812481}{\tt astro-ph/9812481 [astro-ph]}}

\bibitem{Avelino-1999}P. P. Avelino and C. J. A. P. Martins, \emph{Does
a varying speed of light solve the cosmological problems?}, Phys.
Lett. B \textbf{459} (1999), 468-472, \textcolor{purple}{\href{https://arxiv.org/abs/astro-ph/9906117}{\tt astro-ph/9906117 [astro-ph]}}

\bibitem{Avelino-2000}P. P. Avelino, C. J. A. P. Martins and G. Rocha,
\emph{VSL theories and the Doppler peak}, Phys. Lett. B \textbf{483}
(2000) 210, \textcolor{purple}{\href{https://arxiv.org/abs/astro-ph/0001292}{\tt astro-ph/0001292}}

\bibitem{Clayton-2000}M. A. Clayton and J. W. Moffat, \emph{Scalar
tensor gravity theory for dynamical light velocity}, Phys. Lett. B
\textbf{477} (2000) 269-275, \textcolor{purple}{\href{https://arxiv.org/abs/gr-qc/9910112}{\tt gr-qc/9910112 [gr-qc]}}

\bibitem{Magueijo-2000}J. Magueijo, \emph{Covariant and locally Lorentz
invariant varying speed of light theories}, Phys. Rev. D \textbf{62}
(2000) 103521, \textcolor{purple}{\href{https://arxiv.org/abs/gr-qc/0007036}{\tt gr-qc/0007036}}

\bibitem{Clayton-2002}M. A. Clayton and J. W. Moffat,\emph{ Vector
field mediated models of dynamical light velocity}, Int. J. Mod. Phys.
D \textbf{11} (2002) 187-206, \textcolor{purple}{\href{https://arxiv.org/abs/gr-qc/0003070}{\tt gr-qc/0003070 [gr-qc]}}

\bibitem{Bassett-2000}B. A. Bassett, S. Liberati, C. Molina-Par\'is
and M. Visser, \emph{Geometrodynamics of Variable-Speed-of-Light Cosmologies},
Phys. Rev. D\textbf{ 62}, 103518 (2000), \textcolor{purple}{\href{https://arxiv.org/abs/astro-ph/0001441}{\tt astro-ph/0001441}}

\bibitem{Liberati-2000}S. Liberati, B. A. Bassett, C. Molina-Par\'is
and M. Visser, \emph{Chi-Variable-Speed-of-Light Cosmologies}, Nucl.
Phys. Proc. Suppl. \textbf{88}, 259 (2000), \textcolor{purple}{\href{https://arxiv.org/abs/astro-ph/0001481}{\tt astro-ph/0001481}}

\bibitem{Drummond-1999}I. T. Drummond, \emph{Variable Light-Cone
Theory of Gravity}, \textcolor{purple}{\href{https://arxiv.org/abs/gr-qc/9908058}{\tt gr-qc/9908058}}

\bibitem{Drummond-1980}I. T. Drummond and S. J. Hathrell, \emph{QED
vacuum polarization in a background gravitational field and its effect
on the velocity of photons}, Phys. Rev. D \textbf{22}, 343 (1980)

\bibitem{Novello-1989}M. Novello and S. D. Jorda, \emph{Does there
exist a cosmological horizon problem?}, Mod. Phys. Lett. A \textbf{4},
1809 (1989)

\bibitem{Volovik-2023}G. E. Volovik, \emph{Planck constants in the
symmetry breaking quantum gravity}, Symmetry \textbf{15}, 991 (2023),
\textcolor{purple}{\href{https://arxiv.org/abs/2304.04235}{\tt 2304.04235 [cond-mat.other]}}

\bibitem{Salzano-2016b}A. Balcerzak, M. P. D\c{a}browski and V.
Salzano, \emph{Modelling spatial variations of the speed of light},
Annalen der Physik \textbf{29}, 1600409 (2017), \textcolor{purple}{\href{https://arxiv.org/abs/1604.07655}{\tt 1604.07655 [astro-ph.CO]}}

\bibitem{Gupta-2020}R. P. Gupta, \emph{Cosmology with relativistically
varying physical constants}, Mon. Not. Roy. Astron. Soc. \textbf{498}
(2020) 3, 4481-4491, \textcolor{purple}{\href{https://arxiv.org/abs/2009.08878}{\tt 2009.08878 [astro-ph.CO]}}

\bibitem{Gupta-2021}R. P. Gupta, \emph{Varying physical constants
and the lithium problem}, Astroparticle Physics \textbf{129}, 102578
(2021), \textcolor{purple}{\href{https://arxiv.org/abs/2010.13628}{\tt 2010.13628 [gr-qc]}}

\bibitem{Cuzinatto-2022}R. R. Cuzinatto, R. P. Gupta, R. F. L. Holanda,
J. F. Jesus and S. H. Pereira, \emph{Testing a varying-\textgreek{L}
model for dark energy within Co-varying Physical Couplings framework},
Mon. Not. Roy. Astron. Soc. \textbf{515}, 5981-5992 (2022), \textcolor{purple}{\href{https://arxiv.org/abs/2204.10764}{\tt 2204.10764 [gr-qc]}}

\bibitem{Abdo-2009}A. A. Abdo et al, \emph{A limit on the variation
of the speed of light arising from quantum gravity effects}, Nature
\textbf{462}, 331-334 (2009)

\bibitem{Agrawal-2021}R. Agrawal, H. Singirikonda and S. Desai, \emph{Search
for Lorentz Invariance Violation from stacked Gamma-Ray Burst spectral
lag data}, JCAP\textbf{ 05} (2021) 029, \textcolor{purple}{\href{https://arxiv.org/abs/2102.11248}{\tt 2102.11248 [astro-ph.HE]}}

\bibitem{Santos-2024}J. Santos, C. Bengaly, B. J. Morais and R. S.
Goncalves, \emph{Measuring the speed of light with cosmological observations:
current constraints and forecasts}, JCAP \textbf{11} (2024) 062, \textcolor{purple}{\href{https://arxiv.org/abs/2409.05838}{\tt 2409.05838 [astro-ph.CO]}}

\bibitem{Uzan-2003}J. P. Uzan, \emph{The Fundamental Constants and
Their Variation: Observational Status and Theoretical Motivations},
Rev. Mod. Phys. \textbf{75} (2003) 403, \textcolor{purple}{\href{https://arxiv.org/abs/hep-ph/0205340}{\tt hep-ph/0205340}}

\bibitem{Uzan-2011}J. P. Uzan, \emph{Varying Constants, Gravitation
and Cosmology}, Living Rev. Rel. \textbf{14} (2011) 2, \textcolor{purple}{\href{https://arxiv.org/abs/1009.5514}{\tt 1009.5514 [astro-ph.CO]}}

\bibitem{Buchalter-2004}A. Buchalter, \emph{On the time variation
of c, G, and h and the dynamics of the cosmic expansion}, \textcolor{purple}{\href{https://arxiv.org/abs/astro-ph/0403202}{\tt astro-ph/0403202}}

\bibitem{Martins-2017}C. J. A. P. Martins, \emph{The status of varying
constants: a review of the physics, searches and implications}, Rep.
Prog. Phys. \textbf{80}, 126902 (2017), \textcolor{purple}{\href{https://arxiv.org/abs/1709.02923}{\tt 1709.02923 [astro-ph.CO]}}

\bibitem{Ellis-2005}G. F. R. Ellis and J. P. Uzan, \emph{\textquoteleft c\textquoteright{}
is the speed of light, isn\textquoteright t it?}, Am. J. Phys. 73
(2005) 240-247, \textcolor{purple}{\href{https://arxiv.org/abs/gr-qc/0305099}{\tt gr-qc/0305099}}

\bibitem{Ellis-2007}G. F. R. Ellis, \emph{Note on Varying Speed of
Light Cosmologies}, Gen. Rel. Grav. \textbf{39} (2007) 511-520, \href{https://arxiv.org/abs/astro-ph/0703751}{\tt astro-ph/0703751}

\bibitem{Magueijo-2008}J. Magueijo and J. W. Moffat, \emph{Comments
on \textquotedbl Note on varying speed of light theories\textquotedbl},
Gen. Rel. Grav. \textbf{40}, 1797-1806 (2008), \textcolor{purple}{\href{https://arxiv.org/abs/0705.4507}{\tt 0705.4507 [gr-qc]}}

\bibitem{Cruz-2012}C. N. Cruz and A. C. A. de Faria Jr., \emph{Variation
of the speed of light with temperature of the expanding universe},
Phys. Rev. D \textbf{86} (2012) 027703, \textcolor{purple}{\href{https://arxiv.org/abs/1205.2298}{\tt 1205.2298 [gr-qc]}}

\bibitem{Moffat-2016}J. W. Moffat, \emph{Variable Speed of Light
Cosmology, Primordial Fluctuations and Gravitational Waves}, Eur.
Phys. J. C \textbf{76}, 130 (2016), \textcolor{purple}{\href{https://arxiv.org/abs/1404.5567}{\tt 1404.5567 [astro-ph.CO]}}

\bibitem{Franzmann-2017}G. Franzmann, \emph{Varying fundamental constants:
a full covariant approach and cosmological applications}, \textcolor{purple}{\href{https://arxiv.org/abs/1704.07368}{\tt 1704.07368 [gr-qc]}}

\bibitem{Cruz-2018}C. N. Cruz and F. A. da Silva, \emph{Variation
of the speed of light and a minimum speed in the scenario of an inflationary
universe with accelerated expansion}, Phys. Dark Univ. \textbf{22}
(2018) 127-136, \textcolor{purple}{\href{https://arxiv.org/abs/2009.05397}{\tt 2009.05397 [physics.gen-ph]}}

\bibitem{Costa-2019}R. Costa, R. R. Cuzinatto, E. M. G. Ferreira
and G. Franzmann, \emph{Covariant c-flation: a variational approach},
Int. J. Mod. Phys. D \textbf{28}, 1950119 (2019), \textcolor{purple}{\href{https://arxiv.org/abs/1705.03461}{\tt 1705.03461 [gr-qc]}}

\bibitem{Lee-2021a}S. Lee, \emph{The minimally extended Varying Speed
of Light (meVSL)}, JCAP \textbf{08} (2021) 054, \textcolor{purple}{\href{https://arxiv.org/abs/2011.09274}{\tt 2011.09274 [astro-ph.CO]}}

\bibitem{Balcerzak-2014a}A. Balcerzak and M. P. D\c{a}browski, \emph{Redshift
drift in varying speed of light cosmology}, Phys. Lett. B \textbf{728}
(2014) 15, \textcolor{purple}{\href{https://arxiv.org/abs/1310.7231}{\tt 1310.7231 [astro-ph.CO]}}

\bibitem{Balcerzak-2014b}A. Balcerzak and M. P. D\c{a}browski, \emph{A
statefinder luminosity distance formula in varying speed of light
cosmology}, JCAP \textbf{06} (2014) 035, \textcolor{purple}{\href{https://arxiv.org/abs/1406.0150}{\tt 1406.0150 [astro-ph.CO]}}

\bibitem{Salzano-2015}V. Salzano, M. P. D\c{a}browski and R. Lazkoz,
\emph{Measuring the speed of light with Baryon Acoustic Oscillations,}
Phys. Rev. Lett. \textbf{114}, 101304 (2015), \textcolor{purple}{\href{https://arxiv.org/abs/1412.5653}{\tt 1412.5653 [astro-ph.CO]}}

\bibitem{Salzano-2016c}V. Salzano, M. P. D\c{a}browski and R. Lazkoz,
\emph{Probing the constancy of the speed of light with future galaxy
survey: The case of SKA and Euclid}, Phys. Rev. D \textbf{93}, 063521
(2016), \textcolor{purple}{\href{https://arxiv.org/abs/1511.04732}{\tt 1511.04732 [astro-ph.CO]}}

\bibitem{Cao-2017}S. Cao, M. Biesiada, J. Jackson, X. Zheng, Y. Zhao
and Z. H. Zhu, \emph{Measuring the speed of light with ultra-compact
radio quasars}, JCAP \textbf{02} (2017) 012, \textcolor{purple}{\href{https://arxiv.org/abs/1609.08748}{\tt 1609.08748 [astro-ph.CO]}}

\bibitem{Salzano-2017a}V. Salzano, \emph{How to Reconstruct a Varying
Speed of Light Signal from Baryon Acoustic Oscillations Surveys},
Universe 3 (2017) no.2, 35

\bibitem{Lang-2018}R. G. Lang, H. Mart\'inez-Huerta and V. de Souza,
\emph{Limits on the Lorentz Invariance Violation from UHECR astrophysics},
Astrophys. J. \textbf{853} (2018) no.1, 23, \textcolor{purple}{\href{https://arxiv.org/abs/1701.04865}{\tt 1701.04865 [astro-ph.HE]}}

\bibitem{Zou-2018}X. B. Zou, H. K. Deng, Z. Y. Yin and H. Wei, \emph{Model-Independent
Constraints on Lorentz Invariance Violation via the Cosmographic Approach},
Phys. Lett. B \textbf{776} (2018) 284-294, \textcolor{purple}{\href{https://arxiv.org/abs/1707.06367}{\tt 1707.06367 [gr-qc]}}

\bibitem{HWAC-2018}H. Mart\'inez-Huerta {[}HAWC{]}, \emph{Potential
constrains on Lorentz invariance violation from the HAWC TeV gamma-rays},
PoS ICRC2017 (2018) 868, \textcolor{purple}{\href{https://arxiv.org/abs/1708.03384}{\tt 1708.03384 [astro-ph.HE]}}

\bibitem{Cao-2018}S. Cao, J. Qi, M. Biesiada, X. Zheng, T. Xu and
Z. H. Zhu, \emph{Testing the Speed of Light over Cosmological Distances:
The Combination of Strongly Lensed and Unlensed Type Ia Supernovae},
Astrophys. J. \textbf{867} (2018) no.1, 50, \textcolor{purple}{\href{https://arxiv.org/abs/1810.01287}{\tt 1810.01287 [astro-ph.CO]}}

\bibitem{Liu-2021}T. Liu, S. Cao, M. Biesiada, Y. Liu, Y. Lian and
Y. Zhang, \emph{Consistency testing for invariance of the speed of
light at different redshifts: the newest results from strong lensing
and Type Ia supernovae observations}, Mon. Not. Roy. Astron. Soc.
\textbf{506} (2021) 2, 2181-2188, \textcolor{purple}{\href{https://arxiv.org/abs/2106.15145}{\tt 2106.15145 [astro-ph.CO]}}

\bibitem{Wang-2019}D. Wang, H. Zhang, J. Zheng, Y. Wang and G. B.
Zhao, \emph{Reconstructing the temporal evolution of the speed of
light in a flat FRW Universe}, Res. Astron. Astrophys. \textbf{19}
(2019) 10, 152,\textcolor{purple}{{} \href{https://arxiv.org/abs/1904.04041}{\tt 1904.04041 [astro-ph.CO]}}

\bibitem{HAWC-2020}A. Albert et al. {[}HAWC{]}, \emph{Constraints
on Lorentz Invariance Violation from HAWC Observations of Gamma Rays
above 100 TeV}, Phys. Rev. Lett. \textbf{124} (2020) no.13, 131101,
\textcolor{purple}{\href{https://arxiv.org/abs/1911.08070}{\tt 1911.08070 [astroph.HE]}}

\bibitem{Pan-2020}Y. Pan, J. Qi, S. Cao, T. Liu, Y. Liu, S. Geng,
Y. Lian and Z. H. Zhu, \emph{Model-independent constraints on Lorentz
invariance violation: implication from updated Gamma-ray burst observations},
Astrophys. J. \textbf{890} (2020) 169, \textcolor{purple}{\href{https://arxiv.org/abs/2001.08451}{\tt 2001.08451 [astro-ph.CO]}}

\bibitem{Mendonca-2021}I. E. C. R. Mendonca, K. Bora, R. F. L. Holanda,
S. Desai and S. H. Pereira, \emph{A search for the variation of speed
of light using galaxy cluster gas mass fraction measurements}, JCAP
\textbf{11} (2021) 034, \textcolor{purple}{\href{https://arxiv.org/abs/2109.14512}{\tt 2109.14512 [astro-ph.CO]}}

\bibitem{Lee-2023a}S. Lee, \emph{Constraining minimally extended
varying speed of light by cosmological chronometers}, Mon. Not. Roy.
Astron. Soc. \textbf{522} (2023) no.3, 3248-3255, \textcolor{purple}{\href{https://arxiv.org/abs/2301.06947}{\tt 2301.06947 [astro-ph.CO]}}

\bibitem{Eaves-2022}R. E. Eaves, \emph{Redshift in varying speed
of light cosmology}, Mon. Not. Roy. Astron. Soc. \textbf{516}, 4136-4145
(2022)

\bibitem{Mukherjee-2024}P. Mukherjee, G. Rodrigues and C. Bengaly,
\emph{Examining the validity of the minimal varying speed of light
model through cosmological observations: Relaxing the null curvature
constraint}, Phys. Dark Univ. \textbf{43} (2024) 101380, \textcolor{purple}{\href{https://arxiv.org/abs/2302.00867}{\tt 2302.00867 [astro-ph.CO]}}

\bibitem{Lee-2023b}S. Lee, \emph{Constraint on the minimally extended
varying speed of light using time dilations in Type Ia supernovae},
Mon. Not. Roy. Astron. Soc. \textbf{524} (2023) no.3, 4019-4023, \textcolor{purple}{\href{https://arxiv.org/abs/2302.09735}{\tt arXiv:2302.09735 [astro-ph.CO]}}

\bibitem{Lee-2021b}S. Lee, \emph{Constraints on the time variation
of the speed of light using Strong lensing}, \textcolor{purple}{\href{https://arxiv.org/abs/2104.09690}{\tt arXiv:2104.09690 [astro-ph.CO]}}

\bibitem{Cuzinatto-2023}R. R. Cuzinatto, C. A. M. de Melo and J.
C. S. Neves, \emph{Shadows of black holes at cosmological distances
in the co-varying physical couplings framework}, Mon. Not. Roy. Astron.
Soc. \textbf{526} (2023) no.3, 3987-3993, \textcolor{purple}{\href{https://arxiv.org/abs/2305.11118}{\tt arXiv:2305.11118 [gr-qc]}}

\bibitem{Zhang-2024}C. Y. Zhang, W. Hong, Y. C. Wang and T. J. Zhang,
\emph{A Stochastic Approach to Reconstructing the Speed of Light in
Cosmology}, Mon. Not. Roy. Astron. Soc. \textbf{534} (2024) 56-69,
\textcolor{purple}{\href{https://arxiv.org/abs/2409.03248}{\tt 2409.03248 [astro-ph.CO]}}

\bibitem{Colaco-2022}L. R. Cola\c{c}o, S. J. Landau, J. E. Gonzalez,
J. Spinelly and G. L. F. Santos, \emph{Constraining a possible time-variation
of the speed of light along with the fine-structure constant using
strong gravitational lensing and Type Ia supernovae observations},
JCAP\textbf{ 08} (2022) 062, \textcolor{purple}{\href{https://arxiv.org/abs/2204.06459}{\tt 2204.06459 [astro-ph.CO]}}

\bibitem{Liu-2018}Y. Liu and B-Q. Ma, \emph{Light speed variation
from gamma ray bursts: criteria for low energy photons}, Eur. Phys.
J. C \textbf{78} (2018) 825, \textcolor{purple}{\href{https://arxiv.org/abs/1810.00636}{\tt 1810.00636 [astro-ph.HE]}}

\bibitem{Xu-2016a}H. Xu and B-Q. Ma, \emph{Light speed variation
from gamma-ray bursts}, Astropart. Phys. \textbf{82}, 72 (2016), \textcolor{purple}{\href{https://arxiv.org/abs/1607.03203}{\tt 1607.03203 [hep-ph]}}

\bibitem{Xu-2016b}H. Xu and B-Q. Ma, \emph{Light speed variation
from gamma ray burst GRB 160509A}, Phys. Lett. B \textbf{760} (2016)
602-604, \textcolor{purple}{\href{https://arxiv.org/abs/1607.08043}{\tt 1607.08043 [hep-ph]}}

\bibitem{Zhu-2021}J. Zhu and B-Q. Ma, \emph{Pre-burst events of gamma-ray
bursts with light speed variation}, Phys. Lett. B \textbf{820} (2021)
136518, \textcolor{purple}{\href{https://arxiv.org/abs/2108.05804}{\tt 2108.05804 [astro-ph.HE]}}

\bibitem{Mangano-2015}G. Mangano, F. Lizzi, and A. Porzio, \emph{Inconstant
Planck\textquoteright s constant}, Int. J. Mod. Phys. A \textbf{30}
(2015) 34, 1550209, \textcolor{purple}{\href{https://arxiv.org/abs/1509.02107}{\tt 1509.02107 [quant-ph]}}

\bibitem{Webb-1999}J. K. Webb, V. V. Flambaum, C. W. Churchill, M.
J. Drinkwater, and J. D. Barrow,\emph{ Evidence for time variation
of the fine structure constant}, Phys. Rev. Lett. \textbf{82}, 884
(1999),\textcolor{purple}{{} \href{https://arxiv.org/abs/astro-ph/9803165}{\tt astro-ph/9803165}}

\bibitem{Webb-2001}J. K. Webb, M. T. Murphy, V. V. Flambaum, V. A.
Dzuba, J. D. Barrow, C. W. Churchill, J. X. Prochaska and A. M. Wolfe,
\emph{Further Evidence for Cosmological Evolution of the Fine Structure
Constant}, Phys. Rev. Lett. \textbf{87} (2001) 091301, \href{https://arxiv.org/abs/astro-ph/0012539}{\tt astro-ph/0012539}

\bibitem{Jiang-2024}L. Jiang et al, \emph{Constraints on the variation
of the fine-structure constant at $3<z<10$ with JWST emission-line
galaxies}, Astrophys. J. \textbf{980} (2025) 1, 93,\textcolor{purple}{{}
\href{https://arxiv.org/abs/2405.08977}{\tt 2405.08977 [astro-ph.CO]}}

\bibitem{Banks-2002}T. Banks, M. Dine and M. R. Douglas, \emph{Time-Varying
alpha and Particle Physics}, Phys. Rev. Lett. \textbf{88} (2002) 131301,
\textcolor{purple}{\href{https://arxiv.org/abs/hep-ph/0112059}{\tt hep-ph/0112059}}

\bibitem{Clifton-review}T. Clifton, P. G. Ferreira, A. Padilla, and
C. Skordis, \emph{Modified Gravity and Cosmology}, Phys. Rep. \textbf{513},
1 (2012),\textcolor{purple}{{} \href{https://arxiv.org/abs/1106.2476}{\tt 1106.2476 [astro-ph.CO]}}

\bibitem{Capo-review}S. Capozziello and M. De Laurentis, \emph{Extended
Theories of Gravity}, Phys. Rept. \textbf{509} (2011) 167-321,\textcolor{purple}{{}
\href{https://arxiv.org/abs/1108.6266}{\tt 1108.6266 [gr-qc]}}

\bibitem{Odintsov-review}S. Nojiri, S. D. Odintsov and V. K. Oikonomou,
\emph{Modified Gravity Theories on a Nutshell: Inflation, Bounce and
Late-time Evolution}, Phys. Rept. \textbf{692} (2017) 1-104,\textcolor{purple}{{}
\href{https://arxiv.org/abs/1705.11098}{\tt 1705.11098 [gr-qc]}}

\bibitem{Bohmer-review}C. G. B\"ohmer and E. Jensko, \emph{Modified
gravity: A unified approach}, Phys. Rev. D \textbf{104}, 024010 (2021),
\textcolor{purple}{\href{https://arxiv.org/abs/2103.15906}{\tt 2103.15906 [gr-qc]}}

\bibitem{CANTANA-2021}E. N. Saridakis, R. Lazkoz, V. Salzano, P.
V. Moniz, S. Capozziello, J. B. Jim\'enez, M. De Laurentis, G. J.
Olmo, Y. Akrami, S. Bahamonde, J. L. Bl\'azquez-Salcedo, C. G. B\"ohmer,
C. Bonvin, M. Bouhmadi-L\'opez, P. Brax, G. Calcagni, R. Casadio,
J. A. R. Cembranos, \'A. de la Cruz-Dombriz, A-C. Davis, A. Delhom,
E. Di Valentino, K. F. Dialektopoulos, B. Elder, J. M. Ezquiaga, N.
Frusciante, R. Garattini, L. \'A. Gergely, A. Giusti, L. Heisenberg,
M. Hohmann, D. Iosifidis, L. Kazantzidis, B. Kleihaus, T. S. Koivisto,
J. Kunz, F. S. N. Lobo, M. Martinelli, P. Mart\'in-Moruno, J. P.
Mimoso, D. F. Mota, S. Peirone, L. Perivolaropoulos, V. Pettorino,
C. Pfeifer, L. Pizzuti, D. Rubiera-Garcia, J. Levi Said, M. Sakellariadou,
I. D. Saltas, A. S. Mancini, N. Voicu and A. Wojnar, \emph{Modified
Gravity and Cosmology: An Update by the CANTATA Network}, Springer
2021, \textcolor{purple}{\href{https://arxiv.org/abs/2105.12582}{\tt 2105.12582 [gr-qc]}}

\bibitem{Buchdahl-1970}H. A. Buchdahl, \emph{Non-linear Lagrangians
and cosmological theory}, Mon. Not. Roy. Astron. Soc. \textbf{150},
1 (1970)

\bibitem{deFelice-review}A. De Felice and S. Tsujikawa, \emph{$f(R)$
theories}, Living Rev. Rel. \textbf{13}, 3 (2010),\textcolor{purple}{{}
\href{https://arxiv.org/abs/1002.4928}{\tt 1002.4928 [gr-qc]}}

\bibitem{Sotiriou-review}T. P. Sotiriou and V. Faraoni, \emph{$f(R)$
Theories Of Gravity}, Rev. Mod. Phys. \textbf{82}, 451-497 (2010),
\textcolor{purple}{\href{https://arxiv.org/abs/0805.1726}{\tt 0805.1726 [gr-qc]}}

\bibitem{Blas-2011}D. Blas, M. Shaposhnikov and D. Zenhausern, \emph{Scale-invariant
alternatives to general relativity}, Phys. Rev. D\textbf{ 84}, 044001
(2011), \textcolor{purple}{\href{https://arxiv.org/abs/1104.1392}{\tt 1104.1392 [hep-th]}}

\bibitem{Ghilencea-2023}D. Ghilencea and C. T. Hill, \emph{Renormalization
Group for Non-minimal $\phi^{2}\,R$ Couplings and Gravitational Contact
Interactions}, Phys. Rev. D \textbf{107} (2023) 8, 085013, \textcolor{purple}{\href{https://arxiv.org/abs/2210.15640}{\tt 2210.15640 [gr-qc]}}

\bibitem{Ferreira-2018}P. G. Ferreira, C. T. Hill, J. Noller and
G. G. Ross, \emph{Inflation in a scale invariant universe}, Phys.
Rev. D \textbf{97}, 123516 (2018), \textcolor{purple}{\href{https://arxiv.org/abs/1802.06069}{\tt 1802.06069 [astro-ph.CO]}}

\bibitem{Salvio-2014}A. Salvio and A. Strumia, \emph{Agravity}, J.
High Energy Phys. \textbf{06}, 080 (2014), \textcolor{purple}{\href{https://arxiv.org/abs/1403.4226}{\tt 1403.4226 [hep-ph]}}

\bibitem{Einhorn-2015}M. B. Einhorn and D. R. T. Jones, \emph{Naturalness
and dimensional transmutation in classically scale-invariant gravity},
J. High Energy Phys. \textbf{03}, 047 (2015), \textcolor{purple}{\href{https://arxiv.org/abs/1410.8513}{\tt 1410.8513 [hep-th]}}

\bibitem{Edery-2014}A. Edery and Y. Nakayama, \emph{Restricted Weyl
invariance in four-dimensional curved spacetime}, Phys. Rev. D \textbf{90},
043007 (2014), \textcolor{purple}{\href{https://arxiv.org/abs/1406.0060}{\tt 1406.0060 [hep-th]}}

\bibitem{Horndeski-1974}G. W. Horndeski, \emph{Second-order scalar-tensor
field equations in a four-dimensional space}, Int. J. Theor. Phys.
\textbf{10} (6): 363 (1974)

\bibitem{Cuzinatto-2021}R. R. Cuzinatto, E. M. De Morais, and B.
M. Pimentel, \emph{Lyra scalar-tensor theory: A scalar-tensor theory
of gravity on Lyra manifold}, Phys. Rev. D \textbf{103}, 124002 (2021),
\textcolor{purple}{\href{https://arxiv.org/abs/2104.06295}{\tt 2104.06295 [gr-qc]}}

\bibitem{Ferreira-2016}P. G. Ferreira, C. T. Hill and G. G. Ross,
\emph{No fifth force in a scale invariant universe}, Phys. Rev. D
\textbf{95}, 064038 (2017), \textcolor{purple}{\href{https://arxiv.org/abs/1612.03157}{\tt 1612.03157 [gr-qc]}}

\bibitem{AlvarezGaume-2016}L. Alvarez-Gaume, A. Kehagias, C. Kounnas,
D. L\"ust, and A. Riotto, \emph{Aspects of Quadratic Gravity}, Fortsch.
Phys. \textbf{64}, 176 (2016),\textcolor{purple}{{} \href{https://arxiv.org/abs/1505.07657}{\tt 1505.07657 [hep-th]}}

\bibitem{Alvarez-2018}E. Alvarez, J. Anero, S. Gonzalez-Martin and
R. Santos-Garcia, \emph{Physical content of quadratic gravity}, Eur.
Phys. J. C \textbf{78} (2018) 10, 794, \textcolor{purple}{\href{https://arxiv.org/abs/1802.05922}{\tt 1802.05922 [hep-th]}}

\bibitem{Donoghue-2021}J. F. Donoghue and G. Menezes, \emph{On Quadratic
Gravity}, Nuovo Cim. C \textbf{45} (2022) 2, 26, \textcolor{purple}{\href{https://arxiv.org/abs/2112.01974}{\tt 2112.01974 [hep-th]}}

\bibitem{Salvio-2018}A. Salvio, \emph{Quadratic Gravity}, Front.
in Phys. 6 (2018) 77, \textcolor{purple}{\href{https://arxiv.org/abs/1804.09944}{\tt 1804.09944 [hep-th]}}

\bibitem{Nguyen-Newtonian}H. K. Nguyen, \emph{Emerging Newtonian
potential in pure $R^{2}$ gravity on a de Sitter background}, J.
High Energy Phys. \textbf{08}, 127 (2023), \textcolor{purple}{\href{https://arxiv.org/abs/2306.03790}{\tt 2306.03790 [gr-qc]}}

\bibitem{Karananas-2024}G. K. Karananas, \emph{The particle content
of $R^{2}$ gravity revisited},\textcolor{purple}{{} \href{https://arxiv.org/abs/2407.09598}{\tt 2407.09598 [hep-th]}}

\bibitem{SSB-Anderson}P. W. Anderson, Phys. Rev. \textbf{130}, 439
(1963)

\bibitem{SSB-Higgs}P. W. Higgs, Phys. Rev. Lett. \textbf{13}, 508
(1964)

\bibitem{SSB-Englert}F. Englert and R. Brout, Phys. Rev. Lett. \textbf{13},
321 (1964)

\bibitem{SSB-Guralnik}G. S. Guralnik, C. R. Hagen, and T. W. B. Kibble,
Phys. Rev. Lett. \textbf{13}, 585 (1964)

\bibitem{Gorbar-2003}E. V. Gorbar and I. L. Shapiro, \emph{Renormalization
Group and Decoupling in Curved Space: III. The Case of Spontaneous
Symmetry Breaking}, JHEP \textbf{02} (2004) 060, \textcolor{purple}{\href{https://arxiv.org/abs/hep-ph/0311190}{\tt hep-ph/0311190}}

\bibitem{VSL2024-dilaton}H. K. Nguyen, \emph{Dilaton-induced variations
in Planck constant and speed of light: An alternative to Dark Energ}y,
Phys. Lett. B \textbf{862} (2025) 139357, \textcolor{purple}{\href{https://arxiv.org/abs/2412.04257}{\tt 2412.04257 [gr-qc]}}

\bibitem{Greiner-RQM-book}E.g., see W. Greiner, \emph{Relativistic
Quantum Mechanics--Wave Equations}, 2nd edition, Springer (1997),
page 230

\bibitem{QM-book}E.g., see D. J. Griffiths, \emph{Introduction to
Quantum Mechanics}, Prentice Hall (1995), page 312

\bibitem{HafeleKeating-1972-a}J. Hafele and R. Keating, \emph{Around-the-world
atomic clocks: Predicted relativistic time gains,} Science \textbf{177},
166 (1972)

\bibitem{HafeleKeating-1972-b}J. Hafele and R. Keating, \emph{Around-the-world
atomic clocks: Observed relativistic time gains}, Science \textbf{177},
168 (1972)

\bibitem{Scolnic-2018}D. M. Scolnic et al, \emph{The complete light-curve
sample of spectroscopically confirmed Type Ia supernovae from Pan-STARRS1
and cosmological constraints from the Combined Pantheon Sample}, Astrophys.
J. \textbf{859} (2), 101 (2018),\textcolor{purple}{{} \href{https://arxiv.org/abs/1710.00845}{\tt 1710.00845 [astro-ph.CO]}}

\bibitem{Riess-1998}A. Riess et al, \emph{Observational evidence
from supernovae for an accelerating universe and a cosmological constant},
Astron. J. \textbf{116}, 1009 (1998),\textcolor{purple}{{} \href{https://arxiv.org/abs/astro-ph/9805201}{\tt astro-ph/9805201}}

\bibitem{Perlmutter-1999}S. Perlmutter et al, \emph{Measurements
of $\Omega$ and $\Lambda$ from 42 high-redshift supernovae}, Astron.
J. \textbf{517}, 565 (1999),\textcolor{purple}{{} \href{https://arxiv.org/abs/astro-ph/9812133}{\tt astro-ph/9812133}}

\bibitem{Lusso-2020}E. Lusso et al,\emph{ Quasars as standard candles
III. Validation of a new sample for cosmological studies}, A\&A \textbf{642},
A150 (2020),\textcolor{purple}{{} \href{https://arxiv.org/abs/2008.08586}{\tt 2008.08586 [astro-ph.GA]}}

\bibitem{Bassett-2013}B. A. Bassett, Y. Fantaye, R. Hlo\v{z}ek,
C. Sabiu and M. Smith, \emph{Observational Constraints on Redshift
Remapping}, \textcolor{purple}{\href{https://arxiv.org/abs/1312.2593}{\tt 1312.2593 [astro-ph.CO]}}

\bibitem{Wojtak-2016}R. Wojtak and F. Prada, \emph{Testing the mapping
between redshift and cosmic scale factor}, Mon. Not. Roy. Astron.
Soc. \textbf{458}, 3331 (2016), \textcolor{purple}{\href{https://arxiv.org/abs/1602.02231}{\tt 1602.02231 [astro-ph.CO]}}

\bibitem{Wojtak-2017}R. Wojtak and F. Prada, \emph{Redshift remapping
and cosmic acceleration in dark-matter-dominated cosmological models},
Mon. Not. Roy. Astron. Soc. \textbf{470}, 4493 (2017),\textcolor{purple}{{}
\href{https://arxiv.org/abs/1610.03599}{\tt 1610.03599 [astro-ph.CO]}}

\bibitem{Blanchard-2003}A. Blanchard, M. Douspis, M. Rowan-Robinson,
and S. Sarkar, \emph{An alternative to the cosmological ``concordance
model''}, Astron. Astrophys. \textbf{412}, 35-44 (2003),\textcolor{purple}{{}
\href{https://arxiv.org/abs/astro-ph/0304237}{\tt astro-ph/0304237}}

\bibitem{Hunt-2007}P. Hunt and S. Sarkar, \emph{Multiple inflation
and the WMAP ``glitches''. II. Data analysis and cosmological parameter
extraction}, Phys. Rev. D\textbf{ 76}, 123504 (2007),\textcolor{purple}{{}
\href{https://arxiv.org/abs/0706.2443}{\tt 0706.2443 [astro-ph]}}

\bibitem{Shanks-2004}T. Shanks, \emph{Problems with the Current Cosmological
Paradigm}, IAU Symp. \textbf{216} (2005) 398,\textcolor{purple}{{} \href{https://arxiv.org/abs/astro-ph/0401409}{\tt astro-ph/0401409}}

\bibitem{Blanchard-2006}A. Blanchard, M. Douspis, M. Rowan-Robinson
and S. Sarkar, \emph{Large-scale galaxy correlations as a test for
dark energy}, Astron. Astrophys. \textbf{449}, 925 (2006),\textcolor{purple}{{}
\href{https://arxiv.org/abs/astro-ph/0512085}{\tt astro-ph/0512085}}

\bibitem{Vagnozzi-2023}S. Vagnozzi, \emph{Seven hints that early-time
new physics alone is not sufficient to solve the Hubble tension},
Universe 9 (2023) 393, \textcolor{purple}{\href{https://arxiv.org/abs/2308.16628}{\tt 2308.16628 [astro-ph.CO]}}

\bibitem{Krishnan-2020}C. Krishnan, E. \'O Colg\'ain, Ruchika,
A. A. Sen, M. M. Sheikh-Jabbari and T. Yang, \emph{Is there an early
Universe solution to Hubble tension?}, Phys. Rev. D \textbf{102},
103525 (2020), \textcolor{purple}{\href{https://arxiv.org/abs/2002.06044}{\tt 2002.06044 [astro-ph.CO]}}

\bibitem{Krishnan-2021}C. Krishnan, E. \'O Colg\'ain, M. M. Sheikh-Jabbari
and T. Yang, \emph{Running Hubble Tension and a $H_{0}$ Diagnostic},
Phys. Rev. D \textbf{103}, 103509 (2021), \textcolor{purple}{\href{https://arxiv.org/abs/2011.02858}{\tt 2011.02858 [astro-ph.CO]}}

\bibitem{Dainotti-2020}M. G. Dainotti, B. De Simone, T. Schiavone,
G. Montani, E. Rinaldi and G. Lambiase, \emph{On the Hubble constant
tension in the SNe Ia Pantheon sample}, Astrophys. J. \textbf{912},
150 (2021), \textcolor{purple}{\href{https://arxiv.org/abs/2103.02117}{\tt 2103.02117 [astro-ph.CO]}}

\bibitem{Bardeen-1995}W. A. Bardeen, \emph{On Naturalness in the
Standard Model}, FERMILAB-CONF-95-391-T, \textcolor{purple}{\href{http://lss.fnal.gov/archive/1995/conf/Conf-95-391-T.pdf}{lss.fnal.gov/archive/1995/conf/Conf-95-391-T.pdf}}

\bibitem{Shaposhnikov:2008xi}M. Shaposhnikov and D. Zenhausern, \emph{Quantum
scale invariance, cosmological constant and hierarchy problem}, Phys.
Lett. B\textbf{ 671} (2009) 162-166, \textcolor{purple}{\href{http://arxiv.org/abs/0809.3406}{\tt 0809.3406 [hep-th]}}

\bibitem{Shaposhnikov:2008xb}M. Shaposhnikov and D. Zenhausern, \emph{Scale
invariance, unimodular gravity and dark energy}, Phys. Lett. B\textbf{
671} (2009) 187-192, \textcolor{purple}{\href{http://arxiv.org/abs/0809.3395}{\tt 0809.3395 [hep-th]}}

\bibitem{Shaposhnikov:2018gul}M. Shaposhnikov and A. Shkerin, \emph{Gravity,
Scale Invariance and the Hierarchy Problem}, JHEP \textbf{10} (2018)
024, \textcolor{purple}{\href{https://arxiv.org/abs/1804.06376}{\tt 1804.06376 [hep-th]}}

\bibitem{Shaposhnikov:2020oln}M. Shaposhnikov, A. Shkerin, I. Timiryasov
and S. Zell, \emph{Einstein-Cartan gravity, matter, and scale-invariant
generalization}, JHEP \textbf{10} (2020) 177, \textcolor{purple}{\href{https://arxiv.org/abs/2007.16158}{\tt 2007.16158 [hep-th]}}

\bibitem{GarciaBellido-2011}J. Garcia-Bellido, J. Rubio, M. Shaposhnikov
and D. Zenhausern, \emph{Higgs-Dilaton Cosmology: From the Early to
the Late Universe}, Phys. Rev. D \textbf{84}, 123504 (2011), \textcolor{purple}{\href{https://arxiv.org/abs/1107.2163}{\tt 1107.2163 [hep-ph]}}

\bibitem{Rubio-2014}J. Rubio and M. Shaposhnikov, \emph{Higgs-Dilaton
cosmology: Universality vs. criticality}, Phys. Rev. D \textbf{90},
027307 (2014), \textcolor{purple}{\href{https://arxiv.org/abs/1406.5182}{\tt 1406.5182 [hep-ph]}}

\bibitem{Bezrukov-2015}F. Bezrukov, J. Rubio, and M. Shaposhnikov,
\emph{Living beyond the edge: Higgs inflation and vacuum metastability},
Phys. Rev. D \textbf{92}, 083512 (2015), \href{https://arxiv.org/abs/1412.3811}{\tt 1412.3811 [hep-ph]}

\bibitem{Mooij:2019rgh}S. Mooij, M. Shaposhnikov and T. Voumard,
\emph{Hidden and explicit quantum scale invariance}, Phys. Rev. D
\textbf{99}, 085013 (2019), \textcolor{purple}{\href{https://arxiv.org/abs/1812.07946}{\tt 1812.07946 [hep-th]}}

\bibitem{Karananas-2016}G. K. Karananas and M. Shaposhnikov, \emph{Scale
invariant alternatives to general relativity. II. Dilaton properties},
Phys. Rev. D \textbf{93}, 084052 (2016), \textcolor{purple}{\href{https://arxiv.org/abs/1603.01274}{\tt 1603.01274 [hep-th]}}

\bibitem{Karananas:2021ner}G. K. Karananas, M. Shaposhnikov, A. Shkerin
and S. Zell, \emph{Scale and Weyl Invariance in Einstein-Cartan Gravity},
Phys. Rev. D \textbf{104}, 124014 (2021), \textcolor{purple}{\href{https://arxiv.org/abs/2108.05897}{\tt 2108.05897 [hep-th]}}

\bibitem{Karananas:2023dth}G. K. Karananas, M. Shaposhnikov and S.
Zell, \emph{Scale invariant Einstein-Cartan gravity and flat space
conformal symmetry}, JHEP \textbf{11} (2023) 171, \textcolor{purple}{\href{https://arxiv.org/abs/2307.11151}{\tt 2307.11151 [hep-th]}}

\bibitem{Karananas:2018oiu}G. K. Karananas and M. Shaposhnikov, \emph{CFT
data and spontaneously broken conformal invariance}, Phys. Rev. D
\textbf{97}, 045009 (2018), \href{https://arxiv.org/abs/1708.02220}{\tt 1708.02220 [hep-th]}

\bibitem{Karananas:2017ikj}G. K. Karananas and M. Shaposhnikov, Gauge
coupling unification without leptoquarks, Phys. Lett. B \textbf{771},
332 (2017), \href{https://arxiv.org/abs/1703.02964}{\tt 1703.02964 [hep-ph]}

\bibitem{Karananas:2016kyt}G. K. Karananas and J. Rubio, \emph{On
the geometrical interpretation of scale-invariant models of inflation},
Phys. Lett. B\textbf{ 761} (2016) 223-228, \textcolor{purple}{\href{http://arxiv.org/abs/1606.08848}{\tt 1606.08848 [hep-ph]}}

\bibitem{Casas:2019mku}S. Casas, G. K. Karananas, M. Pauly and J.
Rubio, \emph{Scale-invariant alternatives to general relativity. III.
The inflation-{}-dark-energy connection}, Phys. Rev. D \textbf{99},
063512 (2019), \textcolor{purple}{\href{https://arxiv.org/abs/1811.05984}{\tt 1811.05984 [astro-ph.CO]}}

\bibitem{Litim:2017nru}D. F. Litim, E. Marchais and P. Mati, \emph{Fixed
points and the spontaneous breaking of scale invariance}, Phys. Rev.
D \textbf{95}, 125006 (2017), \textcolor{purple}{\href{https://arxiv.org/abs/1702.05749}{\tt 1702.05749 [hep-th]}}

\bibitem{Rubio-2017}J. Rubio and C. Wetterich, \emph{Emergent scale
symmetry: Connecting inflation and dark energy}, Phys. Rev. D \textbf{96},
063509 (2017), \textcolor{purple}{\href{https://arxiv.org/abs/1705.00552}{\tt 1705.00552 [gr-qc]}}

\bibitem{Wetterich:2002wm}C. Wetterich, \emph{Conformal fixed point,
cosmological constant and quintessence}, Phys. Rev. Lett. \textbf{90}
(2003) 231302, \textcolor{purple}{\href{http://arxiv.org/abs/hep-th/0210156}{\tt hep-th/0210156}}

\bibitem{Wetterich:2019lmi}C. Wetterich, \emph{Quantum scale symmetry},
\textcolor{purple}{\href{https://arxiv.org/abs/1901.04741}{\tt 1901.04741 [hep-th]}}

\bibitem{Wetterich:2021kil}C. Wetterich, \emph{Fundamental Scale
Invariance}, Nucl. Phys. B \textbf{964} (2021) 115326, \textcolor{purple}{\href{https://arxiv.org/abs/2007.08805}{\tt 2007.08805 [hep-th]}}

\bibitem{Gabrielli:2013hma}E. Gabrielli, M. Heikinheimo, K. Kannike,
A. Racioppi, M. Raidal and C. Spethmann, \emph{Towards Completing
the Standard Model: Vacuum Stability, EWSB and Dark Matter}, Phys.
Rev. D \textbf{89} (2014) 015017, \textcolor{purple}{\href{http://arxiv.org/abs/1309.6632}{\tt 1309.6632 [hep-ph]}}

\bibitem{Kannike-2017}K. Kannike, M. Raidal, C. Spethmann and H.
Veermae, \emph{Evolving Planck Mass in Classically Scale-Invariant
Theories}, JHEP \textbf{04}, 026 (2017), \textcolor{purple}{\href{https://arxiv.org/abs/1610.06571}{\tt 1610.06571 [hep-ph]}}

\bibitem{Kannike:2014mia}K. Kannike, A. Racioppi and M. Raidal, \emph{Embedding
inflation into the Standard Model - more evidence for classical scale
invariance}, JHEP \textbf{06} (2014) 154, \textcolor{purple}{\href{http://arxiv.org/abs/1405.3987}{\tt 1405.3987 [hep-ph]}}

\bibitem{Kannike:2015kda}K. Kannike, A. Racioppi and M. Raidal, \emph{Linear
inflation from quartic potential}, JHEP \textbf{01} (2016) 035, \textcolor{purple}{\href{http://arxiv.org/abs/1509.05423}{\tt 1509.05423 [hep-ph]}}

\bibitem{Kannike:2015apa}K. Kannike, G. H\"utsi, L. Pizza, A. Racioppi,
M. Raidal, A. Salvio and A. Strumia, \emph{Dynamically Induced Planck
Scale and Inflation}, JHEP \textbf{05} (2015) 065, \textcolor{purple}{\href{http://arxiv.org/abs/1502.01334}{\tt 1502.01334 [astro-ph.CO]}}

\bibitem{Kannike:2016bny}K. Kannike, G. M. Pelaggi, A. Salvio and
A. Strumia, \emph{The Higgs of the Higgs and the diphoton channel},
JHEP \textbf{07} (2016) 101, \textcolor{purple}{\href{http://arxiv.org/abs/1605.08681}{\tt 1605.08681 [hep-ph]}}

\bibitem{Kannike:2021nrd}K. Kannike, K. Loos and L. Marzola, \emph{Minima
of Classically Scale-Invariant Potentials}, JHEP \textbf{06} (2021)
128, \textcolor{purple}{\href{https://arxiv.org/abs/2011.12304}{\tt 2011.12304 [hep-ph]}}

\bibitem{Dirgantara:2023jhu}B. Dirgantara, K. Kannike and W. Sreethawong,
\emph{Vacuum Stability and Radiative Symmetry Breaking of the Scale-Invariant
Singlet Extension of Type II Seesaw Model}, Eur. Phys. J. C \textbf{83}
(2023) 253, \textcolor{purple}{\href{https://arxiv.org/abs/2301.00487}{\tt 2301.00487 [hep-ph]}}

\bibitem{Einhorn-2017}M. B. Einhorn and D. R. T. Jones, \emph{Renormalizable,
asymptotically free gravity without ghosts or tachyons}, Phys. Rev.
D \textbf{96}, 124025 (2017), \textcolor{purple}{\href{https://arxiv.org/abs/1710.03795}{\tt 1710.03795 [hep-th]}}

\bibitem{Einhorn:2019lin}M. B. Einhorn and D. R. Timothy Jones, \emph{Grand
Unified Theories in Renormalisable, Classically Scale Invariant Gravity},
JHEP \textbf{10} (2019) 012, \textcolor{purple}{\href{https://arxiv.org/abs/1908.01400}{\tt 1908.01400 [hep-th]}}

\bibitem{Einhorn:2015lzy}M. B. Einhorn and D. R. T. Jones, \emph{Induced
Gravity I: Real Scalar Field}, JHEP \textbf{01} (2016) 019, \textcolor{purple}{\href{http://arxiv.org/abs/1511.01481}{\tt 1511.01481 [hep-th]}}

\bibitem{Einhorn:2016mws}M. B. Einhorn and D. R. T. Jones, \emph{Induced
Gravity II: Grand Unification}, JHEP \textbf{05} (2016) 185, \textcolor{purple}{\href{http://arxiv.org/abs/1602.06290}{\tt 1602.06290 [hep-th]}}

\bibitem{Hill:2014mqa}C. T. Hill, \emph{Is the Higgs Boson Associated
with Coleman-Weinberg Dynamical Symmetry Breaking?}, Phys. Rev. D\textbf{
89} (2014) 073003, \textcolor{purple}{\href{http://arxiv.org/abs/1401.4185}{\tt 1401.4185 [hep-ph]}}

\bibitem{Allison:2014zya}K. Allison, C. T. Hill and G. G. Ross, \emph{Ultra-weak
sector, Higgs boson mass, and the dilaton}, Phys. Lett. B \textbf{738}
(2014) 191-195, \textcolor{purple}{\href{http://arxiv.org/abs/1404.6268}{\tt 1404.6268 [hep-ph]}}

\bibitem{Allison:2014hna}K. Allison, C. T. Hill and G. G. Ross, \emph{An
ultra-weak sector, the strong CP problem and the pseudo-Goldstone
dilaton}, Nucl. Phys. B \textbf{891} (2015) 613-626, \textcolor{purple}{\href{http://arxiv.org/abs/1409.4029}{\tt 1409.4029 [hep-ph]}}

\bibitem{Ferreira:2016jhu}P. G. Ferreira, C. T. Hill, and G. G. Ross,
\emph{Scale-Independent Inflation and Hierarchy Generation}, Phys.
Lett. B (2016), 10.1016/j.physletb.2016.10.036, \href{https://arxiv.org/abs/1603.05983}{\tt 1603.05983 [hep-th]}

\bibitem{Ferreira-2017}P. G. Ferreira, C. T. Hill, and G. G. Ross,
\emph{Weyl Current, Scale-Invariant Inflation and Planck Scale Generation},
Phys. Rev. D \textbf{95}, 043507 (2017), \href{https://arxiv.org/abs/1610.09243}{\tt 1610.09243 [hep-th]}

\bibitem{Ferreira:2018rth}P. G. Ferreira, C. T. Hill and G. G. Ross,
\emph{Inertial Spontaneous Symmetry Breaking and Quantum Scale Invariance},
Phys. Rev. D \textbf{98}, 116012 (2018), \textcolor{purple}{\href{https://arxiv.org/abs/1801.07676}{\tt 1801.07676 [hep-th]}}

\bibitem{Ferreira:2020ghj}P. G. Ferreira and O. J. Tattersall, \emph{Scale
Invariant Gravity and Black Hole Ringdown}, Phys.Rev.D \textbf{101}
(2020) 2, 024011, \href{https://arxiv.org/abs/1910.04480}{\tt 1910.04480 [gr-qc]}

\bibitem{Quiros:2014ngh}I. Quiros,\emph{ Scale invariance: fake appearances},
\href{https://arxiv.org/abs/1405.6668}{\tt 1405.6668 [gr-qc]}

\bibitem{Quiros:2024fgd}I. Quiros, \emph{Revisiting local-scale invariant
gravitational theory}, \textcolor{purple}{\href{https://arxiv.org/abs/2402.03184}{\tt 2402.03184 [gr-qc]}}

\bibitem{Kurkov:2016nhj}M. Kurkov, \emph{Emergent spontaneous symmetry
breaking and emergent symmetry restoration in rippling gravitational
background}, Eur. Phys. J. C \textbf{76}, 329 (2016), \href{https://arxiv.org/abs/1601.00622}{\tt 1601.00622 [hep-th]}

\bibitem{Salvio:2018lkj}A. Salvio and A. Strumia, \emph{Agravity
up to infinite energy}, Eur. Phys. J. C78, 124 (2018), \textcolor{purple}{\href{https://arxiv.org/abs/1705.03896}{\tt 1705.03896 [hep-th]}}

\bibitem{Salvio:2017hju}A. Salvio, \emph{Inflationary Perturbations
in No-Scale Theories}, Eur. Phys. J. C \textbf{77}, 267 (2017), \href{https://arxiv.org/abs/1703.08012}{\tt 1703.08012 [astro-ph.CO]}

\bibitem{Ghilencea:2019nhg}D. M. Ghilencea, \emph{Spontaneous breaking
of Weyl quadratic gravity to Einstein action and Higgs potential},
JHEP \textbf{03}, 049 (2019), \href{https://arxiv.org/abs/1812.08613}{\tt 1812.08613 [hep-th]}

\bibitem{Ghilencea:2019iuj}D. M. Ghilencea, \emph{Stueckelberg breaking
of Weyl conformal geometry with applications to gravity}, Phys. Rev.
D \textbf{101}, 045010 (2020), \href{https://arxiv.org/abs/1904.06596}{\tt 1904.06596 [hep-th]}

\bibitem{Ghilencea-2019}D. M. Ghilencea and H. M. Lee, \emph{Weyl
gauge symmetry and its spontaneous breaking in Standard Model and
inflation}, Phys. Rev. D \textbf{99}, 115007 (2019), \textcolor{purple}{\href{https://arxiv.org/abs/1809.09174}{\tt 1809.09174 [hep-th]}}

\bibitem{Ghilencea:2023hyt}D. M. Ghilencea, \emph{Non-metric geometry
as the origin of mass in gauge theories of scale invariance}, Eur.
Phys. J. C \textbf{83} (2023) 176, \textcolor{purple}{\href{https://arxiv.org/abs/2203.05381}{\tt 2203.05381 [hep-th]}}

\bibitem{Ghilencea:2015mza}D. M. Ghilencea, \emph{Manifestly scale-invariant
regularization and quantum effective operators}, Phys. Rev. D \textbf{93}
(2016) 105006, \textcolor{purple}{\href{http://arxiv.org/abs/1508.00595}{\tt 1508.00595 [hep-ph]}}

\bibitem{Ghilencea:2024kuy}D. M. Ghilencea and C. T. Hill, \emph{Standard
Model in conformal geometry: local vs gauged scale invariance}, Annals
Phys. \textbf{460} (2024) 169562, \textcolor{purple}{\href{https://arxiv.org/abs/2303.02515}{\tt 2303.02515 [hep-th]}}

\bibitem{Ghilencea:2017afg}D. M. Ghilencea, Z. Lalak and P. Olszewski,
\emph{Standard Model with spontaneously broken quantum scale invariance},
Phys. Rev. D \textbf{96}, 055034 (2017), \textcolor{purple}{\href{https://arxiv.org/abs/1612.09120}{\tt 1612.09120 [hep-ph]}}

\bibitem{Ghilencea:2016ckm}D. M. Ghilencea, Z. Lalak and P. Olszewski,
\emph{Two-loop scale-invariant scalar potential and quantum effective
operators}, Eur. Phys. J. C \textbf{76} (2016) 12, 656, \textcolor{purple}{\href{http://arxiv.org/abs/1608.05336}{\tt 1608.05336 [hep-th]}}

\bibitem{Weisswange:2023dft}M. Wei{\ss}wange, D. M. Ghilencea and
D. St\"ockinger, \emph{Quantum scale invariance in gauge theories
and applications to muon production}, Phys. Rev. D \textbf{107} (2023)
085008, \textcolor{purple}{\href{https://arxiv.org/abs/2208.01293}{\tt 2208.01293 [hep-ph]}}

\bibitem{Utiyama:1973mnk}R. Utiyama, \emph{On Weyl's Gauge Field},
Prog. Theor. Phys. \textbf{50}, 2080 (1973)

\bibitem{Utiyama:1975egy}R. Utiyama, \emph{On Weyl's Gauge Field.
II}, Prog. Theor. Phys. \textbf{53}, 565 (1975)

\bibitem{Nishino-2011}H. Nishino and S. Rajpoot, \emph{Weyl's scale
invariance for the standard model, renormalizability and the zero
cosmological constant}, Class. Quant. Grav. 28, 145014 (2011)

\bibitem{Nishino:2007wer}H. Nishino and S. Rajpoot, \emph{Comment
on Electroweak Higgs as a Pseudo-Goldstone Boson of Broken Scale Invariance},
\href{https://arxiv.org/abs/0704.1836}{\tt 0704.1836 [hep-ph]}

\bibitem{Nishino:2009nju}H. Nishino and S. Rajpoot, \emph{Weyl's
scale invariance: Inflation, dark matter and dark energy connections},
Proceedings, 4th International Workshop on the Dark Side of the Universe
(DSU 2008): Cairo, Egypt, June 1-5, 2008, AIP Conf. Proc. 1115, 33
(2009)

\bibitem{Hempfling:1996ht}R.~Hempfling, \emph{The Next-to-minimal
Coleman-Weinberg model}, Phys. Lett. B \textbf{379} (1996) 153-158,
\textcolor{purple}{\href{http://arxiv.org/abs/hep-ph/9604278}{\tt hep-ph/9604278}}

\bibitem{Chang:2007ki}W.-F. Chang, J. N. Ng and J. M. S. Wu, \emph{Shadow
Higgs from a scale-invariant hidden U(1)(s) model}, Phys. Rev. D\textbf{
75} (2007) 115016, \textcolor{purple}{\href{http://arxiv.org/abs/hep-ph/0701254}{\tt hep-ph/0701254}}

\bibitem{Meissner:2006zh}K. A. Meissner and H. Nicolai, \emph{Conformal
Symmetry and the Standard Model}, Phys. Lett. B \textbf{648} (2007)
312-317, \textcolor{purple}{\href{http://arxiv.org/abs/hep-th/0612165}{{\tt hep-th/0612165}}}

\bibitem{Foot:2007as}R. Foot, A. Kobakhidze and R. R. Volkas, \emph{Electroweak
Higgs as a pseudo-Goldstone boson of broken scale invariance}, Phys.
Lett. B \textbf{655} (2007) 156-161, \textcolor{purple}{\href{http://arxiv.org/abs/0704.1165}{\tt 0704.1165 [hep-ph]}}

\bibitem{Foot:2007ay}R. Foot, A. Kobakhidze, K. McDonald and R. Volkas,
\emph{Neutrino mass in radiatively-broken scale-invariant models},
Phys. Rev. D \textbf{76} (2007) 075014, \textcolor{purple}{\href{http://arxiv.org/abs/0706.1829}{\tt 0706.1829 [hep-ph]}}

\bibitem{Foot:2007iy}R. Foot, A. Kobakhidze, K. L. McDonald and R.
R. Volkas, \emph{A Solution to the hierarchy problem from an almost
decoupled hidden sector within a classically scale invariant theory},
Phys. Rev. D \textbf{77} (2008) 035006, \textcolor{purple}{\href{http://arxiv.org/abs/0709.2750}{\tt 0709.2750 [hep-ph]}}

\bibitem{Foot:2011et}R. Foot and A. Kobakhidze, \emph{Electroweak
Scale Invariant Models with Small Cosmological Constant}, Int. J.
Mod. Phys. A \textbf{30} (2015) 1550126, \textcolor{purple}{\href{http://arxiv.org/abs/1112.0607}{\tt 1112.0607 [hep-ph]}}

\bibitem{Foot:2010av}R. Foot, A. Kobakhidze and R. R. Volkas, \emph{Stable
mass hierarchies and dark matter from hidden sectors in the scale-invariant
standard model}, Phys. Rev. D \textbf{82} (2010) 035005, \textcolor{purple}{\href{http://arxiv.org/abs/1006.0131}{\tt 1006.0131 [hep-ph]}}

\bibitem{Foot:2010et}R. Foot, A. Kobakhidze and R. R. Volkas, \emph{Cosmological
constant in scale-invariant theories}, Phys. Rev. D \textbf{84} (2011)
075010, \textcolor{purple}{\href{http://arxiv.org/abs/1012.4848}{\tt 1012.4848 [hep-ph]}}

\bibitem{Kobakhidze:2017grt}A. Kobakhidze and S. Liang, \emph{Standard
Model with hidden scale invariance and light dilaton}, \textcolor{purple}{\href{https://arxiv.org/abs/1701.04927}{\tt 1701.04927 [hep-ph]}}

\bibitem{Iso:2009nw}S. Iso, N. Okada and Y. Orikasa, \emph{The minimal
B--L model naturally realized at TeV scale}, Phys. Rev. D \textbf{80}
(2009) 115007, \textcolor{purple}{\href{http://arxiv.org/abs/0909.0128}{\tt 0909.0128 [hep-ph]}}

\bibitem{Iso:2009ss}S. Iso, N. Okada and Y. Orikasa, \emph{Classically
conformal B--L extended Standard Model}, Phys. Lett. B \textbf{676}
(2009) 81-87, \textcolor{purple}{\href{http://arxiv.org/abs/0902.4050}{\tt 0902.4050 [hep-ph]}}

\bibitem{Iso:2009prn}S. Iso and Y. Orikasa, \emph{TeV Scale B-L model
with a flat Higgs potential at the Planck scale - in view of the hierarchy
problem}, PTEP \textbf{2013} (2013) 023B08, \textcolor{purple}{\href{http://arxiv.org/abs/1210.2848}{\tt 1210.2848 [hep-ph]}}

\bibitem{Ishiwata:2011aa}K. Ishiwata, \emph{Dark Matter in Classically
Scale-Invariant Two Singlets Standard Model}, Phys. Lett. B \textbf{710}
(2012) 134-138, \textcolor{purple}{\href{http://arxiv.org/abs/1112.2696}{\tt 1112.2696 [hep-ph]}}

\bibitem{Holthausen:2009uc}M. Holthausen, M. Lindner and M. A. Schmidt,
\emph{Radiative Symmetry Breaking of the Minimal Left-Right Symmetric
Model}, Phys. Rev. D \textbf{82} (2010) 055002, \textcolor{purple}{\href{http://arxiv.org/abs/0911.0710}{\tt 0911.0710 [hep-ph]}}

\bibitem{AlexanderNunneley:2010nw}L. Alexander-Nunneley and A. Pilaftsis,
\emph{The Minimal Scale Invariant Extension of the Standard Model},
JHEP \textbf{09} (2010) 021, \textcolor{purple}{\href{http://arxiv.org/abs/1006.5916}{\tt 1006.5916 [hep-ph]}}

\bibitem{Lee:2012jn}J. S. Lee and A. Pilaftsis, \emph{Radiative Corrections
to Scalar Masses and Mixing in a Scale Invariant Two Higgs Doublet
Model}, Phys. Rev. D \textbf{86} (2012) 035004, \textcolor{purple}{\href{http://arxiv.org/abs/1201.4891}{\tt 1201.4891 [hep-ph]}}

\bibitem{Heikinheimo:2013fta}M. Heikinheimo, A. Racioppi, M. Raidal,
C. Spethmann and K. Tuominen, \emph{Physical Naturalness and Dynamical
Breaking of Classical Scale Invariance}, Mod. Phys. Lett. A \textbf{29}
(2014) 1450077, \textcolor{purple}{\href{http://arxiv.org/abs/1304.7006}{\tt 1304.7006 [hep-ph]}}

\bibitem{Heikinheimo:2014xza}M. Heikinheimo and C. Spethmann, \emph{Galactic
Centre GeV Photons from Dark Technicolor}, JHEP \textbf{12} (2014)
084, \textcolor{purple}{\href{http://arxiv.org/abs/1410.4842}{\tt 1410.4842 [hep-ph]}}

\bibitem{Hur:2011sv}T. Hur and P. Ko, \emph{Scale invariant extension
of the standard model with strongly interacting hidden sector}, Phys.
Rev. Lett. \textbf{106} (2011) 141802, \textcolor{purple}{\href{http://arxiv.org/abs/1103.2571}{\tt 1103.2571 [hep-ph]}}

\bibitem{Carone:2013wla}C. D. Carone and R. Ramos, \emph{Classical
scale-invariance, the electroweak scale and vector dark matter}, Phys.
Rev. D \textbf{88} (2013) 055020, \textcolor{purple}{\href{http://arxiv.org/abs/1307.8428}{\tt 1307.8428 [hep-ph]}}

\bibitem{Farzinnia:2015fka}A. Farzinnia and S. Kouwn, \emph{Classically
scale invariant inflation, supermassive WIMPs, and adimensional gravity},
Phys. Rev. D \textbf{93} (2016) 063528, \textcolor{purple}{\href{http://arxiv.org/abs/1512.05890}{\tt 1512.05890 [hep-ph]}}

\bibitem{Farzinnia:2013pga}A. Farzinnia, H.-J. He and J. Ren, \emph{Natural
Electroweak Symmetry Breaking from Scale Invariant Higgs Mechanism},
Phys. Lett. B \textbf{727} (2013) 141-150, \textcolor{purple}{\href{http://arxiv.org/abs/1308.0295}{\tt 1308.0295 [hep-ph]}}

\bibitem{Farzinnia:2015uma}A. Farzinnia, \emph{Prospects for Discovering
the Higgs-like Pseudo-Nambu-Goldstone Boson of the Classical Scale
Symmetry}, Phys. Rev. D \textbf{92} (2015) 095012, \textcolor{purple}{\href{http://arxiv.org/abs/1507.06926}{\tt 1507.06926 [hep-ph]}}

\bibitem{Farzinnia:2014xia}A. Farzinnia and J. Ren, \emph{Higgs Partner
Searches and Dark Matter Phenomenology in a Classically Scale Invariant
Higgs Boson Sector}, Phys. Rev. D \textbf{90} (2014) 015019, \textcolor{purple}{\href{http://arxiv.org/abs/1405.0498}{\tt 1405.0498 [hep-ph]}}

\bibitem{Chivukula:2013xka}R. S. Chivukula, A. Farzinnia, J. Ren
and E. H. Simmons, \emph{Constraints on the Scalar Sector of the Renormalizable
Coloron Model}, Phys. Rev. D \textbf{88} (2013) 075020, \textcolor{purple}{\href{http://arxiv.org/abs/1307.1064}{\tt 1307.1064 [hep-ph]}}

\bibitem{Chun:2013soa}E. J. Chun, S. Jung and H. M. Lee, \emph{Radiative
generation of the Higgs potential}, Phys. Lett. B \textbf{725} (2013)
158-163, \textcolor{purple}{\href{http://arxiv.org/abs/1304.5815}{\tt 1304.5815 [hep-ph]}}

\bibitem{Hambye:2013sna}T. Hambye and A. Strumia, \emph{Dynamical
generation of the weak and Dark Matter scale}, Phys. Rev. D \textbf{88}
(2013) 055022, \textcolor{purple}{\href{http://arxiv.org/abs/1306.2329}{\tt 1306.2329 [hep-ph]}}

\bibitem{Englert:2013gz}C. Englert, J. Jaeckel, V. V. Khoze and M.
Spannowsky, \emph{Emergence of the Electroweak Scale through the Higgs
Portal}, JHEP \textbf{04} (2013) 060, \textcolor{purple}{\href{http://arxiv.org/abs/1301.4224}{\tt 1301.4224 [hep-ph]}}

\bibitem{Khoze:2013oga}V. V. Khoze and G. Ro, \emph{Leptogenesis
and Neutrino Oscillations in the Classically Conformal Standard Model
with the Higgs Portal}, JHEP \textbf{10} (2013) 075, \textcolor{purple}{\href{http://arxiv.org/abs/1307.3764}{\tt 1307.3764 [hep-ph]}}

\bibitem{Khoze:2013uia}V. V. Khoze, \emph{Inflation and Dark Matter
in the Higgs Portal of Classically Scale Invariant Standard Model},
JHEP \textbf{11} (2013) 215, \textcolor{purple}{\href{http://arxiv.org/abs/1308.6338}{\tt 1308.6338 [hep-ph]}}

\bibitem{Khoze:2014xha}V. V. Khoze, C. McCabe and G. Ro, \emph{Higgs
vacuum stability from the dark matter portal}, JHEP \textbf{08} (2014)
026, \textcolor{purple}{\href{http://arxiv.org/abs/1403.4953}{\tt 1403.4953 [hep-ph]}}

\bibitem{Khoze:2016zfi}V. V. Khoze and A. D. Plascencia, \emph{Dark
Matter and Leptogenesis Linked by Classical Scale Invariance}, JHEP
\textbf{11} (2016) 025, \textcolor{purple}{\href{http://arxiv.org/abs/1605.06834}{\tt 1605.06834 [hep-ph]}}

\bibitem{Khoze:2023sth}V. V. Khoze and D. L. Milne, \emph{Gravitational
waves and dark matter from classical scale invariance}, Phys. Rev.
D \textbf{107} (2023) 095012, \textcolor{purple}{\href{https://arxiv.org/abs/2212.04784}{\tt 2212.04784 [hep-ph]}}

\bibitem{Antipin:2013exa}O. Antipin, M. Mojaza and F. Sannino, \emph{Conformal
Extensions of the Standard Model with Veltman Conditions}, Phys. Rev.
D \textbf{89} (2014) 085015, \textcolor{purple}{\href{http://arxiv.org/abs/1310.0957}{\tt 1310.0957 [hep-ph]}}

\bibitem{Davoudiasl:2014pya}H. Davoudiasl and I. M. Lewis, \emph{Right-Handed
Neutrinos as the Origin of the Electroweak Scale}, Phys. Rev. D \textbf{90}
(2014) 033003, \href{http://arxiv.org/abs/1404.6260}{\tt 1404.6260 [hep-ph]}

\bibitem{Chway:2013pta}D. Chway, T. H. Jung, H. D. Kim and R. Dermisek,
\emph{Radiative Electroweak Symmetry Breaking Model Perturbative All
the Way to the Planck Scale}, Phys. Rev. Lett. \textbf{113} (2014)
051801, \textcolor{purple}{\href{http://arxiv.org/abs/1308.0891}{\tt 1308.0891 [hep-ph]}}

\bibitem{Hashimoto:2013hta}M. Hashimoto, S. Iso and Y. Orikasa, \emph{Radiative
symmetry breaking at the Fermi scale and flat potential at the Planck
scale}, Phys. Rev. D \textbf{89} (2014) 016019, \textcolor{purple}{\href{http://arxiv.org/abs/1310.4304}{\tt 1310.4304 [hep-ph]}}

\bibitem{Kubo:2014ida}J. Kubo, K. S. Lim and M. Lindner, \emph{Gamma-ray
Line from Nambu-Goldstone Dark Matter in a Scale Invariant Extension
of the Standard Model}, JHEP \textbf{09} (2014) 016, \textcolor{purple}{\href{http://arxiv.org/abs/1405.1052}{\tt 1405.1052 [hep-ph]}}

\bibitem{Kubo:2014ova}J. Kubo, K. S. Lim and M. Lindner, \emph{Electroweak
Symmetry Breaking via QCD}, Phys. Rev. Lett. \textbf{113} (2014) 091604,
\textcolor{purple}{\href{http://arxiv.org/abs/1403.4262}{\tt 1403.4262 [hep-ph]}}

\bibitem{Kubo:2015cna}J. Kubo and M. Yamada, \emph{Genesis of electroweak
and dark matter scales from a bilinear scalar condensate}, Phys. Rev.
D \textbf{93} (2016) 075016, \textcolor{purple}{\href{http://arxiv.org/abs/1505.05971}{\tt 1505.05971 [hep-ph]}}

\bibitem{Aoki:2024kty}M. Aoki, J. Kubo and J. Yang, \emph{Scale Invariant
Extension of the Standard Model: A Nightmare Scenario in Cosmology},
JCAP \textbf{05} (2024) 096, \textcolor{purple}{\href{https://arxiv.org/abs/2401.12442}{\tt 2401.12442 [hep-ph]}}

\bibitem{Holthausen:2013ota}M. Holthausen, J. Kubo, K. S. Lim and
M. Lindner, \emph{Electroweak and Conformal Symmetry Breaking by a
Strongly Coupled Hidden Sector}, JHEP \textbf{12} (2013) 076, \textcolor{purple}{\href{http://arxiv.org/abs/1310.4423}{\tt 1310.4423 [hep-ph]}}

\bibitem{Lindner:2014oea}M. Lindner, S. Schmidt and J. Smirnov, \emph{Neutrino
Masses and Conformal Electro-Weak Symmetry Breaking}, JHEP \textbf{10}
(2014) 177, \textcolor{purple}{\href{http://arxiv.org/abs/1405.6204}{\tt 1405.6204 [hep-ph]}}

\bibitem{Radovcic:2014rea}S. Benic and B. Radovcic, \emph{Electroweak
breaking and Dark Matter from the common scale}, Phys. Lett. B \textbf{732}
(2014) 91-94, \textcolor{purple}{\href{http://arxiv.org/abs/1401.8183}{\tt 1401.8183 [hep-ph]}}

\bibitem{Benic:2014aga}S. Benic and B. Radovcic, \emph{Majorana dark
matter in a classically scale invariant model}, JHEP \textbf{01} (2015)
143, \textcolor{purple}{\href{http://arxiv.org/abs/1409.5776}{\tt 1409.5776 [hep-ph]}}

\bibitem{Altmannshofer:2015ppp}W. Altmannshofer, W. A. Bardeen, M.
Bauer, M. Carena and J. D. Lykken, \emph{Light Dark Matter, Naturalness,
and the Radiative Origin of the Electroweak Scale}, JHEP \textbf{01}
(2015) 032, \textcolor{purple}{\href{http://arxiv.org/abs/1505.00128}{\tt 1505.00128 [hep-ph]}}

\bibitem{Ametani:2015jla}Y. Ametani, M. Aoki, H. Goto and J. Kubo,
\emph{Nambu-Goldstone Dark Matter in a Scale Invariant Bright Hidden
Sector}, Phys. Rev. D \textbf{91} (2015) 115007, \textcolor{purple}{\href{http://arxiv.org/abs/1505.00128}{\tt 1505.00128 [hep-ph]}}

\bibitem{Carone:2015jra}C. D. Carone and R. Ramos, \emph{Dark chiral
symmetry breaking and the origin of the electroweak scale}, Phys.
Lett. B\textbf{ 746} (2015) 424-429, \textcolor{purple}{\href{http://arxiv.org/abs/1505.04448}{\tt 1505.04448 [hep-ph]}}

\bibitem{Das:2015nwk}A. Das, N. Okada and N. Papapietro, \emph{Electroweak
vacuum stability in classically conformal B-L extension of the Standard
Model}, Eur. Phys. J. C \textbf{77} (2017) 2, 122, \textcolor{purple}{\href{http://arxiv.org/abs/1509.01466}{\tt 1509.01466 [hep-ph]}}

\bibitem{Endo:2015ifa}K. Endo and Y. Sumino, \emph{A Scale-invariant
Higgs Sector and Structure of the Vacuum}, JHEP \textbf{05} (2015)
030, \textcolor{purple}{\href{http://arxiv.org/abs/1503.02819}{\tt 1503.02819 [hep-ph]}}

\bibitem{Endo:2015nba}K. Endo and K. Ishiwata, \emph{Direct detection
of singlet dark matter in classically scale-invariant standard model},
Phys. Lett. B \textbf{749} (2015) 583-588, \textcolor{purple}{\href{http://arxiv.org/abs/1507.01739}{\tt 1507.01739 [hep-ph]}}

\bibitem{Guo:2014bha}J. Guo and Z. Kang, \emph{Higgs Naturalness
and Dark Matter Stability by Scale Invariance}, Nucl. Phys. B\textbf{
898} (2015) 415-430, \textcolor{purple}{\href{http://arxiv.org/abs/1401.5609}{\tt 1401.5609 [hep-ph]}}

\bibitem{Guo:2015lxa}J. Guo, Z. Kang, P. Ko and Y. Orikasa, \emph{Accidental
dark matter: Case in the scale invariant local B-L model}, Phys. Rev.
D \textbf{91} (2015) 115017, \textcolor{purple}{\href{http://arxiv.org/abs/1502.00508}{\tt 1502.00508 [hep-ph]}}

\bibitem{Humbert:2015epa}P. Humbert, M. Lindner and J. Smirnov, \emph{The
Inverse Seesaw in Conformal Electro-Weak Symmetry Breaking and Phenomenological
Consequences}, JHEP \textbf{06} (2015) 035, \textcolor{purple}{\href{http://arxiv.org/abs/1503.03066}{\tt 1503.03066 [hep-ph]}}

\bibitem{Kang:2014cia}Z. Kang, \emph{Upgrading sterile neutrino dark
matter to FI$m$P using scale invariance}, Eur. Phys. J. C \textbf{75}
(2015) 471, \textcolor{purple}{\href{http://arxiv.org/abs/1411.2773}{\tt 1411.2773 [hep-ph]}}

\bibitem{Kang:2015aqa}Z. Kang, \emph{View FImP miracle (by scale
invariance) \`a la self-interaction}, Phys. Lett. B \textbf{751}
(2015) 201-204, \textcolor{purple}{\href{http://arxiv.org/abs/1505.06554}{\tt 1505.06554 [hep-ph]}}

\bibitem{Okada:2015gia}H. Okada, Y. Orikasa and K. Yagyu, \emph{Higgs
Triplet Model with Classically Conformal Invariance}, \textcolor{purple}{\href{http://arxiv.org/abs/1510.00799}{\tt 1510.00799 [hep-ph]}}

\bibitem{Pelaggi:2014wba}G. M. Pelaggi, \emph{Predictions of a model
of weak scale from dynamical breaking of scale invariance}, Nucl.
Phys. B\textbf{ 893} (2015) 443-458, \textcolor{purple}{\href{http://arxiv.org/abs/1406.4104}{\tt 1406.4104 [hep-ph]}}

\bibitem{Plascencia:2015xwa}A. D. Plascencia, \emph{Classical scale
invariance in the inert doublet model}, JHEP \textbf{09} (2015) 026,
\textcolor{purple}{\href{http://arxiv.org/abs/1507.04996}{\tt 1507.04996 [hep-ph]}}

\bibitem{Sannino:2015wka}F. Sannino and J. Virkaj\"arvi, \emph{First
Order Electroweak Phase Transition from (Non)Conformal Extensions
of the Standard Model}, Phys. Rev. D\textbf{ 92} (2015) 045015, \textcolor{purple}{\href{http://arxiv.org/abs/1505.05872}{\tt 1505.05872 [hep-ph]}}

\bibitem{Wang:2015sxe}Z.-W. Wang, F. S. Sage, T. G. Steele and R.
B. Mann, \emph{Asymptotic Safety in the Conformal Hidden Sector?},
J. Phys. G \textbf{45} (2018) 9, 095002, \textcolor{purple}{\href{http://arxiv.org/abs/1511.02531}{\tt 1511.02531 [hep-ph]}}

\bibitem{Ahriche:2015loa}A. Ahriche, K. L. McDonald and S. Nasri,
\emph{A Radiative Model for the Weak Scale and Neutrino Mass via Dark
Matter}, JHEP \textbf{02} (2016) 038, \textcolor{purple}{\href{http://arxiv.org/abs/1508.02607}{\tt 1508.02607 [hep-ph]}}

\bibitem{Ahriche:2016cio}A. Ahriche, K. L. McDonald and S. Nasri,
\emph{The Scale-Invariant Scotogenic Model}, JHEP \textbf{06} (2016)
182, \textcolor{purple}{\href{http://arxiv.org/abs/1604.05569}{\tt 1604.05569 [hep-ph]}}

\bibitem{Ahriche:2016ixu}A. Ahriche, A. Manning, K. L. McDonald and
S. Nasri, \emph{Scale-Invariant Models with One-Loop Neutrino Mass
and Dark Matter Candidates}, Phys. Rev. D \textbf{94}, 053005 (2016),
\textcolor{purple}{\href{http://arxiv.org/abs/1604.05995}{\tt 1604.05995 [hep-ph]}}

\bibitem{Ahriche:2022ngh}A. Ahriche, \emph{Purely Radiative Higgs
Mass in Scale invariant models}, Nucl. Phys. B \textbf{982} (2022)
115896, \textcolor{purple}{\href{https://arxiv.org/abs/2110.10301}{\tt 2110.10301 [hep-ph]}}

\bibitem{Das:2016zue}A. Das, S. Oda, N. Okada and D.-s. Takahashi,
\emph{Classically conformal U(1) extended standard model, electroweak
vacuum stability, and LHC Run-2 bounds}, \emph{Phys. Rev.} \textbf{D93}
(2016) 115038, \textcolor{purple}{\href{http://arxiv.org/abs/1605.01157}{\tt 1605.01157 [hep-ph]}}

\bibitem{Oda:2015gna}S. Oda, N. Okada and D.-s. Takahashi, \emph{Classically
conformal U(1) extended standard model and Higgs vacuum stability},
Phys. Rev. D\textbf{ 92} (2015) 015026, \textcolor{purple}{\href{http://arxiv.org/abs/1504.06291}{\tt 1504.06291 [hep-ph]}}

\bibitem{Ghorbani:2015xvz}K. Ghorbani and H. Ghorbani, \emph{Scalar
Dark Matter in Scale Invariant Standard Model}, JHEP \textbf{04} (2016)
024, \textcolor{purple}{\href{http://arxiv.org/abs/1511.08432}{\tt 1511.08432 [hep-ph]}}

\bibitem{Haba:2015lka}N. Haba, H. Ishida, N. Okada and Y. Yamaguchi,
\emph{Bosonic seesaw mechanism in a classically conformal extension
of the Standard Model}, Phys. Lett. B \textbf{754} (2016) 349-352,
\textcolor{purple}{\href{http://arxiv.org/abs/1508.06828}{\tt 1508.06828 [hep-ph]}}

\bibitem{Haba:2015nwl}N. Haba, H. Ishida, R. Takahashi and Y. Yamaguchi,
\emph{Gauge coupling unification in a classically scale invariant
model}, JHEP \textbf{02} (2016) 058, \textcolor{purple}{\href{http://arxiv.org/abs/1511.02107}{\tt 1511.02107 [hep-ph]}}

\bibitem{Haba:2015rha}N. Haba and Y. Yamaguchi, \emph{Vacuum stability
in the $U(1)_{\chi}$ extended model with vanishing scalar potential
at the Planck scale}, PTEP \textbf{2015} (2015) 093B05, \textcolor{purple}{\href{http://arxiv.org/abs/1504.05669}{\tt 1504.05669 [hep-ph]}}

\bibitem{Haba:2015qbz}N. Haba, H. Ishida, N. Kitazawa and Y. Yamaguchi,
\emph{A new dynamics of electroweak symmetry breaking with classically
scale invariance}, Phys. Lett. B \textbf{755} (2016) 439-443, \textcolor{purple}{\href{http://arxiv.org/abs/1512.05061}{\tt 1512.05061 [hep-ph]}}

\bibitem{Helmboldt:2016mpi}A. J. Helmboldt, P. Humbert, M. Lindner
and J. Smirnov, \emph{Minimal Conformal Extensions of the Higgs Sector},
JHEP \textbf{07} (2017) 113, \textcolor{purple}{\href{http://arxiv.org/abs/1603.03603}{\tt 1603.03603 [hep-ph]}}

\bibitem{Ishida:2016ogu}H. Ishida, S. Matsuzaki and Y. Yamaguchi,
\emph{Invisible Axion-Like Dark Matter from Electroweak Bosonic Seesaw},
Phys. Rev. D \textbf{94}, 095011 (2016), \textcolor{purple}{\href{http://arxiv.org/abs/1604.07712}{\tt 1604.07712 [hep-ph]}}

\bibitem{Jinno:2016knw}R. Jinno and M. Takimoto, \emph{Probing classically
conformal B--L model with gravitational waves}, Phys. Rev. D \textbf{95},
015020 (2017), \textcolor{purple}{\href{http://arxiv.org/abs/1604.05035}{\tt 1604.05035 [hep-ph]}}

\bibitem{Karam:2015jta}A. Karam and K. Tamvakis, \emph{Dark matter
and neutrino masses from a scale-invariant multi-Higgs portal}, Phys.
Rev. D\textbf{ 92} (2015) 075010, \textcolor{purple}{\href{http://arxiv.org/abs/1508.03031}{\tt 1508.03031 [hep-ph]}}

\bibitem{Karam:2016rsz}A. Karam and K. Tamvakis, \emph{Dark Matter
from a Classically Scale-Invariant $SU(3)_{X}$}, Phys. Rev. D \textbf{94}
(2016) 055004, \textcolor{purple}{\href{http://arxiv.org/abs/1607.01001}{\tt 1607.01001 [hep-ph]}}

\bibitem{Karam:2017rty}A. Karam, T. Pappas, and K. Tamvakis, \emph{Frame-dependence
of higher-order inflationary observables in scalar-tensor theories},
Phys. Rev. D \textbf{96}, 064036 (2017), \href{https://arxiv.org/abs/1707.00984}{\tt 1707.00984 [gr-qc]}

\bibitem{Marzola:2016xgb}L. Marzola and A. Racioppi, \emph{Minimal
but non-minimal inflation and electroweak symmetry breaking}, JCAP
\textbf{10} (2016) 010, \textcolor{purple}{\href{http://arxiv.org/abs/1606.06887}{\tt 1606.06887 [hep-ph]}}

\bibitem{Wang:2015cda}Z.-W. Wang, T. G. Steele, T. Hanif and R. B.
Mann, \emph{Conformal Complex Singlet Extension of the Standard Model:
Scenario for Dark Matter and a Second Higgs Boson}, JHEP \textbf{08}
(2016) 065, \textcolor{purple}{\href{http://arxiv.org/abs/1510.04321}{\tt 1510.04321 [hep-ph]}}

\bibitem{Wu:2016jdo}F. Wu, \emph{Aspects of a Non-minimal Conformal
Extension of the Standard Model}, Phys. Rev. D \textbf{94}, 055011
(2016), \textcolor{purple}{\href{http://arxiv.org/abs/1606.08112}{\tt 1606.08112 [hep-ph]}}

\bibitem{Hatanaka:2016rek}H. Hatanaka, D.-W. Jung and P. Ko, \emph{AdS/QCD
approach to the scale-invariant extension of the standard model with
a strongly interacting hidden sector}, JHEP \textbf{08} (2016) 094,
\textcolor{purple}{\href{http://arxiv.org/abs/1606.02969}{\tt 1606.02969 [hep-ph]}}

\bibitem{Minkowski:1977aj}P. Minkowski, \emph{On the Spontaneous
Origin of Newton's Constant}, Phys. Lett. B \textbf{71} (1977) 419-421

\bibitem{Zee:1978wi}A. Zee, \emph{A Broken Symmetric Theory of Gravity},
Phys. Rev. Lett. \textbf{42} (1979) 417

\bibitem{Smolin:1979uz}L. Smolin, \emph{Towards a Theory of Space-Time
Structure at Very Short Distances}, Nucl. Phys. B\textbf{ 160} (1979)
253-268

\bibitem{Adler:1980bx}S. L. Adler, \emph{Order R Vacuum Action Functional
in Scalar Free Unified Theories with Spontaneous Scale Breaking},
Phys. Rev. Lett. \textbf{44} (1980) 1567

\bibitem{Adler:2021fgr}S. L. Adler, \emph{Hubble parameter and related
formulas for a Weyl scaling invariant dark energy action}, Int. J.
Mod. Phys. D \textbf{30} (2021) 2150044, \textcolor{purple}{\href{https://arxiv.org/abs/2008.07598}{\tt 2008.07598 [astro-ph.CO]}}

\bibitem{Adler:2023thu}S. L. Adler, \emph{Solar system relativity
tests, formulas for light deflection by a central mass, and modification
of the lens equation, for a Weyl scaling invariant dark energy}, Gen.
Rel. Grav. \textbf{55} (2023) 1, \textcolor{purple}{\href{https://arxiv.org/abs/2204.09132}{\tt 2204.09132 [gr-qc]}}

\bibitem{Adler:2024dtn}S. L. Adler, \emph{Equation of state, and
atomic electron effective potential, for a Weyl scaling invariant
dark energy}, Phys. Rev. D \textbf{110}, 024051 (2024), \textcolor{purple}{\href{https://arxiv.org/abs/2209.14484}{\tt 2209.14484 [gr-qc]}}

\bibitem{Lin:2014mua}C. Lin, \emph{Large Hierarchy from Non-minimal
Coupling}, Commun. Theor. Phys. \textbf{68} (2017) 223-226, \textcolor{purple}{\href{http://arxiv.org/abs/1405.4821}{\tt 1405.4821 [hep-th]}}

\bibitem{Cooper:1982du}F. Cooper and G. Venturi, \emph{Cosmology
and Broken Scale Invariance}, Phys. Rev. D\textbf{ 24} (1981) 3338

\bibitem{Finelli:2007wb}F. Finelli, A. Tronconi and G. Venturi, \emph{Dark
Energy, Induced Gravity and Broken Scale Invariance}, Phys. Lett.
B\textbf{ 659} (2008) 466-470, \textcolor{purple}{\href{http://arxiv.org/abs/0710.2741}{\tt 0710.2741 [astro-ph]}}

\bibitem{Tronconi:2010pq}A. Tronconi and G. Venturi, \emph{Quantum
Back-Reaction in Scale Invariant Induced Gravity Inflation}, Phys.
Rev. D \textbf{84} (2011) 063517, \textcolor{purple}{\href{http://arxiv.org/abs/1011.3958}{\tt 1011.3958 [gr-qc]}}

\bibitem{Tronconi:2025thy}A. Tronconi and G. Venturi, \emph{Scale
Invariant Dark Energy}, \textcolor{purple}{\href{https://arxiv.org/abs/2502.08334}{\tt 2502.08334 [gr-qc]}}

\bibitem{Cerioni:2010ke}A. Cerioni, F. Finelli, A. Tronconi and G.
Venturi, \emph{Inflation and Reheating in Spontaneously Generated
Gravity}, Phys. Rev. D \textbf{81} (2010) 123505, \textcolor{purple}{\href{http://arxiv.org/abs/1005.0935}{\tt 1005.0935 [gr-qc]}}

\bibitem{Kamenshchik:2012rs}A. Y. Kamenshchik, A. Tronconi and G.
Venturi, \emph{Dynamical Dark Energy and Spontaneously Generated Gravity},
Phys. Lett. B\textbf{ 713} (2012) 358-364, \textcolor{purple}{\href{http://arxiv.org/abs/1204.2625}{\tt 1204.2625 [gr-qc]}}

\bibitem{Kaiser:2010ps}D. I. Kaiser, \emph{Conformal Transformations
with Multiple Scalar Fields}, Phys. Rev. D \textbf{81} (2010) 084044,
\textcolor{purple}{\href{http://arxiv.org/abs/1003.1159}{\tt 1003.1159 [gr-qc]}}

\bibitem{Callan:1970ze}C. G. Callan, Jr., S. R. Coleman and R. Jackiw,
\emph{A New improved energy - momentum tensor}, Annals Phys. \textbf{59}
(1970) 42-73

\bibitem{Polchinski:1987dy}J. Polchinski, \emph{Scale and Conformal
Invariance in Quantum Field Theory}, Nucl. Phys. B \textbf{303} (1988)
226-236

\bibitem{Coleman:1970je}S. R. Coleman and R. Jackiw, \emph{Why dilatation
generators do not generate dilatations?}, Annals Phys. \textbf{67}
(1971) 552-598

\bibitem{Fortin:2011sz}J.-F. Fortin, B. Grinstein and A. Stergiou,
\emph{Scale without Conformal Invariance: Theoretical Foundations},
JHEP \textbf{07} (2012) 025, \textcolor{purple}{\href{http://arxiv.org/abs/1107.3840}{\tt 1107.3840 [hep-th]}}

\bibitem{Jack:1990eb}I. Jack and H. Osborn, \emph{Analogs for the
$c$ Theorem for Four-dimensional Renormalizable Field Theories},
Nucl. Phys. B \textbf{343} (1990) 647-688

\bibitem{Gildener:1976ih}E. Gildener and S. Weinberg, \emph{Symmetry
Breaking and Scalar Bosons}, Phys. Rev. D\textbf{ 13} (1976) 3333

\bibitem{Rothman:1982lki}T. Rothman and R. Matzner, \emph{Scale-covariant
gravitation and primordial nucleosynthesis}, Astrophys. J. \textbf{257}
(1982) 450

\bibitem{Maitiniyazi:2025thg}Y. Maitiniyazi, S. Matsuzaki, K. Oda
and M. Yamada, \emph{Spacetime and Planck mass generation from scale-invariant
degenerate gravity}, Phys. Rev. D \textbf{111} (2025) 4, 046002, \textcolor{purple}{\href{https://arxiv.org/abs/2411.17238}{\tt 2411.17238 [hep-th]}}

\bibitem{Girmohanta:2024ijn}S. Girmohanta, Y. Nakai, Y-C. Qiu and
Z. Zhang, \emph{Wiggly dilaton: a landscape of spontaneously broken
scale invariance}, \textcolor{purple}{\href{https://arxiv.org/abs/2411.16304}{\tt 2411.16304 [hep-th]}}

\bibitem{Papadopoulos:2024jhg}G. Papadopoulos, \emph{Scale and Conformal
Invariance in Heterotic $\sigma$-Models}, \textcolor{purple}{\href{https://arxiv.org/abs/2409.01818}{\tt 2409.01818 [hep-th]}}

\bibitem{Frasca:2024pon}M. Frasca, A. Ghoshal and N. Okada, \emph{Non-perturbative
Origin of Electroweak Scale via Higgs-portal: Dyson-Schwinger in Conformally
Invariant Scalar Sector}, \textcolor{purple}{\href{https://arxiv.org/abs/2408.00093}{\tt 2408.00093 [hep-ph]}}

\bibitem{Maeder:2016kyn}A. Maeder, \emph{An alternative to the LCDM
model: the case of scale invariance}, Astrophys. J. \textbf{834},
194 (2016), \textcolor{purple}{\href{https://arxiv.org/abs/1701.03964}{\tt 1701.03964 [astro-ph.CO]}}

\bibitem{Maeder:2024fgd}A. Maeder, \emph{Observational tests in scale
invariance I: galaxy clusters and rotation of galaxies}, \textcolor{purple}{\href{https://arxiv.org/abs/2403.08759}{\tt 2403.08759 [astro-ph.GA]}}

\bibitem{Maeder:2024ngh}A. Maeder and F. Courbin, \emph{Observational
tests in scale invariance II: gravitational lensing}, \textcolor{purple}{\href{https://arxiv.org/abs/2403.08379}{\tt 2403.08379 [astro-ph.GA]}}

\bibitem{Maeder:2024kid}A. Maeder and F. Courbin, \emph{A Survey
of Dynamical and Gravitational Lensing Tests in Scale Invariance:
The Fall of Dark Matter?}, \textcolor{purple}{\href{https://arxiv.org/abs/2410.21379}{\tt 2410.21379 [astro-ph.CO]}}

\bibitem{Gueorguiev:2025njh}V. G. Gueorguiev and A. Maeder, \emph{Elucidating
the Dark Energy and Dark Matter Phenomena Within the Scale-Invariant
Vacuum (SIV) Paradigm}, Universe \textbf{2025}, 11(2) 48, \textcolor{purple}{\href{https://arxiv.org/abs/2502.02282}{\tt 2502.02282 [astro-ph.CO]}}

\bibitem{Banados:2024bnh}M. Ba\~nados, \emph{Gauging the scale invariance
of Einstein equations: Weyl invariant equations for gravity}, \textcolor{purple}{\href{https://arxiv.org/abs/2402.15675}{\tt 2402.15675 [gr-qc]}}

\bibitem{Farnsworth:2022guk}K. Farnsworth, K. Hinterbichler and O.
Hulik, \emph{Scale vs. Conformal Invariance at the IR Fixed Point
of Quantum Gravity}, Phys. Rev. D \textbf{105}, 066026 (2022), \textcolor{purple}{\href{https://arxiv.org/abs/2110.10160}{\tt 2110.10160 [hep-th]}}

\bibitem{Farnsworth:2024dft}K. Farnsworth, K. Hinterbichler and O.
Hulik, \emph{Scale and Conformal Invariance on (A)dS}, Phys. Rev.
D \textbf{110}, 045011 (2024), \textcolor{purple}{\href{https://arxiv.org/abs/2402.12430}{\tt 2402.12430 [hep-th]}}

\bibitem{Bertini:2024dfr}N. R. Bertini, D. C. Rodrigues and I. L.
Shapiro, \emph{Scale-dependent cosmology from effective quantum gravity
in the invariant framework}, Phys. Dark Univ. \textbf{45} (2024) 101502,
\textcolor{purple}{\href{https://arxiv.org/abs/2401.11559}{\tt 2401.11559 [gr-qc]}}

\bibitem{Wang:2023tfh}Q-Y. Wang, Y. Tang and Y-L. Wu, \emph{Inflation
in Weyl Scaling Invariant Gravity with $R^{3}$ Extensions}, Phys.
Rev. D \textbf{107} (2023) 083511, \textcolor{purple}{\href{https://arxiv.org/abs/2301.03744}{\tt 2301.03744 [astro-ph.CO]}}

\bibitem{Adak:2023vhd}M. Adak, N. Ozdemir and O. Sert, \emph{Scale
invariant Einstein-Cartan theory in three dimensions}, Eur. Phys.
J. C \textbf{83} (2023) 106, \textcolor{purple}{\href{https://arxiv.org/abs/2212.02917}{\tt 2212.02917 [gr-qc]}}

\bibitem{Boudet:2023hrd}S. Boudet, M. Rinaldi and S. M. Silveravalle,
\emph{On the stability of scale-invariant black holes}, JHEP \textbf{01}
(2023) 133, \textcolor{purple}{\href{https://arxiv.org/abs/2211.06110}{\tt 2211.06110 [gr-qc]}}

\bibitem{Ghoshal:2023jtv}A. Ghoshal, D. Mukherjee and M. Rinaldi,
\emph{Inflation and primordial gravitational waves in scale-invariant
quadratic gravity with Higgs}, JHEP \textbf{05} (2023) 023, \textcolor{purple}{\href{https://arxiv.org/abs/2205.06475}{\tt 2205.06475 [gr-qc]}}

\bibitem{DelCima:2022knr}O. M. Del Cima, D. H. T. Franco, L. S. Lima
and E. S. Miranda, \emph{The quantum scale invariance in graphene-like
quantum electrodynamics}, Phys. Lett. B \textbf{835} (2022) 137544,
\textcolor{purple}{\href{https://arxiv.org/abs/2209.10611}{\tt 2209.10611 [hep-th]}}

\bibitem{Ota:2023kuy}A. Ota, M. Sasaki and Y. Wang, \emph{Scale-invariant
enhancement of gravitational waves during inflation}, Mod. Phys. Lett.
A \textbf{38} (2023) 12n13, 2350063, \textcolor{purple}{\href{https://arxiv.org/abs/2209.02272}{\tt 2209.02272 [astro-ph.CO]}}

\bibitem{Dias:2022ent}A. G. Dias, J. Leite and B. L. S\'anchez-Vega,
\emph{Scale-invariant 3-3-1-1 model with B--L symmetry}, Phys. Rev.
D \textbf{106} (2022) 115008, \textcolor{purple}{\href{https://arxiv.org/abs/2207.06276}{\tt 2207.06276 [hep-ph]}}

\bibitem{Shimon:2021lkt}M. Shimon, \emph{Locally Scale-Invariant
Gravity}, \textcolor{purple}{\href{https://arxiv.org/abs/2108.11788}{\tt 2108.11788 [gr-qc]}}

\bibitem{Shimon:2022rth}M. Shimon, \emph{Cosmology in a locally scale
invariant gravity}, \textcolor{purple}{\href{https://arxiv.org/abs/2205.07251}{\tt 2205.07251 [gr-qc]}}

\bibitem{Borissova:2023dth}J. N. Borissova, A. Held and N. Afshordi,
\emph{Scale-invariance at the core of quantum black holes}, Class.
Quant. Grav. \textbf{40} (2023) 075011, \textcolor{purple}{\href{https://arxiv.org/abs/2203.02559}{\tt 2203.02559 [gr-qc]}}

\bibitem{Oda:2013uca}I. Oda, \emph{Higgs Mechanism in Scale-Invariant
Gravity}, Adv. Stud. Theor. Phys. \textbf{8} (2014) 215-249, \textcolor{purple}{\href{http://arxiv.org/abs/1308.4428}{\tt 1308.4428 [hep-ph]}}

\bibitem{Oda:2021lun}I. Oda, \emph{Scale Invariance and Dilaton Mass},
\textcolor{purple}{\href{https://arxiv.org/abs/2110.15408}{\tt 2110.15408 [hep-th]}}

\bibitem{Oda:2022mhu}I. Oda, \emph{Quantum Scale Invariant Gravity
with de Donder Gauge}, Phys. Rev. D \textbf{105} (2022) 066001, \textcolor{purple}{\href{https://arxiv.org/abs/2201.07354}{\tt 2201.07354 [hep-th]}}

\bibitem{Safari:2022kum}M. Safari, A. Stergiou, G. P. Vacca and O.
Zanusso, \emph{Scale and Conformal Invariance in Higher Derivative
Shift Symmetric Theories}, JHEP \textbf{02} (2022) 034, \textcolor{purple}{\href{https://arxiv.org/abs/2112.01084}{\tt 2112.01084 [hep-th]}}

\bibitem{Martins:2021olp}C. J. A. P. Martins, C. M. J. Marques, C.
B. D. Fernandes, J. S. J. S. Oliveira, D. A. R. Pinheiro and B. A.
R. Rocha, \emph{Alternatives to $\Lambda$: Torsion, Generalized Couplings,
and Scale Invariance}, MG16, 907-920, \textcolor{purple}{\href{https://arxiv.org/abs/2111.08086}{\tt 2111.08086 [astro-ph.CO]}}

\bibitem{Fernandes:2021adr}C. B. D. Fernandes, C. J. A. P. Martins
and B. A. R. Rocha, \emph{Constraining alternatives to a cosmological
constant: generalized couplings and scale invariance}, Phys. Dark
Univ. \textbf{31} (2021) 100761, \textcolor{purple}{\href{https://arxiv.org/abs/2012.10513}{\tt 2012.10513 [astro-ph.CO]}}

\bibitem{Braathen:2021ful}J. Braathen, S. Kanemura and M. Shimoda,
\emph{Two-loop analysis of classically scale-invariant models with
extended Higgs sectors}, JHEP \textbf{03} (2021) 297, \textcolor{purple}{\href{https://arxiv.org/abs/2011.07580}{\tt 2011.07580 [hep-ph]}}

\bibitem{Braathen:2022ngy}J. Braathen, S. Kanemura and M. Shimoda,
\emph{Two-loop corrections to the Higgs trilinear coupling in classically
scale-invariant theories}, PoS EPS-HEP2021 (2022) 605, \textcolor{purple}{\href{https://arxiv.org/abs/2110.11270}{\tt 2110.11270 [hep-ph]}}

\bibitem{Barman:2022htu}B. Barman and A. Ghoshal, \emph{Scale invariant
FIMP miracle}, JCAP \textbf{03} (2022) 003, \textcolor{purple}{\href{https://arxiv.org/abs/2109.03259}{\tt 2109.03259 [hep-ph]}}

\bibitem{vanDeBruck:2021juk}C. van de Bruck and R. Daniel, \emph{Inflation
and Scale-invariant $R^{2}$-Gravity}, Phys. Rev. D \textbf{103},
123506 (2021), \textcolor{purple}{\href{https://arxiv.org/abs/2102.11719}{\tt 2102.11719 [gr-qc]}}

\bibitem{Koivisto:2021rhy}T. Koivisto and L. Zheng, \emph{Scale-invariant
cosmology in de Sitter gauge theory}, Phys. Rev. D \textbf{103}, 124063
(2021), \textcolor{purple}{\href{https://arxiv.org/abs/2101.07638}{\tt 2101.07638 [gr-qc]}}

\bibitem{HerreroValea:2020dht}M. Herrero-Valea, \emph{A Path (Integral)
to Scale Invariance}, \textcolor{purple}{\href{https://arxiv.org/abs/2007.04335}{\tt 2007.04335 [hep-th]}}

\bibitem{Banik:2020lrt}I. Banik and P. Kroupa, \emph{Scale-invariant
dynamics in the Solar System}, Mon. Not. Roy. Astron. Soc. \textbf{497}
(2020) 1, L62-L66, \textcolor{purple}{\href{https://arxiv.org/abs/2007.00654}{\tt 2007.00654 [astro-ph.CO]}}

\bibitem{Tang:2020ktu}Y. Tang and Y-L. Wu, \emph{Weyl Scaling Invariant
$R^{2}$ Gravity for Inflation and Dark Matter}, Phys. Lett. B \textbf{809},
135716 (2020), \textcolor{purple}{\href{https://arxiv.org/abs/2006.02811}{\tt 2006.02811 [hep-ph]}}

\bibitem{Kugo:20220htd}T. Kugo, \emph{Necessity and Insufficiency
of Scale Invariance for solving Cosmological Constant Problem}, PoS
CORFU2019 (2020) 071, \textcolor{purple}{\href{https://arxiv.org/abs/2004.01868}{\tt 2004.01868 [hep-th]}}

\bibitem{Burrage:2018krt}C. Burrage, E. J. Copeland, P. Millington
and M. Spannowsky, \emph{Fifth forces, Higgs portals and broken scale
invariance}, JCAP \textbf{11} (2018) 036, \textcolor{purple}{\href{https://arxiv.org/abs/1804.07180}{\tt 1804.07180 [hep-th]}}

\bibitem{Banerjee:2018ngt}A. Banerjee, A. Kundu and A. Ray, \emph{Scale
invariance with fundamental matters and anomaly: A holographic description},
JHEP \textbf{06} (2018) 144, \textcolor{purple}{\href{https://arxiv.org/abs/1802.05069}{\tt 1802.05069 [hep-th]}}

\bibitem{Devecioglu:2018jut}D. O. Devecioglu, N. Ozdemir, M. Ozkan
and U. Zorba, \emph{Scale Invariance in Newton-Cartan and Ho\v{r}ava-Lifshitz
Gravity}, Class. Quant. Grav. \textbf{35} (2018) 115016, \textcolor{purple}{\href{https://arxiv.org/abs/1801.08726}{\tt 1801.08726 [hep-th]}}

\bibitem{Myung:2018fty}Y. S. Myung, \emph{Renormalizability and Newtonian
potential in scale-invariant gravity}, Int. J. Mod. Phys. D \textbf{27}
(2018) 12, 1850105, \textcolor{purple}{\href{https://arxiv.org/abs/1708.03451}{\tt 1708.03451 [gr-qc]}}

\bibitem{MannheimKazanas-1989}P. D. Mannheim and D. Kazanas, \emph{Exact
vacuum solution to conformal Weyl gravity and galactic rotation curves},
Astrophys. J. \textbf{342}, 635 (1989)

\bibitem{MannheimKazanas-1994}P. D. Mannheim and D. Kazanas, \emph{Newtonian
limit of conformal gravity and the lack of necessity of the second
order Poisson equation}, Gen. Rel. Grav. \textbf{26}, 337 (1994)

\bibitem{MannheimOBrien-2012}P. D. Mannheim and J. G. O'Brien, \emph{Fitting
galactic rotation curves with conformal gravity and a global quadratic
potential}, Phys. Rev. D \textbf{85}, 124020 (2012), \textcolor{purple}{\href{https://arxiv.org/abs/1011.3495}{\tt 1011.3495 [astro-ph.CO]}}

\bibitem{Kazanas-2021}D. Kazanas, D. Papadopoulos and D. Christodoulou,
\emph{Gravity Beyond Einstein? Yes, but in Which Direction?}, Phil.
Trans. R. Soc. A \textbf{380}, 0367 (2021), \textcolor{purple}{\href{https://arxiv.org/abs/2302.03001}{\tt 2302.03001 [gr-qc]}}

\bibitem{Burgess-2013}C. P. Burgess, \emph{The Cosmological Constant
Problem: Why it\textquoteright s hard to get Dark Energy from Micro-physics},
Post-Planck Cosmology: Lecture Notes of the Les Houches Summer School:
Volume \textbf{100}, July 2013, 149-197, \textcolor{purple}{\href{https://arxiv.org/abs/1309.4133}{\tt 1309.4133 [hep-th]}}

\bibitem{Padilla-2015}A. Padilla, \emph{Lectures on the Cosmological
Constant Problem}, \textcolor{purple}{\href{https://arxiv.org/abs/1502.05296}{\tt 1502.05296 [hep-th]}}

\bibitem{Martin-2012}J. Martin, \emph{Everything You Always Wanted
To Know About The Cosmological Constant Problem (But Were Afraid To
Ask)}, Comptes Rendus Physique \textbf{13} (2012) 566-665, \textcolor{purple}{\href{https://arxiv.org/abs/1205.3365}{\tt 1205.3365 [astro-ph.CO]}}

\bibitem{Weinberg-2000}S. Weinberg, \emph{The Cosmological Constant
Problems}, 4th International Symposium on Sources and Detection of
Dark Matter in the Universe (DM 2000), 18-26, \textcolor{purple}{\href{https://arxiv.org/abs/astro-ph/0005265}{\tt astro-ph/0005265}}

\bibitem{Csaki-2013}C. Cs\'aki and P. Tanedo, \emph{Beyond the Standard
Model}, 2013 European School of High-Energy Physics, pp.169-268, \textcolor{purple}{\href{https://arxiv.org/abs/1602.04228}{\tt 1602.04228 [hep-ph]}}

\bibitem{Matsas-2023}G. E. A. Matsas, V. Pleitez, A. Saa and D. A.
T. Vanzella, \emph{The number of fundamental constants from a spacetime-based
perspective}, Sci. Rep. \textbf{14} (2024) 1, 22594, \textcolor{purple}{\href{https://arxiv.org/abs/2311.09249}{\tt 2311.09249 [gr-qc]}}

\bibitem{Hartree-1928}D. R. Hartree, \emph{The Wave Mechanics of
an Atom with a Non-Coulomb Central Field. Part I. Theory and Methods},
Mathematical Proceedings of the Cambridge Philosophical Society (1928),
\textcolor{purple}{\href{https://scispace.com/pdf/the-wave-mechanics-of-an-atom-with-a-non-coulomb-central-3f3upgdftm.pdf}{https://scispace.com/pdf/the-wave-mechanics-of-an-atom-with-a-non-coulomb-central-3f3upgdftm.pdf}}
\end{thebibliography}
\end{document}